\documentclass{aa}

\usepackage{graphicx}
\usepackage{txfonts}
\usepackage{natbib}
\bibpunct{(}{)}{;}{a}{}{,} % to follow the A&A style
\usepackage{hyperref}
\hypersetup{
    colorlinks=true,
    citecolor=blue,
    linkcolor=blue,
    filecolor=magenta,
    urlcolor=blue,
}

\usepackage{xspace}

\defcitealias{2025Jankowski}{Paper~I}

\newcommand{\dmunit}{pc~$\text{cm}^{-3}$\xspace}

\newcommand{\mean}[1]{\left< #1 \right>\xspace}

\begin{document}

\title{Science Using Single-Pulse Exploration with Combined Telescopes}
\subtitle{II. Pulse profile evolution and single-pulse modulation}
\titlerunning{Science Using Single-Pulse Exploration with Combined Telescopes II}

\author{
F.~Jankowski\inst{1}\thanks{Corresponding author; \texttt{fabian.jankowski@cnrs-orleans.fr}}
\and
J.-M.~Grie{\ss}meier\inst{1,2}
\and
T.~Roy\inst{3}
\and
M.~Surnis\inst{4}
\and
G.~Theureau\inst{1,2}
\and
L.~Bondonneau\inst{2}
\and
P.~No\'e\inst{1}
\and
J.~P\'etri\inst{5}
\and
L.~Guillemot\inst{1,2}
\and
I.~Cognard\inst{1,2}
\and
M.~Serylak\inst{6}
\and
I.~Kravtsov\inst{7,1}
\and
V.~Zakharenko\inst{7}
\and
O.~Ulyanov\inst{7}
}

\institute{
LPC2E, OSUC, Univ Orleans, CNRS, CNES, Observatoire de Paris, F-45071 Orleans, France
\and
Observatoire Radioastronomique de Nan\c{c}ay, Observatoire de Paris, Universit\'e PSL, Université d'Orl\'eans, CNRS, 18330 Nan\c{c}ay, France
\and
Nicolaus Copernicus Astronomical Center, Rabianska 8, Torun 87-100, Poland
\and
Department of Physics, IISER Bhopal, Bhauri Bypass Road, Bhopal, 462066, India
\and
Universit\'e de Strasbourg, CNRS, Observatoire astronomique de Strasbourg, UMR 7550, F-67000 Strasbourg, France
\and
SKA Observatory, Jodrell Bank, Lower Withington, Macclesfield SK11 9FT, UK
\and
Institute of Radio Astronomy of NAS of Ukraine, 4 Mystetstv St., 61002, Kharkiv, Ukraine
}

\date{Received XXX; accepted XXX}

% \abstract{}{}{}{}{} 
% 5 {} token are mandatory
 
\abstract
{
The details of the pulsars' radio emission remain uncertain. The pulse profiles are the most readily available observables, whose morphology reflects the pulsar geometry, emission altitude, and magnetospheric configuration.
}
{
We aim to illuminate how the pulsars' integrated pulse profiles and single-pulse modulation evolve with radio frequency, while separating intrinsic and propagation effects.
}
{
We present integrated pulse profiles and phase-resolved modulation indices for 12 radio pulsars at up to five frequency bands. Our wideband dataset was systematically acquired with the Nan\c{c}ay Radio Observatory telescopes and the uGMRT, covering the frequency range 10~MHz to 2.8~GHz, including NenuFAR, LOFAR FR606, uGMRT, and NRT observations. We carefully measured the pulse widths and scattering times using multi-component profile fitting. We employed a profile width scaling model to separate the profile evolution into scattering and intrinsic contributions. We quantified the pulsars' single-pulse variability across frequency.
}
{ 
We describe the pulse profile evolution. Two pulsars have scattering indices compatible with Kolmogorov turbulence, while the majority have flatter scattering indices between $-3$ and $-0.5$. This suggests complex scattering environments, dominated by localised turbulence. Complex scattering behaviour is seen in some pulsars, and most profiles are scattering-dominated below 200~MHz. The measured intrinsic power law indices are small, with absolute values between 0 and 0.3. Five pulsars exhibit decreasing intrinsic widths, two or three pulsars show flat behaviour, and four pulsars have increasing profile widths with frequency. Accounting for biases, most pulsar modulation parameters decrease with frequency, while three pulsars show flat behaviour.
}
{
The pulse width evolution is explained by intrinsic radio beam narrowing, the emergence of profile components, and the balance between the two. Our modulation analysis indicates that the single-pulse emission from most of our pulsars becomes more erratic towards lower radio frequencies, consistent with increasing amplitude modulation.
}


\maketitle
\nolinenumbers
%
%-------------------------------------------------------------------

%
% Intoduction
%

\section{Introduction}
\label{sec:introduction}

In contemporary astrophysics, the radio pulsars’ frequency evolution is recognised as an important problem, posing challenges for both observers and theorists in understanding their complex profile morphology. \citet{1970Komesaroff} first studied the evolution of the pulse profiles with frequency. The profile widths and component separation of canonical pulsars typically decrease with increasing observing frequency, which is believed to be due to emission originating from different heights in the pulsar magnetosphere, a paradigm commonly known as radius-to-frequency mapping (RFM; \citealt{1970Komesaroff, 1978Cordes}). PSR~B1133+16 is a good example. However, such frequency evolution is not well pronounced for millisecond pulsars such as PSR~J2145$-$0750, as their emission regions are comparatively narrow. The pulse component separation is commonly interpreted within the empirical RFM framework proposed by \citet{1991Thorsett}, which describes a systematic frequency dependence on emission altitude. In this model, the component separation or overall profile width decreases systematically with increasing observing frequency and asymptotically approaches a minimum value at the highest frequencies. This behaviour is commonly parametrised as $W_{\nu} = A \nu^{-\alpha} + W_\text{min}$, where $W_{\nu}$ denotes the profile width (or component separation) at frequency $\nu$, $A$ is a normalization constant, $\alpha$ is the power law scaling index, and $W_\text{min}$ is the limiting high-frequency width. For most canonical pulsars, $\alpha$ typically lies in the range 0.1 to 0.5, indicating a gradual narrowing of the pulse profile with increasing frequency, while $W_\text{min}$ is generally interpreted as the intrinsic beam width associated with emission originating close to the stellar surface or from the innermost accessible field lines. The model provides a simple phenomenological realisation of RFM, in which lower-frequency radiation is emitted at higher magnetospheric altitudes and therefore subtends a wider angular region owing to the divergence of dipolar magnetic field lines. Its principal success is the ability to reproduce the smooth broadband profile narrowing observed in many pulsars. However, the formalism remains empirical and does not constrain the underlying emission mechanism. Deviations from simple power law behaviour in some pulsars further indicate that propagation effects, non-dipolar magnetic geometry, patchy emission structure, or multiple emission altitudes may also influence the observed profile evolution \citep{1991Thorsett}. More importantly, the \citeauthor{1991Thorsett} model does not account for profile broadening due to scattering, which is crucial at low frequencies.

Some recent profile evolution studies include \citet{2012Hassall} (15~MHz--8~GHz), \citet{2016Pilia} (15--167~MHz), \citet{2019Olszanski} (327~MHz--4.6~GHz), and \citet{2021Posselt} (897--1672~MHz). Additionally, several authors have used the pulse profiles from the European Pulsar Network (EPN) database\footnote{\url{https://psrweb.jb.man.ac.uk/epndb/}} for their profile evolution analyses \citep{2014Chen, 2021Xu, 2024Vohl}, or in combination with their own data \citep{2019Zhao}. While these studies are useful in their own right and have broad frequency coverage, the heterogeneous EPN dataset contains unknown and complex data processing systematics. Motivated by this, we systematically acquired a wideband dataset (10~MHz to 2.8~GHz). The advantage is that we entirely control the data processing and that our NenuFAR, LOFAR FR606, and NRT data were recorded with the same family of well-tested pulsar backends, which share a common codebase. A further novelty of our study is that we include low-frequency data below 100~MHz and that we specifically measure and account for scattering.

Aside from the integrated pulse profiles, the single-pulse modulation and its frequency dependence characterise the pulsar radio emission mechanism. This is because pulsar radiation resembles amplitude-modulated noise consisting of nano-shots, most prominently observed in the Crab pulsar. For this, the phase-resolved modulation index $m$ is an important tool, defined as the ratio of the standard deviation to the mean total intensity per phase bin. It measures the degree of single-pulse modulation across the pulsar's pulse profile and reveals pulse phases of exceptional behaviour. Catalogues of modulation index curves have been presented as part of drifting subpulse or pulse-energy studies \citep{2006Weltevrede, 2012Burke-Spolaor, 2023Song}. However, they are typically estimated at a single observing frequency, which prevents a systematic study of their frequency evolution. Our multi-frequency dataset presented here aims to resolve this issue.

In this paper, we analyse the pulse profile morphology and the phase-resolved single-pulse modulation, focusing on its frequency dependence. The paper is structured as follows. In Sect.~\ref{sec:observations}, we describe our observations, the resulting dataset, our data processing, and our phase alignment method. In Sect.~\ref{sec:results}, we describe our profile modelling, present the updated pulsar DMs, the integrated pulse profiles, and analyse the profile evolution with frequency, separated into scattering and intrinsic width evolution. We then present the pulsars' single-pulse modulation properties and their frequency-dependent changes, investigating S/N bias and the influence of interstellar scintillation. We discuss our results in Sect.~\ref{sec:discussion} and compare them with the literature. Finally, we give our conclusion in Sect.~\ref{sec:conclusions}. Appendix~\ref{ap:observations} contains a table with the details of our observational data, Appendix~\ref{ap:profiles} presents the band-integrated pulse profiles at up to five frequency bands, and Appendix~\ref{ap:widthestimators} discusses several pulse width estimators, introducing a new one to pulsar astrophysics.

%
% Observations
%

\section{Observations and data processing}
\label{sec:observations}

The data presented in this work were obtained as part of the `Science Using Single-Pulse Exploration with Combined Telescopes' (SUSPECT) project\footnote{\url{https://suspectproject.com}} \citep{2025Jankowski}, which is a new multi-telescope and multi-frequency pulsar observing programme running at the Nan\c{c}ay Radio Observatory (ORN) telescopes in France and the upgraded GMRT in India. More information about the project can be found in the SUSPECT overview paper (\citealt{2025Jankowski}; hereafter \citetalias{2025Jankowski}). While \citetalias{2025Jankowski} only included uGMRT data, this work is the first to present a larger fraction of our wideband dataset.

\subsection{uGMRT data}

We obtained the data as part of the uGMRT observing projects `Understanding the wide-band single-pulse properties of bright radio pulsars with the upgraded GMRT' (project codes 44\_056, 45\_029, 46\_064, and 47\_065; PI: Jankowski) over several observing sessions in 2023, 2024, and early 2025. We performed Band-4 observation with the GMRT Wideband Backend (GWB), recording Stokes I total-intensity data over 200~MHz of digitised bandwidth between 550 and 750~MHz. We saved 8-bit phased array (PA) mode data with a sampling time of $81.92~\mu\text{s}$ and 2048 frequency channels. For the higher DM pulsars, we saved 16-bit coherently dedispersed (CD) mode data together with the PA data. Those data have 1024 frequency channels. For the two fastest spinning pulsars, we reduced the sampling time to $40.96~\mu\text{s}$ and the number of frequency channels to 512. Table~\ref{tab:observations} presents the details of our observations. Our data reduction follows the methodology described in \citetalias{2025Jankowski}, which we summarise here. We converted the raw uGMRT data to \texttt{SIGPROC} filterbank files \citep{2011Lorimer} using the \texttt{rficlean} software \citep{2021Maan}, which we also used for a first level of RFI excision. We then created single-pulse integrations using \texttt{DSPSR} \citep{2011VanStraten}. We used the best-available pulsar ephemerides, updating the DMs before folding based on our low-frequency NenuFAR observations. After folding, we applied standard data cleaning and post-processing tasks using the \texttt{PSRCHIVE} software suite \citep{2004Hotan}. We carried out all further analyses using a custom \texttt{Python}-based software called \texttt{spanalysis} in version 0.6.0.

\subsection{NenuFAR data}

We observed our pulsar sample with the New Extension in Nan\c{c}ay Upgrading LOFAR (NenuFAR) telescope \citep{2020Zarka}, either simultaneously or within close time intervals of the uGMRT observations. We recorded real-time folded and single-pulse data using the Low-frequency Ultimate Pulsar Processing Instrumentation (LUPPI) backend \citep{2021Bondonneau}, covering a total digitised bandwidth of 75~MHz, allocated in two 37.5~MHz frequency lanes between 10 and 85 MHz. The data were coherently dedispersed at the best-known pulsar DMs and coherently Faraday de-rotated at their best-known RMs. We slightly varied the minimum recorded frequency to account for instrumental processing constraints. As a consequence, the centre frequencies of the data differ slightly between pulsars (Table~\ref{tab:observations}). The data have $2 \times 192$ frequency channels, contain full polarisation information, and are stored as 32-bit floating-point numbers in \texttt{PSRFITS} format. The single-pulse data were recorded at a sampling time of $327.68~\mu\text{s}$. Our data processing consisted of merging the frequency lanes, flattening the bandpass, and converting the data to 8-bit integers using a custom \texttt{Python} software \citep{2021Bondonneau}. We then created single-pulse archive files using \texttt{DSPSR} and the same pulsar ephemerides as used for the uGMRT data, including the updated pulsar DMs. We performed standard RFI cleaning and post-processing tasks using \texttt{PSRCHIVE} and all further data analysis using \texttt{spanalysis}.

\subsection{LOFAR FR606 data}

Simultaneously with the NenuFAR pointings, we observed our pulsar sample with the French LOw-Frequency ARray (LOFAR; \citealt{2013VanHaarlem}) station FR606, which is located within NenuFAR's dense core at the Nan\c{c}ay Radio Observatory (ORN). The data were obtained with the high-band antennas (HBA) in single-station mode. We recorded a frequency band between $\sim$102 and 198~MHz, with a digitised bandwidth of $\sim$95~MHz, split across 488 frequency channels in full polarisation. Observations before 2023 December were recorded as baseband data and were offline converted to \texttt{SIGPROC} filterbank files stored as 8-bit integers using \texttt{PSRCHIVE}'s \texttt{digifil} utility. We time-aligned and spliced the four frequency lanes together using our custom \texttt{searchtools} software\footnote{\url{https://github.com/fjankowsk/searchtools}} in version 0.4.0. Data after 2023 December were recorded using the `allegro-ng' backend, a clone of NenuFAR's LUPPI real-time backend, adjusted for the higher-frequency LOFAR data. The resulting single-pulse data were coherently dedispersed at the best-known pulsar DMs and the data taken after 2023 December were additionally coherently Faraday de-rotated using their known RMs. The data are stored as 32-bit floating-point numbers in \texttt{PSRFITS} format, as in the NenuFAR case. We chose sampling times of 327.68 or $655.36~\mu\text{s}$ for the pulsars in this work (Table~\ref{tab:observations}). Our data processing closely matched that of NenuFAR described above.

\subsection{NRT data}

We recorded search mode data of three of our pulsars with the Nan\c{c}ay Radio Telescope (NRT) in early 2026. The data were obtained with the L-band receiver and the NUPPI backend, covering a bandwidth of 512~MHz centred at 1484~MHz split into $128 \times 4$~MHz channels in full Stokes polarisation. We used sampling times of 64 and 124~$\mu \text{s}$ (Table~\ref{tab:observations}). The data were coherently dedispersed. We first spliced the eight frequency lanes together using the \texttt{searchtools} software and converted the 32-bit data to 8-bit integers as before. After single pulse folding using \texttt{DSPSR}, we applied similar RFI excision methods to the data and performed all further data analysis using \texttt{spanalysis}.

We included additional archival NRT fold-mode data obtained at L-band (1.5~GHz) and S-band (2.5~GHz) for some of our pulsars (Grie{\ss}meier \& Guillemot, private communication). Those were obtained earlier and were solely used for our pulse profile evolution analysis. Those data were coherently dedispersed, covering 512~MHz of bandwidth across 128 frequency channels with 2048 phase bins.

\subsection{Phase aligning the multi-frequency dataset}

We folded all single-pulse data for a given pulsar using the same ephemeris, including the updated epoch-wise DMs presented below. Pulsar-specific instrumental offsets (phase jumps) were determined using standard pulsar timing techniques as part of our data preparation. However, precise absolute phase alignment between observing bands was not required for this work; relative profile alignment was sufficient. Thus, we fine-aligned the profiles automatically via the \texttt{spanalysis} software based on their amplitude-weighted mean (centroid or centre of flux) pulse phases as fiducial points, defined as
\begin{equation}
    \phi_m = \phi_\text{cof} = \sum_i \phi_i I_i \: / \: \sum_i I_i,
    \label{eq:meanphase}
\end{equation}
where $\phi_i$ is the pulse phase and $I_i$ is the total intensity profile amplitude at phase bin $i$. We rectified the profile data and truncated them at 5--10~\% amplitude to remove noise bias.

%
% Results
%

\section{Results}
\label{sec:results}

\subsection{Updated dispersion measures and profile modelling}

\begin{figure}
  \centering
  \includegraphics[width=\columnwidth]{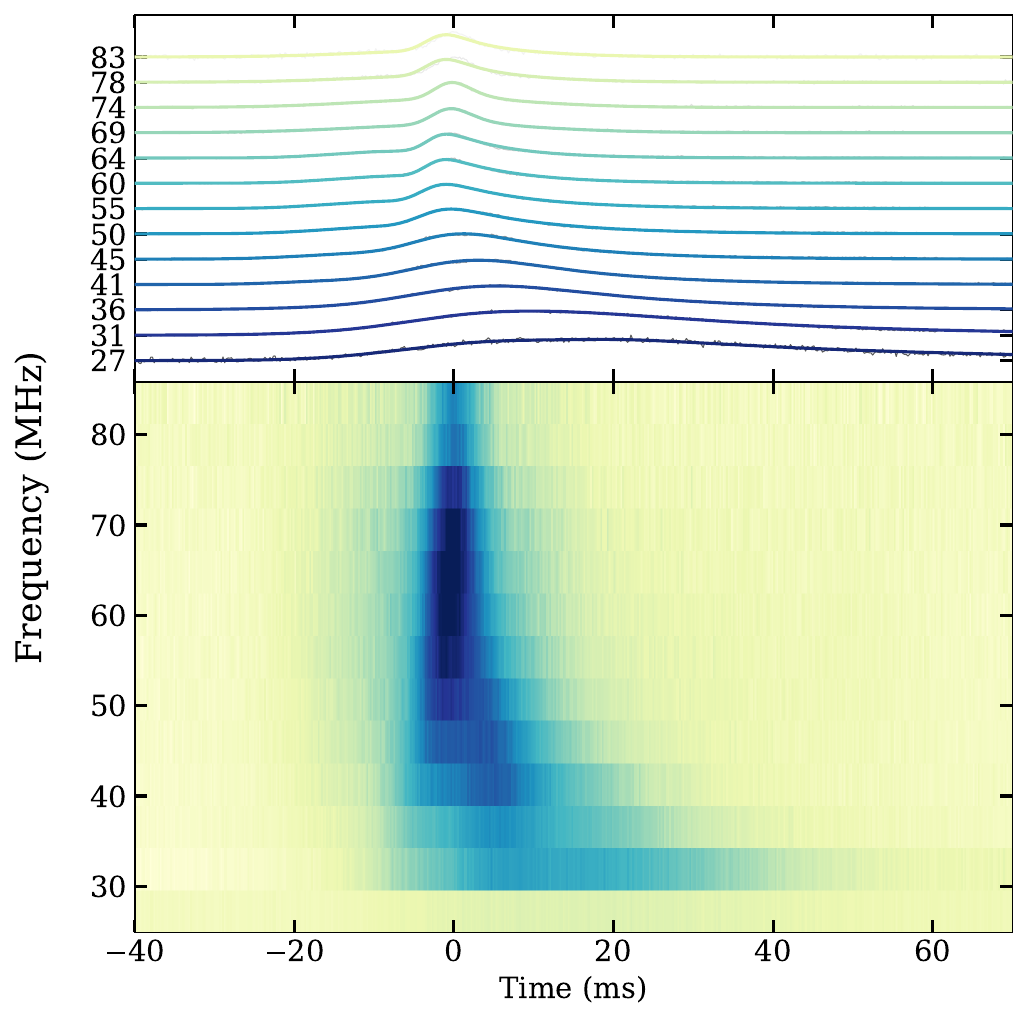}
  \caption{Example plot using our NenuFAR data of PSR~B0919+06, which demonstrates our profile modelling and scattering fit analysis that we used to measure the pulse widths and determine updated scattering-corrected DMs. The panels show the sub-banded pulse profiles with their best-fitting scattering models overlaid (top) and the dedispersed dynamic spectrum (bottom). Clear scatter broadening and intrinsic profile evolution are visible.}
 \label{fig:scatfit}
\end{figure}

\begin{table}
\caption{Scattering-corrected DMs measured from our NenuFAR data.}
\label{tab:dms}
\begin{tabular}{lcccc}
\hline\hline
PSR                         & $N_\text{p}$      & $N_\text{c}$      & DM                        & Date\\
                            &                   &                   & (\dmunit)                 & (yyyy-mm-dd)\\
\hline
\object{B0031$-$07}         & 2                 & 2--3              & 10.878(4)                 & 2023-12-04\\
\object{B0329+54}           & 3                 & 3--5              & 26.7558(2)                & 2023-04-26\\
B0329+54                    & 3                 & 3--5              & 26.7568(1)                & 2025-11-17\\
\object{B0809+74}           & 2                 & 3                 & 5.7507(5)                 & 2023-04-25\\
\object{B0823+26}           & 2                 & 3--4              & 19.476(1)                 & 2023-05-05\\
B0823+26                    & 2                 & 3--4              & 19.4798(1)                & 2023-12-05\\
\object{B0834+06}           & 2                 & 2--3              & 12.8612(2)                & 2023-04-25\\
\object{B0919+06}           & 1                 & 2--3              & 27.2953(1)                & 2024-04-23\\
\object{B0943+10}           & 2                 & 1--2              & 15.3296(1)                & 2023-12-05\\
\object{B0950+08}           & 3                 & 3--4              & 2.970(1)                  & 2024-05-05\\
B0950+08                    & 3                 & 3--4              & 2.9693(3)                 & 2025-11-18\\
B0950+08                    & 3                 & 3--4              & 2.9708(3)                 & 2026-03-12\\
\object{B1112+50}           & 2                 & 2--3              & 9.1838(3)                 & 2023-12-05\\
B1112+50                    & 2                 & 2--3              & 9.1843(3)                 & 2026-04-05\\
\object{B1133+16}           & 2                 & 3                 & 4.8450(1)                 & 2023-04-25\\
\object{B1237+25}           & 3                 & 3--6              & 9.245(1)                  & 2023-12-05\\
B1237+25                    & 3                 & 3--6              & 9.243(1)                  & 2026-02-26\\
\object{B1822$-$09}         & 2                 & 2--4              & 19.3772(3)                & 2023-04-27\\
B1822$-$09                  & 2                 & 2--4              & 19.3748(3)                & 2023-12-05\\
\hline
\end{tabular}
\tablefoot{
We list the number of pulse profile peaks ($N_\text{p}$), the number of individual profile components used in our \texttt{scatfit} analysis ($N_\text{c}$), the best-fitting DM, and the DM measurement epoch.
}
\end{table}

We used the \texttt{scatfit}\footnote{\url{https://github.com/fjankowsk/scatfit}} scattering fit and pulse profile modelling software \citep{2022Jankowski, 2023Jankowski} in version 0.5.2 on our NenuFAR data to determine updated scattering-corrected pulsar DMs before single-pulse folding the data. Estimating updated DMs is an important preparation step, as our low-frequency data are highly sensitive to subtle changes in the observed DMs. For instance, our NenuFAR data easily resolve DM offsets of $\mathcal{O}(10^{-4}) \: \text{pc} \: \text{cm}^{-3}$. Temporal DM changes of at least this magnitude are common in most pulsars because of the motion of the pulsar-Earth sightline across the ionised interstellar medium (ISM; secular DM trends), changes in electron content of localised environments along the line of sight (LOS; e.g.\ pulsar wind nebulae, supernova remnants, HII regions), or the solar wind for pulsars close to the ecliptic.

We used the cleaned real-time folded NenuFAR data for this analysis. For bright pulsars, we split the 75~MHz bandwidth into 8--13 frequency sub-bands of 6--9~MHz each. We implemented a new `band-integrated' profile model in \texttt{scatfit}, which we describe below. This was important because the profile evolution within a sub-band is significant at those low frequencies. That is, the narrow bandwidth approximation fails - the profile changes appreciably within a given frequency sub-band. Our new band-integrated model computes the scattered pulse profiles at typically $N_\text{freq} = 6$ centre frequencies\footnote{Chosen so that there is one sub-profile per MHz of bandwidth.} spaced geometrically within a given sub-band and evolved with frequency $\nu$ in terms of scattering time ($\propto \nu^{-4}$; scattering index) and fluence ($\propto \nu^{-1.5}$; spectral index). The total band-integrated model in a sub-band is the superposition of the individual scattered sub-profiles. We computed it as the fluence-normalised weighted sum across the frequency grid $\left\{ \nu_i \right\}$ within a sub-band centred at $\nu_\text{c}$ as
\begin{equation}
    \bar{f} \left( t, \vec{a} \right) = A \: \sum_{i=0}^{N_\text{freq}} f_i \left( t, F_i, \tau_{\text{s},i}, \vec{b} \right),
    \label{eq:bandintegrated}
\end{equation}
where $A = F_c / \left( dt \sum_t \sum_i f_i \right)$ is the fluence normalisation constant, $dt$ is the phase bin width, $F_i = F_c \left( \nu_i / \nu_c \right)^{-1.5}$ is the $i$-th fluence, $\tau_{\text{s},i} = \tau_\text{s} \left( \nu_i / \nu_c \right)^{-4}$ is the $i$-th scattering time, $\vec{a}$ and $\vec{b}$ are the parameter vectors, and $f_i$ are exponentially modified Gaussian functions as given by Eq.~5 in \citet{2023Jankowski}. The subscript $i$ denotes frequency grid parameters, while the subscript $c$ indicates values measured at the sub-band centre frequency. The band-integrated model super-samples and forward models the profile evolution within a given frequency sub-band, for iterative comparison with the observed profile data during the fitting process. It thereby avoids problems with low-S/N observed profiles that can occur in narrow sub-bands.

Another major improvement in \texttt{scatfit} version 0.5.0 is the addition of a genuine multi-component mode, which lets us reliably model complex multi-modal (double-, triple-, \ldots peaked) pulse profiles consisting of several Gaussian sub-components. In particular, we used the existing analytical expression for a single scattered Gaussian pulse in the thin screen scattering regime (Eq.~5 in \citealt{2023Jankowski}). The convolution operation with a pulse broadening function (PBF) is distributive, meaning that $h * (f + g) = (h * f) + (h * g)$, where $*$ denotes the convolution operator. Here, $h$ is the PBF (e.g. thin screen scattering), and $f$ and $g$ are the independent pulse profile components. More generally,
\begin{equation}
    f_\text{tot} = h * \left( \sum_{i=0}^N f_i \right) = \sum_{i=0}^N (h * f_i) = \sum_{i=0}^N (f_i)_a,
\end{equation}
where $f_\text{tot}$ is the total observed pulse profile after scattering, and the subscript $a$ denotes the analytical expression from above. Thus, the total scattered profile is the superposition of the individual scattered profile components. In practice, we fixed the scattering time parameter $\tau_\text{s}$ for all components to that of the first component (meaning a single scattering screen affecting all components) and allowed only the first component to have a non-zero DC (baseline mean) offset. We also mildly enforced that the component locations increased in pulse phase or time (order prior), which helped the fits converge.

The DMs were estimated separately for each profile component, where we used the mean or error-weighted mean as the total DM measurement. We list the resulting updated pulsar DMs and their corresponding measurement epochs in Table~\ref{tab:dms}. An example fit is shown in Fig.~\ref{fig:scatfit}.

%
% Pulse profiles
%

\subsection{Integrated pulse profile morphology}

After single-pulse folding, RFI excision, and phase alignment, we examined the pulsars' integrated pulse profiles in each frequency band, which are shown in Figs.~\ref{fig:profiles1}, \ref{fig:profiles2}, \ref{fig:profiles3}, \ref{fig:profiles4}, and \ref{fig:profiles5}. The display format is the same as in \citetalias{2025Jankowski}. The profiles are centred on the amplitude-weighted mean phase (centre of flux). We show zoom-ins of the main pulse (MP) phase ranges, with the visually chosen phase gates displayed as coloured vertical lines. For pulsars with several distinct profile components, we label them in the plots. Several pulsars have low-amplitude interpulses (IPs), which we ignore here for clarity. We studied PSR~B1822$-$09's IP characteristics extensively in \citetalias{2025Jankowski}. Additionally, we show separate plots of PSR~B0823+26's bright B-mode (2023-12-05; Fig.~\ref{fig:profiles1}) and quiet Q-mode emission (2023-04-25; Fig.~\ref{fig:profiles2}), as the pulsar was in its Q-mode for the entire observing duration on 2023-04-25.

\subsection{Profile evolution with frequency}

\begin{table*}
\caption{Boxcar-equivalent pulse widths measured from the band-integrated profiles of the pulsars in our dataset.}
\label{tab:pulsewidths}
\begin{tabular}{lcccccc}
\hline\hline
             & \multicolumn{5}{c}{This work}    & \multicolumn{1}{c}{Literature}\\
PSR          & NenuFAR          & FR606         & GMRT              & NRT L         & NRT S         & G\&L98\\
             & 50 MHz           & 150 MHz       & 650 MHz           & 1.5 GHz       & 2.5 GHz       & 408 MHz\\
             & $W_\text{eq}$    & $W_\text{eq}$ & $W_\text{eq}$     & $W_\text{eq}$ & $W_\text{eq}$ & $W_\text{eq}$\\
             & (ms)             & (ms)          & (ms)              & (ms)          & (ms)          & (ms)\\
\hline
B0031$-$07   & 123.1(8)         & 78.0(5)       & 67.8(1)           & 60.7(2)       & 44.9(4)       & 60.1(5)\\
B0329+54     & 25.6(2)          & 8.0(1)        & 15.0(1)           & 15.6(1)       & 17.0(1)       & 9.3(3)\\
B0809+74     & 124.5(6)         & 56.3(4)       & 56.1(2)           & 58.6(2)       & --            & 48.7(3)\\
B0823+26     & 21.9(1)          & 11.2(1)       & 6.5(1)            & 5.1(1)        & 5.0(1)        & 7.0(1)\\
B0823+26 Q   & --               & --            & 7.5(2)            & --            & --            & --\\
B0834+06     & 19.0(1)          & 15.8(1)       & 17.0(3)           & 18.1(1)       & --            & 19.3(2)\\
B0919+06     & 23.7(2)          & 10.9(2)       & 9.0(1)            & 6.5(1)        & 6.9(3)\tablefootmark{a}   & 13.1(3)\\
B0943+10     & 37.8(2)          & 35.8(5)       & 45(1)\tablefootmark{a} & --    & --    & 40(4)\tablefootmark{a}\\
B0950+08     & 25.0(1)          & 15.3(1)       & 10.3(1)           & 11.5(1)       & 10.9(1)       & 11.9(3)\\
B1112+50     & 33.9(4)          & 20.5(1)       & 18.1(1)           & 20.6(2)\tablefootmark{a}& --  & 18.3(3)\\
B1133+16     & 19.2(2)          & 21.3(2)       & 16.7(1)           & 11.3(1)       & 9.5(1)        & 22.9(3)\\
B1237+25     & 24.2(5)          & 18.9(2)       & 22.6(6)           & 30.4(3)       & 29.5(3)       & 26.5(3)\\
B1822$-$09   & 35.8(3)          & 11.8(2)       & 13.7(1)           & --            & --            & 13.9(4)\\
\hline
\end{tabular}
\tablefoot{
    \tablefoottext{a}{Low S/N.}
}
\end{table*}

\begin{figure}
  \centering
  \includegraphics[width=0.49\textwidth]{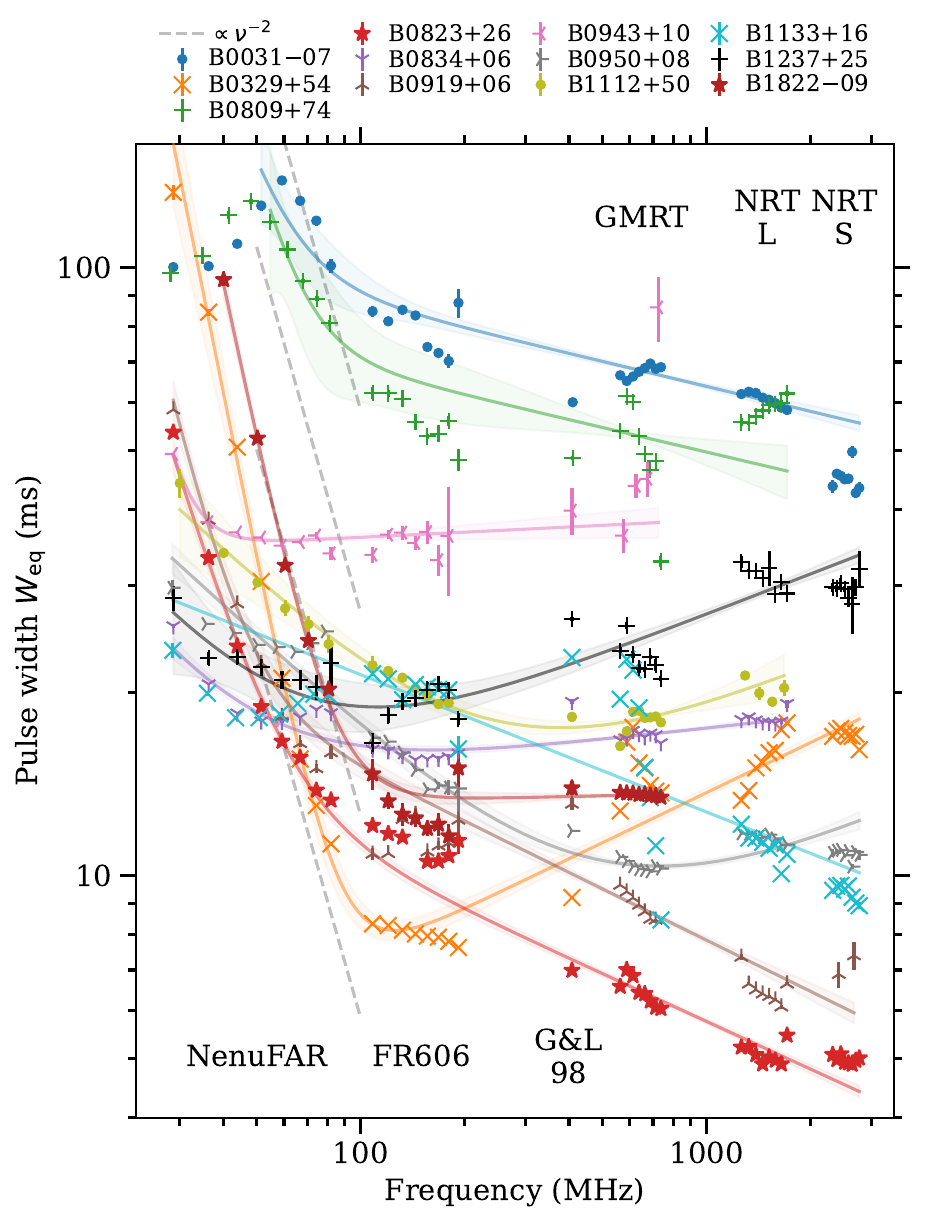}
  \caption{Frequency evolution of the boxcar-equivalent pulse width $W_\text{eq}$ for our pulsar sample. We show our best fits of Eq.~\ref{eq:widthmodel} to the data with coloured solid lines and their $1\sigma$ uncertainties as shaded regions. The grey dashed lines visualise pulse width scaling solely due to scattering.}
 \label{fig:pulsewidthscaling}
\end{figure}

\begin{figure}
  \centering
  \includegraphics[width=0.49\textwidth]{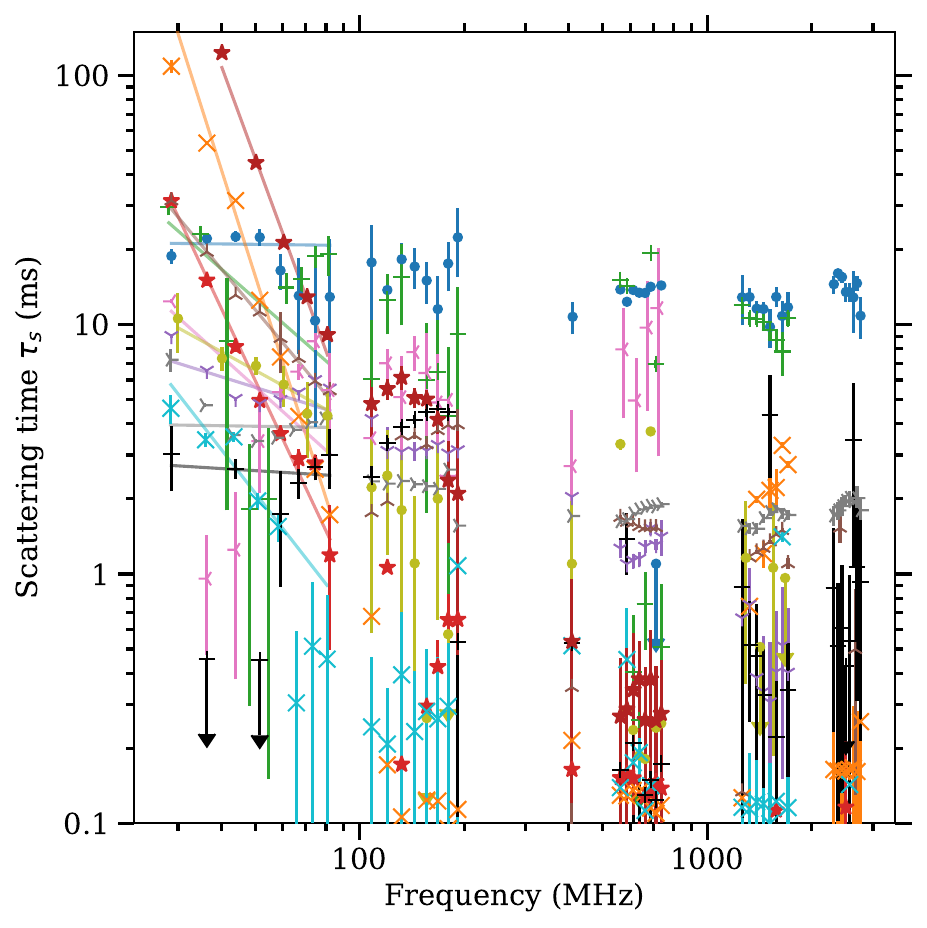}
  \caption{Measured scattering times $\tau_\text{s}$ as a function of frequency for our pulsar sample. The markers match the legend in Fig.~\ref{fig:pulsewidthscaling}. Some low-value upper limits with $\tau_\text{s} \ll 0.1~\text{ms}$ were omitted for clarity. The solid lines depict the best-fit power laws to our NenuFAR measurements.}
 \label{fig:taus}
\end{figure}

\begin{figure}
  \centering
  \includegraphics[width=0.49\textwidth]{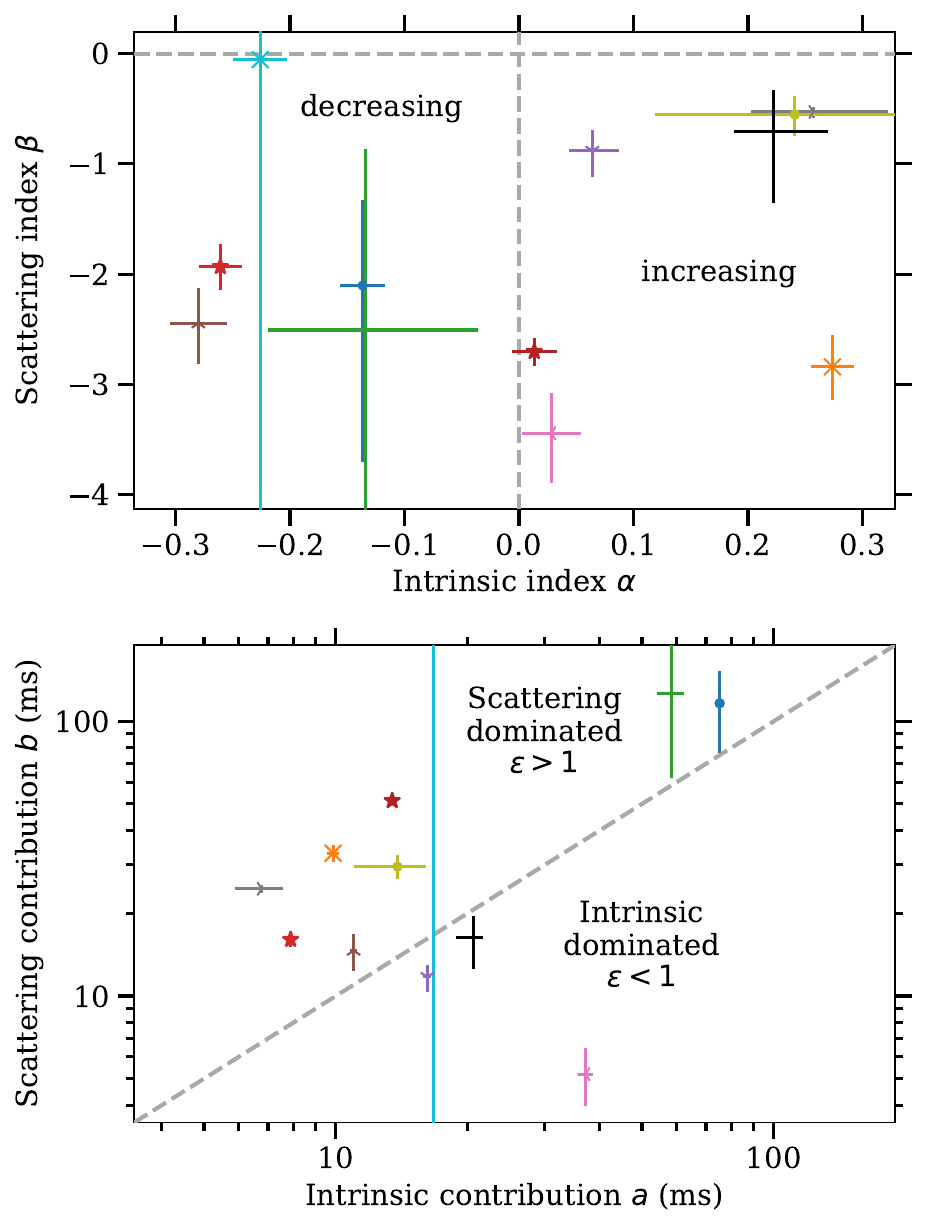}
  \caption{Best-fitting parameters of the fits shown in Fig.~\ref{fig:pulsewidthscaling}. Top: Scattering index $\beta$ plotted against intrinsic power law index $\alpha$. The vertical grey dashed line ($\alpha = 0$) separates the $\alpha$ range into decreasing and increasing profile evolution with frequency. The horizontal grey dashed line marks negligible scattering contribution ($\beta = 0$). Bottom: Scattering contribution $b$ plotted against intrinsic width contribution $a$. The grey dashed line indicates equality, separating the scattering- and intrinsic-width-dominated parts of the parameter space.}
 \label{fig:pulsewidthparams}
\end{figure}

When comparing the pulse profiles in Figs.~\ref{fig:profiles1}, \ref{fig:profiles2}, \ref{fig:profiles3}, \ref{fig:profiles4}, and \ref{fig:profiles5}, it is clear that most of the pulsar profiles in this work evolve significantly with radio frequency. This includes scatter broadening of the pulsars' radio signals in the turbulent, ionised ISM at low frequencies, which typically scales with radio frequency $\nu$ as $\nu^{-4.4}$ for Kolmogorov turbulence. Indeed, scattering becomes increasingly apparent in our FR606 and, especially, in our NenuFAR data. Single-sided exponential scattering tails are most obvious in the NenuFAR profiles of PSRs~B0329+54, B0823+26, B0834+06, B0919+06, B0943+10, and B1822$-$09. We also see clear intrinsic profile evolution with frequency, which manifests itself as a width change of the central (core) components, the appearance of new profile components mostly in the profile wings (conal components), the merging of those wing components, and the change in separation between the leading and trailing wing components. The intrinsic profile evolution can be complex, and there is significant profile evolution within an observing band, particularly at NenuFAR frequencies.

We quantified the profile evolution by fitting a multi-component scattering model (single scattering screen) to our data using the \texttt{scatfit} software, as described above. For each instrument, we used the best-fitting scattering-corrected DM at the given observing epoch (Tab.~\ref{tab:dms}) as determined from our NenuFAR data, which has the highest DM sensitivity. \texttt{scatfit} provided the best-fitting parameters for each individual profile component, such as its relative fluence, location in phase, width, DC offset (baseline mean), and scattering time, in each frequency sub-band (typically eight). We characterised the best-fitting semi-analytical pulse model by its full widths at 50 and 10~\% maximum ($W_{50}$ \& $W_{10}$), and its boxcar-equivalent pulse width
\begin{equation}
    W_\text{eq} = \frac{ dt \sum_i \mu_i  }{ \max \mu_i } = \frac{ \sum_i F_i }{ \max \mu_i },
    \label{eq:weq}
\end{equation}
where $\mu_i = \mean{S}_i$ is the mean Stokes I profile at phase bin $i$, $\max \mu_i$ is its maximum or its peak flux density, $dt$ is the phase bin width, and $F_i$ denotes fluence. $W_\text{eq}$ is the profile's total integrated fluence divided by its peak flux density and represents the width of a boxcar-shaped pulse of the same area. Additionally, we measured the $\text{D}4 \sigma$ pulse width $W_\text{d4s}$ based on the second-moment or variance width defined as
\begin{equation}
    W_\text{d4s} = 4 \sigma = 4 \: \left( \sum_i \mu_i \left( \phi_i - \phi_m \right)^2 / S \right)^{1/2},
    \label{eq:wd4s}
\end{equation}
where $\phi_m$ is the profile amplitude-weighted mean phase (centroid or centre of flux; Eq.~\ref{eq:meanphase}) and $S = \sum_i \mu_i$ is the integrated flux density, pulse energy, or power. Eq.~\ref{eq:wd4s} is a 1D adaptation of the $\text{D}4 \sigma$ pulse width commonly used to quantify laser beam widths in laboratory physics and formalised in the ISO 11146 standard. Our implementation rectifies the profile data and truncates them at 1~\% amplitude before computing $W_\text{d4s}$. \texttt{scatfit} computed the pulse width estimators and their uncertainties from the final Markov chain samples of the fitted (noiseless) scattering model.

We were mainly interested in the evolution of the overall profile width, focusing primarily on $W_\text{eq}$. Using $W_\text{eq}$ is beneficial because it accounts for all profile components (area-based) and is unaffected by the choice of peak level at which to quote the full width. We discuss our fitting methodology and the pulse width estimators we chose in Appendix~\ref{ap:widthestimators}. Table~\ref{tab:pulsewidths} presents our band-integrated $W_\text{eq}$ measurements and their $1\sigma$ uncertainties at the various centre frequencies. We augmented our dataset slightly by measuring $W_\text{eq}$ using \texttt{scatfit} from the pulse profiles of \citet{1998Gould} at 408~MHz, downloaded from the EPN database, which nicely cover the frequency range between our FR606 and uGMRT Band-4 measurements.

Fig.~\ref{fig:pulsewidthscaling} visualises the sub-banded $W_\text{eq}$ evolution with observing frequency. Following \citet{2023Jankowski}, we decomposed the total observed pulse width ($W_\text{eq}$ here) into the quadrature sum
\begin{equation}
\begin{split}
    W_\text{obs} (\nu) & = \sqrt{ {W_\text{i} (\nu)}^2 + {W_\text{s} (\nu)}^2 + dt^2 + {t_\text{dm} (\nu) }^2 }\\
    & \approx \sqrt{ {W_\text{i} (\nu)}^2 + {W_\text{s} (\nu)}^2 },
\end{split}
    \label{eq:widthmodel}
\end{equation}
where $W_\text{i} (\nu) = a \: \left( \nu / \nu_0 \right)^\alpha$ describes the intrinsic profile evolution, $W_\text{s} (\nu) = b \: \left( \nu / \nu_1 \right)^\beta$ models the scatter broadening, $dt \ll W_\text{int}$ is the phase bin width of the data, $t_\text{dm} \approx 0$ is the intra-channel dispersive smearing\footnote{Our NenuFAR, FR606, GMRT CD-mode, and NRT data were coherently dedispersed. The channelisation of our GMRT PA-mode data was chosen to ensure negligible intra-channel DM smearing.}, $\nu_0 = 50 \: \text{MHz}$ and $\nu_1 = 300 \: \text{MHz}$ are two arbitrarily chosen reference frequencies, and $\alpha$ and $\beta$ are the power law indices of the intrinsic and scattering evolution, respectively. $\beta$ matches the ISM scattering index for $\alpha \approx 0$. We fit the scaling model in Eq.~\ref{eq:widthmodel} to our sub-banded $W_\text{eq}$ measurements using the \texttt{LMFIT} software \citep{2025Newville}. The fit parameters were $a, b, \alpha$, and $\beta$. We show the resulting best fits as coloured solid lines in Fig.~\ref{fig:pulsewidthscaling} and visualise their $1\sigma$ uncertainty bands as lightly shaded regions. Fig.~\ref{fig:taus} presents our measured scattering times $\tau_\text{s}$ as a function of frequency. We fit a power law of the form $\tau_\text{s} (\nu) = \tilde{b} \: (\nu / \nu_2)^{\tilde{\beta}}$ to our NenuFAR measurements using \texttt{LMFIT}, with the free parameters $\tilde{b}$ and $\tilde{\beta}$, and $\nu_2$ set to an arbitrary reference frequency within our data. The parameter $\tilde{\beta}$ is a direct measurement of the ISM scattering index. We show the resulting best-fitting power laws as coloured solid lines in Fig.~\ref{fig:taus}.

The pulse width scaling model (Eq.~\ref{eq:widthmodel}) describes our measurements well except for three pulsars, namely PSRs~B0031$-$07, B0809+74, and B1133+16. PSRs~B0031$-$07 and B0809+74 show similar behaviour in that their pulse width rises monotonically in the NenuFAR band down to $\sim$60~MHz and $\sim$50~MHz, after which it declines again towards lower frequencies. Looking at the NenuFAR dynamic spectrum of PSR~B0031$-$07, that is because the fluence of the leading profile component (P1) falls rapidly in the NenuFAR band towards lower frequencies. P1 appears to have a flatter or more inverted spectrum than P2. In other words, P1 disappears into the baseline noise floor at the lowest frequencies; we are effectively `losing' a profile component. The case is almost exactly the same for PSR~B0809+74. P1 and P2 merge at the highest NenuFAR frequency to form a single profile peak. They separate as we move to lower frequencies and become almost fully disjoined below $\sim$40~MHz and clearly separated in phase at $\sim$30~MHz. P1's fluence falls faster than that of P2 due to a flatter or more inverted spectral index. It increasingly disappears into the noise floor at the lowest frequencies, thereby decreasing the total pulse width. To resolve those fitting issues, we exclude PSR~B0031$-$07's data below 60~MHz and PSR~B0809+74's data below 50~MHz from our scaling model fits. Moreover, PSR~B1133+16 is an outlier and exhibits minimal scattering across most of our frequency range. Aside from these special cases, the other pulsars are well-described by our scaling model across the entire frequency range.

\subsection{Scattering}
\label{sec:scattering}

\begin{figure}
  \centering
  \includegraphics[width=0.49\textwidth]{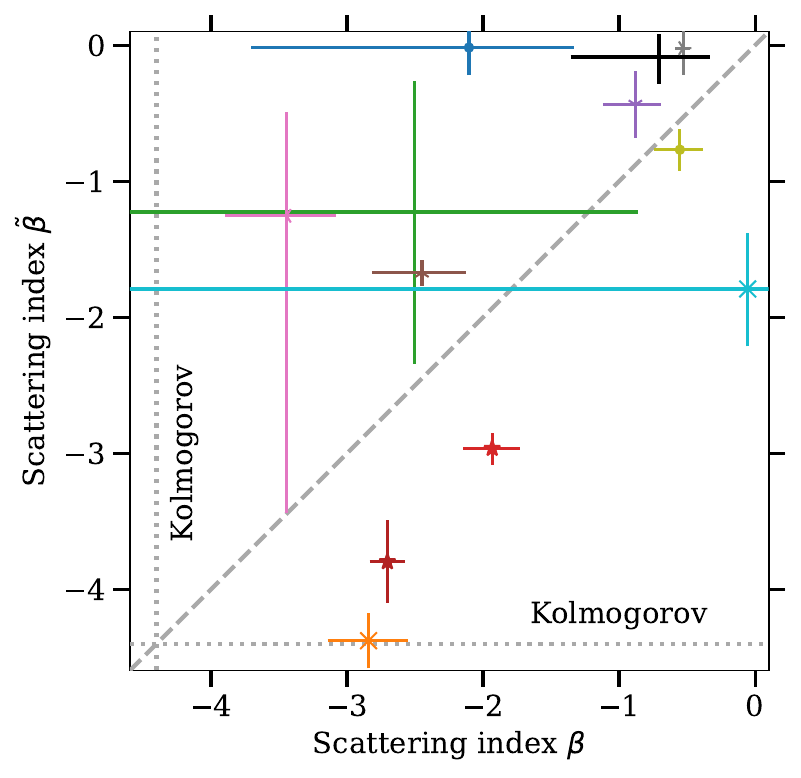}
  \caption{Comparison of the scattering indices measured from the total pulse width ($\beta$) with those measured from a direct fit to the scattering time data ($\tilde{\beta}$). The grey dashed line indicates equality and the grey dotted lines mark the $-4.4$ value expected for Kolmogorov turbulence.}
 \label{fig:scatindexcomparison}
\end{figure}

We now discuss the scattering contributions measured from fitting the pulse width scaling model (Eq.~\ref{eq:widthmodel}) to our $W_\text{eq}$ data. Our simultaneous (or near simultaneous) observations presented here are important for this analysis, as temporal changes in scattering parameters are guaranteed to be negligible.

Fig.~\ref{fig:pulsewidthparams} presents the best-fitting parameters from our pulse width scaling model fit. The top panel visualises the power law scaling by plotting the scattering index $\beta$ against the intrinsic index $\alpha$. The best-fitting scattering indices range from $\beta \approx -3.4$ for PSR~B0943+10 to around $-0.5$ for PSR~B0950+08. There seem to be two groups of pulsars, those with steeper, more negative scattering indices in the range $-2$ to $-3.5$ (in decreasing order: PSRs B0823+26, B0031$-$09, B0919+06, B0809+74, B1822$-$09, B0329+54, and B0943+10), and those with more shallow scattering indices ranging from $-0.9$ to $-0.5$ (in increasing order: PSRs~B0834+06, B1237+25, B1112+50, and B0950+08).

Our power law fits to the scattering times provide an independent and more direct measurement of the ISM scattering indices. Hence, it is useful to compare the best-fitting scattering indices $\beta$ measured from $W_\text{eq}$'s frequency evolution (Fig.~\ref{fig:pulsewidthscaling}) with the best-fitting scattering indices $\tilde{\beta}$ measured from our direct power law fits to the $\tau_\text{s}$ data (Fig.~\ref{fig:taus}). Fig.~\ref{fig:scatindexcomparison} shows the two independent scattering index measurements plotted against each other, with the grey dashed line indicating equality and the Kolmogorov value marked. Most of the scattering index pairs $\left( \tilde{\beta}, \beta \right)$ are compatible with each other at the $2 \sigma$ level, in particular the flatter ones with $\tilde{\beta} \geq -2$. On the other hand, the scattering indices $\tilde{\beta}$ of PSRs~B0329+54, B1822$-$09, and B0823+26 are significantly steeper than their $\beta$ counterparts. Looking at Fig.~\ref{fig:taus}, there appears to be a break in scattering index around $\sim$100~MHz for these pulsars, with a $\tau_\text{s}$ scaling that is significantly flatter above 100~MHz than below. Indeed, if we include the FR606 data in our $\tau_\text{s}$ power law fit, the resulting $\tilde{\beta}$ values match their $\beta$ counterparts within the uncertainties. We conclude that we see genuine deviations from a simple power law scaling in scattering behaviour in those pulsars. Moreover, PSR~B0329+54 shows hints of more complex scattering also at frequencies below 50~MHz, with a suggestive break around 40--50~MHz that flattens the low-frequency scattering index. Above $\sim$130~MHz, the scattering times drop below 0.1~ms and become much more challenging to measure. Given the high-frequency plateaus in the $\tau_\text{s}$ measurements shown in Fig.~\ref{fig:taus}, we estimate our current $\tau_\text{s}$ sensitivity to around 0.01--0.1~ms in absolute terms and to around 10~\% in relative terms\footnote{Estimated from the $\tau_\text{s}$ and $W_\text{d4s}$ values of the widest high-frequency ($\geq$ L-band) profiles of PSRs~B0031$-$07 and B0809+74.} for the S/N and profile complexity of this particular dataset. The sensitivity is better for simpler profiles, such as single-component pulses often seen in fast radio bursts.

PSR~B0329+54's scattering index $\tilde{\beta}$ perfectly matches the $-4.4$ expectation value for Kolmogorov turbulence. Similarly, PSR~B1822$-$09's $\tilde{\beta}$ is consistent with the Kolmogorov value at the $2 \sigma$ level. However, most of our measured scattering indices are discrepant with Kolmogorov turbulence, with either significant or substantial deviations. This hints at more complex scattering environments with localised turbulence in dense filaments or other structures. We suspect that these local structures dominate the scattering in the mostly low-DM pulsars (Tab.~\ref{tab:dms}) and that their scattering behaviour becomes more Kolmogorov as more scattering screen are averaged along the LOS. This is supported by the fact that the pulsars with the steepest, most Kolmogorov-like $\tilde{\beta}$ (PSRs~B0329+54, B1822$-$09, and B0823+26) have amongst the highest DMs in our sample.

PSR~B1133+16 is peculiar in that it exhibits minimal scattering with $\beta \approx 0$ and scattering times $\tau_\text{s} \leq 1~\text{ms}$ across most of our dataset. It only exceeds the 1~ms threshold significantly below $\sim$60~MHz (Fig.~\ref{fig:taus}). This means that its width scaling is consistent with purely intrinsic profile evolution over most of our frequency range. Indeed, at $\sim$4.85~$\text{pc} \: \text{cm}^{-3}$ it has the second lowest DM in our pulsar sample. However, even for this seemingly under-scattered pulsar LOS, we measure significant scattering in the NenuFAR band below $\sim$60~MHz.

Fig.~\ref{fig:pulsewidthparams} bottom panel presents the individual contributions of the scattering and intrinsic scaling to the total pulse width by plotting the normalisation parameters $b$ and $a$ against each other. The diagonal grey dashed line indicates equality. As we have chosen reference frequencies $\nu_0 \neq \nu_1$, $a$ and $b$ describe the individual contributions to different parts of the spectrum, where the scattering (low frequency, NenuFAR band) and intrinsic width contributions (higher frequency) are most readily measured. This is intentional and important so that we can most easily separate the contributions. Most of our pulsars (8) are clearly scattering-dominated with sizeable scattering fractions $\epsilon = b / a > 1$ ranging from 1.3 (PSR~B0919+06) to 3.7 (PSR~B0950+08), indicating that scattering exceeds the intrinsic width contribution, often substantially. On the other hand, three pulsars are intrinsic-width dominated with $\epsilon \approx 0.8$ for PSR~B1237+25 to 0.14 for PSR~B0943+10. PSR~B1133+16 is a special case with $b \approx 0$, as mentioned above. Those four pulsars have significant intrinsic pulse widths of around 20--40~ms below 700~MHz that only mildly evolve with frequency.

\subsection{Intrinsic profile evolution}

Regarding the intrinsic pulse width evolution, five pulsars have negative intrinsic power law indices ranging from $\alpha \approx -0.3$ for PSR~B0919+06 to $-0.14$ for PSRs~B0031$-$09 and B0809+74 (Fig.~\ref{fig:pulsewidthparams}). Their profiles become intrinsically narrower with increasing frequency. Indeed, PSR~B0919+06's profile evolves strongly with frequency. It exhibits clear scattering in the NenuFAR band and shows strong evolution of the relative amplitudes of the leading and trailing wing components. PSR~B0823+26's $W_\text{eq}$ approximately halves from 150~MHz to 650~MHz. Its MP widens and its PC becomes more apparent at lower frequencies (Fig.~\ref{fig:profiles1}). PSR~B1133+16's width roughly halves from 150~MHz to 2.5~GHz. The most obvious change is in the relative amplitude of the leading (P1) and trailing peaks (P2), whose dominance shifts from P2 at low frequencies to P1 at high frequencies due to differing spectral indices (Fig.~\ref{fig:profiles5}). The component separation and overall pulse width decrease appreciably with frequency.

Secondly, three pulsars have power law indices $\alpha$ consistent with zero (PSRs~B0943+10, B1822$-$09, and possibly B0834+06), suggesting negligible intrinsic pulse width evolution. Their pulse widths plateau between 100~MHz and 650~MHz to 1.8~GHz, with PSR~B0834+06's evolution being best constrained. Apart from scattering, B0834+06's profiles look almost identical with most of the evolution happening in the trailing component, which becomes slightly more prominent with increasing frequency.

Finally, four other pulsars have significantly positive intrinsic indices, indicating intrinsic profile widening with increasing frequency. Their best-fitting indices cover a very limited range between $\alpha \approx 0.22$ for PSR~B1237+25 to $\alpha \approx 0.27$ for PSR~B0329+54. PSR~B1237+25's $W_\text{eq}$ increase can be explained by the brightness enhancement of its trailing conal component (P5) relative to its leading dominant component (P1), as well as strengthened emission of the second and fourth components P2 and P4. Aside from the P5 enhancement, the most striking change is the almost complete disappearance of the central core component (P3) from 650~MHz onwards, which is obscured by P4 and parts of P2. P3 has a significantly reduced contribution at this frequency. PSR~B1112+50's $W_\text{eq}$ upturn is due to the reappearance of its leading shoulder component as a distinct feature, which is most clearly visible at L-band with a notch-like separation. It $\alpha$ uncertainty is large. B0329+54's pulse width increase is explained by the enhancement of the leading and trailing outer conal components (P1 and P5) with respect to the core component P3, a strengthening of the inner cone components (P2 and P4), and increased pedestal or baseline emission (P6).

In summary, negative $\alpha$ are associated with an intrinsic narrowing of the radio emission envelope (radio beam size or footprint) and a reduction in component separation with increasing frequency. Physically, this is likely a geometric effect. Positive $\alpha$ occur primarily due to the appearance or re-appearance of profile components that were suppressed at lower frequencies. This is probably caused by differences in component spectral index in the standard nested hollow cone model. $\alpha \approx 0$ seems to be a special case. It could either be that the competing effects (beam size reduction \& component emergence) roughly balance. Alternatively, the component spectral indices could be closely comparable to each other over a wide frequency range and the pulsar viewing geometry must be favourable so that radio beam size reduction effects become negligible (large impact parameter).

%
% Modulation indices
%

\subsection{Phase-resolved single-pulse modulation}

\begin{table*}
\caption{Pulsar modulation indices measured from our band-integrated data.}
\label{tab:modindex}
\begin{tabular}{lccccccccccccc}
\hline\hline
PSR          & \multicolumn{3}{c}{NenuFAR}          & \multicolumn{3}{c}{FR606}     & \multicolumn{3}{c}{GMRT}      & \multicolumn{3}{c}{NRT L}\\
             & \multicolumn{3}{c}{50 MHz}           & \multicolumn{3}{c}{150 MHz}   & \multicolumn{3}{c}{650 MHz}   & \multicolumn{3}{c}{1.5 GHz}\\
             & $m_\text{min}$   & $\bar{m}$ & $\mean{\text{S/N}}$ & $m_\text{min}$    & $\bar{m}$ & $\mean{\text{S/N}}$ & $m_\text{min}$    & $\bar{m}$ & $\mean{\text{S/N}}$   & $m_\text{min}$    & $\bar{m}$ & $\mean{\text{S/N}}$   & Class\\
\hline
B0031$-$07   & 2.35(8)  & 3.8   & 18    & 2.9(1)    & 3.7       & 16    & 1.56(2)           & 1.8   & 222   & --        & --    & --    & D\\
B0329+54     & 1.31(2)  & 1.8   & 34    & 1.086(8)  & 1.5       & 157   & 0.485(5)          & 0.7   & 2787  & --        & --    & --    & D\\
B0809+74     & 2.08(6)  & 2.5   & 42    & 1.52(4)   & 1.5       & 18    & 0.55(1)           & 0.8   & 454   & --        & --    & --    & D\\
B0823+26     & 0.87(1)  & 1.1   & 21    & 0.88(1)   & 1.3       & 33    & 0.99(1)           & 1.3   & 479   & --        & --    & --    & F\\
B0823+26 Q   & --       & --    & --    & --        & --        & --    & 9.6(6)            & 5.0   & 7     & --        & --    & --    & --\\
B0834+06     & 1.28(2)  & 1.3   & 40    & 0.88(1)   & 1.0       & 33    & 0.69(1)           & 0.7   & 177   & --        & --    & --    & D\\
B0919+06     & 0.691(5) & 1.5   & 21    & 0.709(7)  & 1.2       & 12    & 0.66(1)           & 0.9   & 168   & --        & --    & --    & F\\
B0943+10     & 1.05(2)  & 1.4   & 34    & 2.5(1)    & 3.6       & 7     & 8(1)              & 9.7   & 4     & --        & --    & --    & U\\
B0950+08     & 1.13(1)  & 2.3   & 48    & 1.72(2)   & 2.7       & 22    & 1.50(3)           & 1.9   & 581   & 1.38(2)   & 1.6   & 386   & D/F\\
B1112+50     & 3.4(1)   & 2.2   & 10    & 2.58(9)   & 2.3       & 17    & 1.99(7)           & 2.2   & 103   & 3.2(2)    & 4.5   & 17    & U/F\\
B1133+16     & 2.00(4)  & 2.1   & 22    & 1.43(4)   & 1.6       & 29    & 1.03(2)           & 1.5   & 252   & --        & --    & --    & D\\
B1237+25     & 2.19(6)  & 2.3   & 16    & 1.28(3)   & 1.7       & 56    & 0.59(1)           & 1.0   & 148   & 0.83(1)   & 1.1   & 111   & D\\
B1822$-$09   & 2.13(6)  & 3.8   & 8     & 1.91(6)   & 2.3       & 5     & 0.68(1)           & 1.3   & 217   & --        & --    & --    & U\\
\hline
\end{tabular}
\tablefoot{
We list the minimum modulation index ($m_\text{min}$), the profile-weighted mean modulation index ($\bar{m}$), the mean single-pulse S/N, and our classification into decreasing (D), flat (F), and unclear (U) frequency-scaling behaviour.
}
\end{table*}

\begin{figure}
  \centering
  \includegraphics[width=0.49\textwidth]{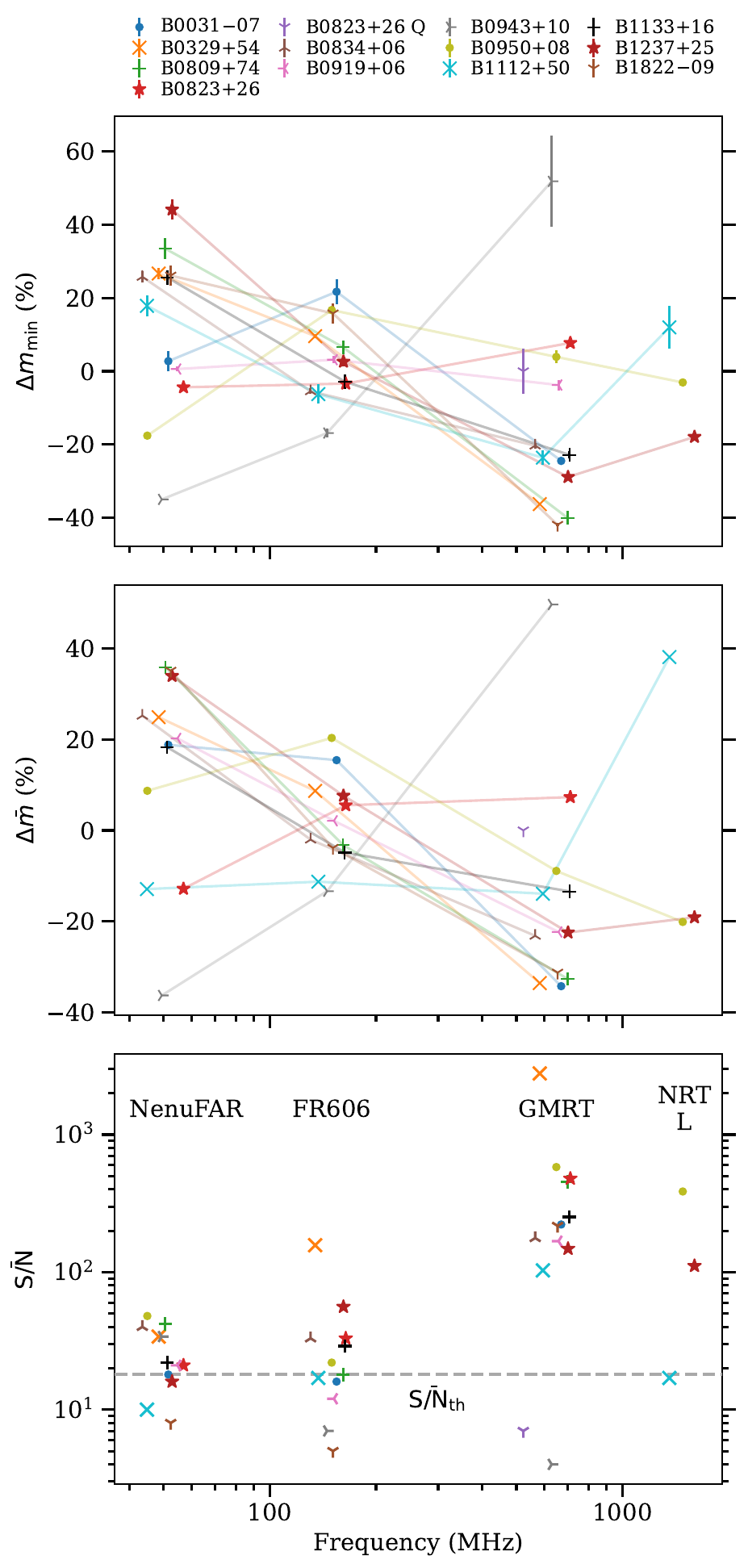}
  \caption{Frequency evolution of the minimum modulation index $\Delta m_\text{min}$ (top), the profile-weighted mean modulation index $\Delta \bar{m}$ (centre), and the mean single-pulse $\bar{\text{S/N}}$ (bottom) for our pulsar sample. A small jitter was added to the frequencies for clarity.}
 \label{fig:modindexscaling}
\end{figure}

\begin{figure}
  \centering
  \includegraphics[width=0.49\textwidth]{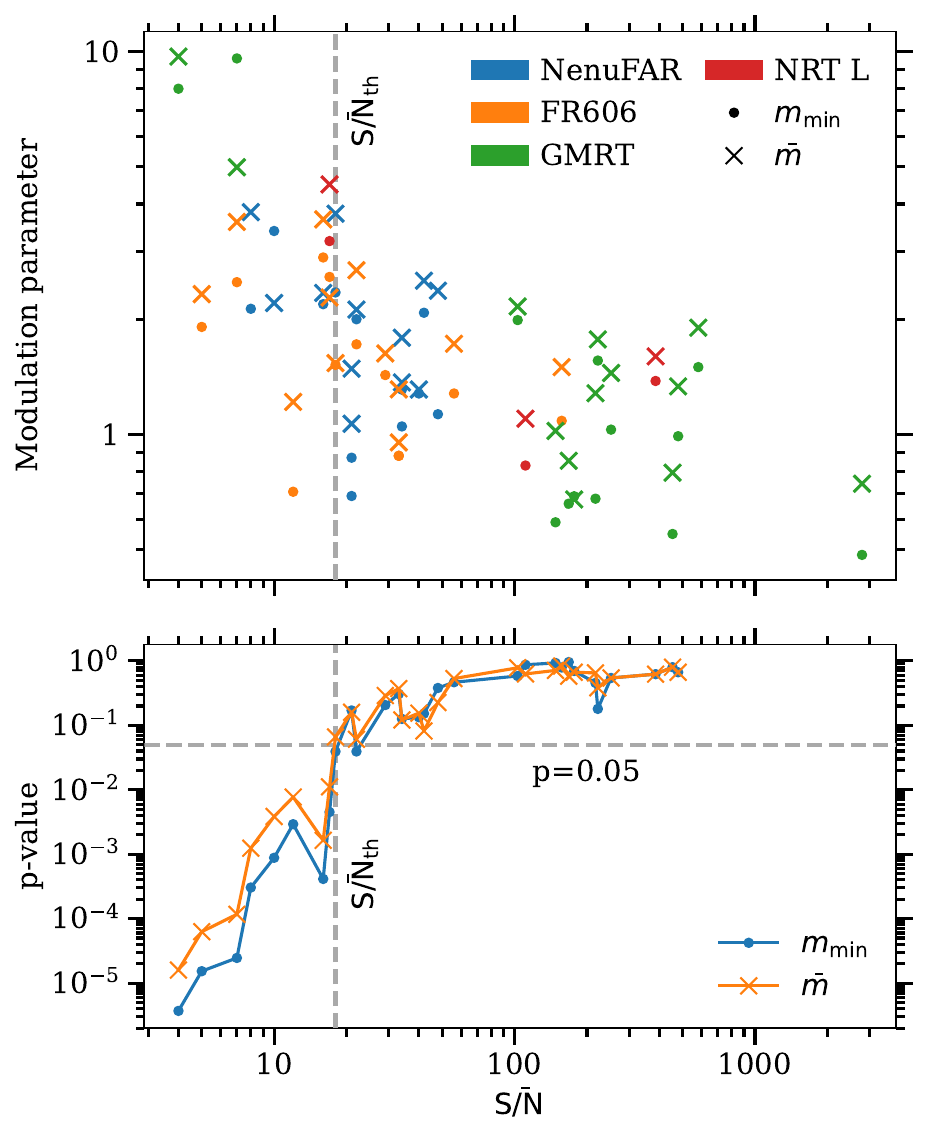}
  \caption{S/N bias estimation. Top: Measured modulation parameters $m_\text{min}$ and $\bar{m}$ plotted against mean single-pulse $\bar{\text{S/N}}$ at the different observing frequencies. Bottom: P-value of the Spearman rank correlation of the modulation parameters in the top panel with $\bar{\text{S/N}}$ as a function of minimum $\bar{\text{S/N}}$. The vertical dashed lines mark the threshold $\bar{\text{S/N}}_\text{th}$ above which correlations with S/N become spurious ($p \geq 0.05$).}
 \label{fig:snrbias}
\end{figure}

\begin{figure}
  \centering
  \includegraphics[width=0.49\textwidth]{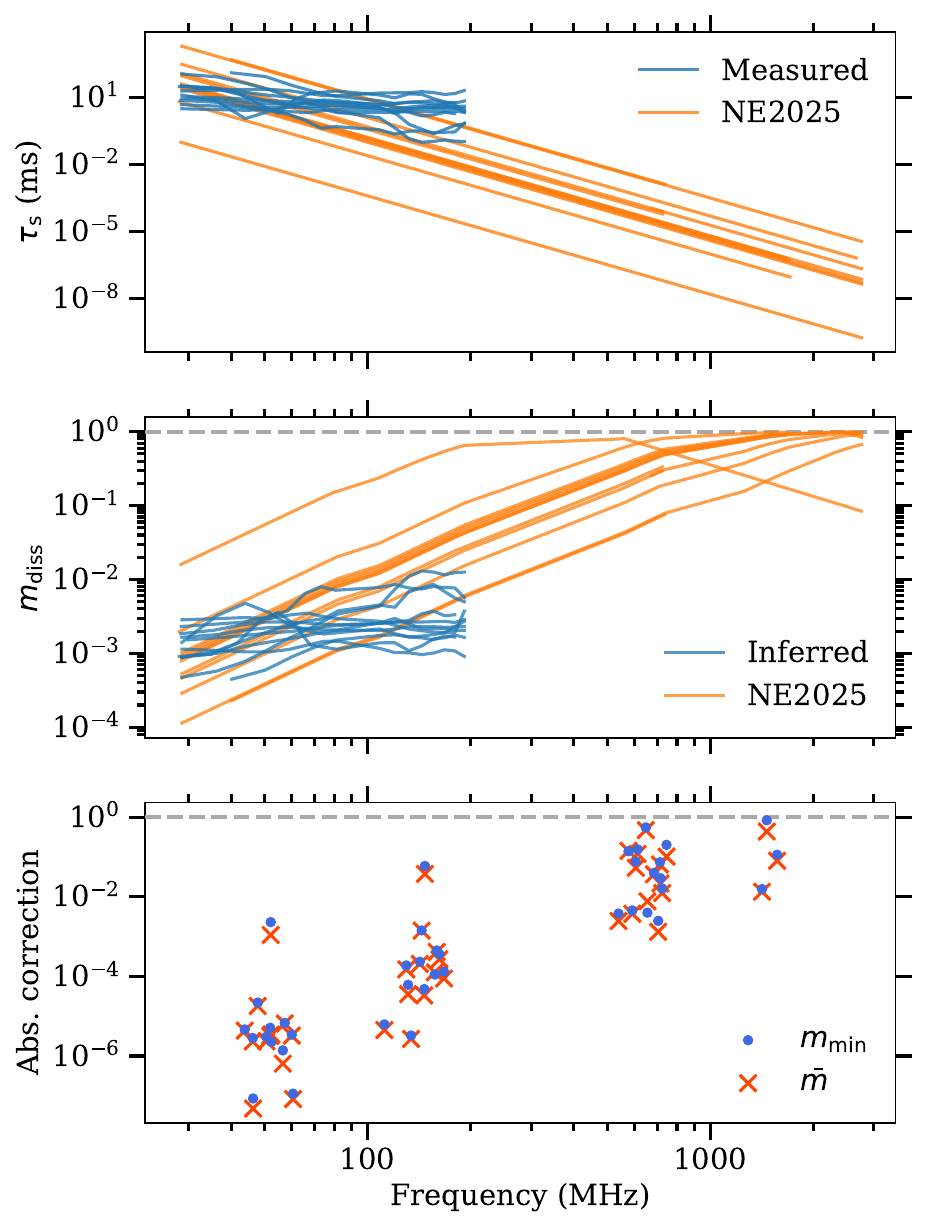}
  \caption{Impact of diffractive scintillation on our modulation analysis. We show running means of the scattering times $\tau_\text{s}$ measured in this work and calculated using the \texttt{NE2025} model (top), the inferred diffractive modulation indices $m_\text{diss}$ (middle), and the absolute correction to our modulation parameters (bottom). The grey dashed horizontal lines mark unity. A small jitter was added to the frequencies for clarity.}
 \label{fig:mdiss}
\end{figure}

We show the band-integrated pulse profiles and their phase-resolved modulation index $m_i$ curves in Figs.~\ref{fig:profiles1}, \ref{fig:profiles2}, \ref{fig:profiles3}, \ref{fig:profiles4}, and \ref{fig:profiles5}. The pulse profiles are referenced to the left flux density scale, while the $m_i$ curves refer to the right scale. We computed the phase-resolved temporal modulation index
\begin{equation}
    m_i = \frac{ \sigma_i }{ \mu_i },
    \label{eq:modulationindex}
\end{equation}
where $\sigma_i$ is the standard deviation and $\mu_i = \mean{S}_i$ is the mean Stokes I flux density in phase bin $i$, with the statistics determined along the pulse number dimension. We show $m_i$ only for phase bins where the integrated Stokes~I pulse profile exceeded typically 5--10~\% of its maximum, and we estimated the $m_i$ uncertainty by bootstrap resampling the data 100 times. To quantify the pulsars' modulation behaviour, we computed two summary statistics or modulation parameters: (1) the minimum of the modulation index curves, $m_\text{min}$, and (2) the flux density or profile amplitude-weighted mean modulation index
\begin{equation}
    \bar{m} = \sum_i m_i \: \mu_i \: / \sum_i \mu_i,
    \label{eq:modindexsum}
\end{equation}
where $\mu_i$ is the integrated Stokes I pulse profile at phase bin $i$. The sums run over all on-pulse phase bins, irrespective of modulation behaviour or relative profile amplitude. Our $\bar{m}$ definition above incorporates on-pulse phase gating to avoid residual RFI. Its uncertainty was estimated by bootstrap resampling the data 100 times, as above. Using $\bar{m}$ is beneficial, as it includes modulation information across a wider phase bin range and appears to be more robust to S/N bias than $m_\text{min}$ \citep{2023Song}. We highlighted $m_\text{min}$ with orange markers in the profile plots.

Table~\ref{tab:modindex} presents our $m_\text{min}$ and $\bar{m}$ measurements at the different frequencies. Fig.~\ref{fig:modindexscaling} visualises their relative changes $\Delta m_\text{min}$ (top panel) and $\Delta \bar{m}$ (middle panel) with frequency, defined as
\begin{equation}
    \Delta y = \frac{ y_i - \mean{ y_i } }{ \max y_i },
    \label{eq:standardisation}
\end{equation}
where $y = \left\{ m_\text{min}, \bar{m} \right\}$, $\mean{ y_i }$ is the mean, and $\max y_i$ is the maximum across the observing bands for each pulsar. Namely, we standardised the data by subtracting the mean and dividing by the maximum, so that all pulsars can be compared on a single scale. Fig.~\ref{fig:modindexscaling} bottom panel shows the mean single-pulse S/N and the threshold S/N for reference. Three general classes can be seen: decreasing (D) modulation parameters with frequency (PSRs~B0329+54, B0809+74, B0834+06, B1133+16, B1237+25, and perhaps B0031$-$07 and B0950+08), approximately flat (F) modulation parameters across frequencies (PSRs~B0823+26, B0919+06, and perhaps B1112+50), and unclear (U) behaviour (PSRs~B0823+26 Q, B0943+10, and B1822$-$09). We list our classifications in Tab.~\ref{tab:modindex}.

Accounting for S/N bias and ISM-induced modulation (see below), the majority of our pulsars' (7/12) modulation parameters decrease with frequency, while three pulsars have roughly flat modulation parameters across frequency, and the behaviour of two pulsars is unclear due to insufficient S/N. This means that the single-pulse emission from most of our pulsars becomes more erratic towards lower radio frequencies, coincident with a general increase in dynamic range.

\subsection{S/N bias}

Fig.~\ref{fig:snrbias} top panel investigates S/N bias in our single-pulse modulation measurements, where we show the measured modulation parameters $m_\text{min}$ and $\bar{m}$ plotted against $\bar{\text{S/N}}$ for all observing frequencies. In the bottom panel, we show the p-value of the Spearman rank correlation computed between each modulation parameter and $\bar{\text{S/N}}$ as we iteratively selected subsamples that included only data points above a given minimum S/N. The threshold S/N above which the correlation became spurious with $p \geq 0.05$ is $\bar{\text{S/N}}_\text{th} \approx 18$. Above this threshold, correlations between the modulation parameters and $\bar{\text{S/N}}$ are insignificant.

The p-value curve for $\bar{m}$ exceeds the one for $m_\text{min}$ by a factor of a few below $\bar{\text{S/N}}_\text{th}$, while they are mostly consistent above the threshold. This seems to confirm our expectation that $\bar{m}$ is a more reliable and less biased estimator for the single-pulse modulation than $m_\text{min}$ in the low-S/N regime.

\subsection{Distinguishing intrinsic and ISM-induced modulation}

The pulsar single pulses exhibit intrinsic (magnetospheric) modulation and modulation induced by external propagation effects, as the radio pulses traverse the turbulent, ionised ISM. Namely, they are affected by diffractive interstellar scintillation (DISS) on short timescales (minutes to hours) and refractive interstellar scintillation (RISS) on much longer timescales (days to months). Our analyses are based on single epoch observations lasting several hours (Tab.~\ref{tab:observations}). Thus, we were mainly interested in the short timescale fluctuations induced by DISS.

Following the treatment in \citetalias{2025Jankowski} and \citet{2018Jankowski}, we derived the diffractive modulation index $m_\text{diss}$ expected for our pulsars at each observing frequency. Namely, we calculated the diffractive scintillation bandwidths $\Delta \nu_\text{d}$ from the scattering times $\tau_\text{s}$ at each frequency using the `uncertainty' relation  $\Delta \nu_\text{d} = C_1 / \left( 2 \pi \: \tau_\text{s} \right)$, with $C_1 = 1.16$ for a uniform, Kolmogorov medium. At NenuFAR frequencies, we used the measured scattering times $\tau_\text{s}$ from our profile fits (Fig.~\ref{fig:taus}), while we computed the $\tau_\text{s}$ values at higher frequencies using the \texttt{NE2025} Galactic free-electron model \citep{2026Ocker}. We assumed negligible time averaging of our single-pulse data and typical effective bandwidths of 50~MHz (NenuFAR), 75~MHz (FR606), 165~MHz (GMRT), and 430~MHz (NRT L-band). The modulation parameters corrected for diffractive scintillation were calculated as
\begin{equation}
    m_{x,\text{c}} = \sqrt{ m_x^2 - m_\text{diss}^2 } \geq 0, \: \forall m_x \geq m_\text{diss},
\end{equation}
where $m_x = \left\{ m_\text{min}, \bar{m} \right\}$. Fig.~\ref{fig:mdiss} shows the measured scattering times $\tau_\text{s}$, inferred diffractive modulation indices $m_\text{diss}$, and the absolute DISS correction $\left| \Delta m_x \right| = \left| m_{x,\text{c}} - m_x \right|$ to our modulation parameters. $\left| \Delta m_x \right|$ becomes only relevant when it exceeds $\sim$0.2 at higher frequencies (GMRT \& NRT L-band), where the scintles are wide and where DISS might not be entirely quenched. The correction is negligible at low radio frequencies (NenuFAR \& FR606) due to the low scintillation bandwidths and sufficient scintle averaging. Thus, we conclude that DISS is irrelevant to our classification of frequency-scaling behaviour. Indeed, as $\Delta m_x \leq 0$, it will only make the modulation parameter curves steeper at high frequencies, thereby reinforcing our decreasing (D) classification. Similarly, the correction is negligible for the flat (F) pulsars.

%
% Discussion
%

\section{Discussion}
\label{sec:discussion}

\subsection{Pulse profiles, frequency evolution, and interpretation}

The pulsar profile morphology is diverse, with duty cycles, peak magnitudes, and distributions of the emission components varying from pulsar to pulsar. This can be attributed to both the emission geometry and the complex plasma processes acting on the emission region of the magnetosphere \citep{2019Roy, 2021Roy}. Understanding pulsars' radio emission is challenging, as they exhibit complex flux density variability and various radio spectra, including single power law, broken power law, log-parabolic, and variable break frequencies \citep{2018Jankowski, 2021Roy}. Understanding those phenomena requires a complex, nonlinear treatment of plasma physics \citep{2021Roy}. Moreover, the polarisation properties of pulsars are recognised as an even more complex problem, showing fractional changes in circular polarisation across pulse longitude, which can be attributed to propagation effects or to some intrinsic non-linear plasma process (see \citet{2025ApJ...994..189R} and references therein). We limit our discussion of polarisation here. However, we have a paper in preparation that incorporates polarisation information.

Conventional radio emission geometry attributes concentric rings of plasma-filled structure levitating over the diverging dipolar magnetosphere, manifesting their core-cone beaming pattern. Some models claim that emission beams are patchy and randomly distributed across the emission region, and that the beam cross-section structure is determined by the LOS cutting geometry. For example, some pulse profiles in this study show predominantly one, two, or three primary components. However, the phase distribution of the component's location and their corresponding magnitudes is not always simple, reflecting the coupling of the plasma's modulation properties with the emission geometry and the presence of relativistic effects such as aberration-retardation (A/R) or plasma current signatures \citep{2019Roy}. Our analysis shows that PSR~B0329+54’s component location is asymmetrically distributed with respect to the main (core) component, clearly reflecting the presence of relativistic effects such as A/R. Details of the treatment of such relativistic effects have far-reaching implications, revealing a strong correlation between phase shift and emission height parameter associated with distinct emission components \citep{2024Roy, 2025ApJ...980..214R}. Additionally, it is clear from the work presented here that scattering contributes a major part to the pulse widths below 200~MHz, even for apparently under-scattered LOSs.

\citet{2024Ardavan} recently conjectured that a uniformly rotating current sheet distribution pattern in the magnetosphere of a non-aligned rotating dipole embraces synergies both from synchrotron radiation and the vacuum version of Cherenkov radiation. Once the emission sources associated with different parts of the magnetosphere are superposed, they form caustics, since the beams from different parts accumulate towards the observer, approaching the speed of light in the limit. Far away from the sources, presumably the waves responsible for generating such caustics, notably at infinity or at cosmological distances, phase cancellation is perfect; they constructively coalesce and form bright enhanced pulses, possibly originating from a coherently distributed emission source across the pulsar magnetosphere. Alternative models include coherent curvature radiation or striped wind emission (gamma-rays), which represent pioneering steps toward probing the emission physics in the pulsar magnetosphere in great detail.

Below, we describe five categories of pulse profile morphologies that are important for the frequency evolution of pulse profiles. Namely, they are: Type~1a, Type~1b, Type~2a, Type~2b, and Type~2c \citep{1998Qiao, 2001Qiao, 2024Qu}. Type~1a represents a core-dominated pulsar, characterised by a small LOS impact angle. They usually exhibit a single core component at lower radio frequencies, but at higher frequencies they evolve to generate a three-component profile. \citet{2026Roy} have recently simulated the Type~1a profile from first principles, showing the component emerging at higher radio frequencies by incorporating inverse Compton scattering (ICS) interactions with a low-frequency pump photon at 10~MHz. Once the scattered beam is formed, it undergoes spectral convolution with the curvature radiation beam, thereby modulating the pulse profile morphology \citep{2026Roy}. PSR~B1933+16 is a standard example of a new component emerging beyond $\sim$1.4~GHz, dominated by a single core. Type~1b is characterised by a higher LOS impact angle, exhibits a triple component at lower frequencies, but evolves to a double cone profile at the higher frequencies, with a standard example being PSR~B1845$-$01 \citep{2001Qiao}. Type~2a belongs to the conal-dominated category and shows up to five components at typical radio frequencies. Due to the lower LOS impact angle, the number of components is reduced to three at low radio frequencies. An example of this type is PSR~B1237+35 \citep{2001Qiao}. Type~2b has a somewhat larger LOS impact angle, shows three components at lower frequencies and evolves to a 4-component profile in higher bands. Type~2c has the highest LOS impact angle, hence only the outer conal rings remain visible, producing a conal double profile at all frequencies. A standard example is PSR~B0525+21 \citep{2001Qiao}. Such diverse profile morphologies carry broad implications, and ICS is a potentially strong mechanism for explaining them and their frequency evolution. To investigate this further, we need a larger sample of pulsars with high-quality, multi-band pulse profiles, which would allow us to understand the flux variability of single pulses, profile frequency evolution, and modulation properties in greater detail. This paper is the first step in that direction.

Our pulse profile evolution analysis shows qualitative agreement with the reported trends \citep{2012Hassall, 2014Chen, 2021Posselt}, particularly in the narrowing of component separation at higher frequencies. Our measured intrinsic power law indices are in agreement. Similar behaviour is also evident from \citet{2016Pilia} (see their Fig.~7), and \citet{2019Olszanski}. Overall, the frequency evolution of the profiles in our dataset is broadly consistent with these earlier results.

\subsection{Mode switching}

Mode switching affects the integrated profiles and single-pulse modulation properties of pulsars, typically on short timescales of several minutes to hours \citepalias{2025Jankowski}, and there is an emerging suspicion that most, or even all, pulsars exhibit more subtle profile variations on long timescales of several years. Indeed, some of our pulsars exhibit mode switching. We account for this in our observation scheduling, focusing on simultaneous or quasi-simultaneous observations across at least two or three bands (Tab.~\ref{tab:observations}). This way, the integrated profiles and modulation parameters are derived from the same mode-switching sequences. Moreover, the differences in profile morphology between the modes are typically small, with PSR~B1822$-$09 having some of the most dramatic changes in the current sample \citepalias{2025Jankowski}. We also presented PSR~B0823+26's Q and B-mode separated profiles in this work, as the pulsar serendipitously spent an entire observing epoch in its Q-mode. A more detailed mode sequencing analysis is in preparation.

%
% Conclusions
%

\section{Conclusions}
\label{sec:conclusions}

We presented integrated pulse profiles and modulation index curves for 12 pulsars at up to five frequency bands. Our wideband dataset was systematically acquired with the ORN telescopes and the uGMRT, covering the frequency range 10~MHz to 2.8~GHz, with most observations performed simultaneously or near simultaneously. Using this multi-frequency dataset, we systematically measured the pulse widths and scattering times in typically eight frequency sub-bands per instrument using multi-component profile fitting, introducing the second-moment width $W_\text{d4s}$. As a by-product of this work, we measured updated scattering-corrected pulsar DMs from our NenuFAR data.

We employed a profile width scaling model to separate the profile evolution into scattering and intrinsic contributions, and we compared the resulting scattering indices with direct fits to our $\tau_\text{s}$ measurements, which are consistent. Two of the pulsars have scattering indices compatible with Kolmogorov turbulence ($-4.4$), while the majority have flatter scattering indices between $-3$ and $-0.5$. Given the low pulsar DMs, this hints at more complex scattering environments with localised turbulence in dense filaments. Complex scattering behaviour is observed in some pulsars, evident as deviations from a simple power law scaling of $\tau_\text{s}$. Most pulsar profiles are scattering-dominated below 200~MHz and substantially in the NenuFAR band.

The measured intrinsic power law indices are small, with $\left| \alpha \right|$ between 0 and $\sim$0.3. Three groups were seen. Five pulsars have $\alpha < 0$ (decreasing intrinsic width), two or three pulsars have $\alpha \approx 0$ (flat), and four pulsars have $\alpha > 0$ (increasing). We explain $\alpha < 0$ as an intrinsic narrowing of the radio emission envelope (radio beam size) and a reduction in component separation with increasing frequency. Physically, this is a geometric and emission height effect in the traditional RFM picture. $\alpha > 0$ occurred primarily due to the appearance of profile components that were suppressed (or merged) at low frequencies. We explain this as differences in the spectral indices of the components in the standard nested hollow cone model. $\alpha \approx 0$ seems to be a special case, where the competing effects roughly balance. Alternatively, these pulsars might have a fine-tuned component spectral index and viewing geometry.

PSRs~B0329+54 and B1237+25 exhibit asymmetrically distributed emission components across pulse phase around their central cores. We conjecture that their emission is strongly influenced by the A/R effect.

We computed two modulation parameters ($m_\text{min}$ \& $\bar{m}$) from the pulsars' band-integrated modulation index curves and determined a mean single-pulse threshold $\bar{\text{S/N}}_\text{th} \approx 18$, above which their S/N bias was insignificant. Accounting for S/N bias and diffractive scintillation, most of our pulsars' (7/12) modulation parameters decrease with frequency (D), while three pulsars have roughly flat modulation parameters (F), and the behaviour of two pulsars is unclear (U). This means that the single-pulse emission from most of our pulsars becomes more erratic towards lower radio frequencies, coincident with a general increase in dynamic range and amplitude modulation.

\section{Data availability}
\label{sec:dataavailability}

We share the integrated pulse profiles from our NenuFAR, LOFAR FR606, uGMRT, and NRT observations in the Zenodo repository \url{https://doi.org/10.5281/zenodo.18603227}. Other high-level data underlying this article will be shared on reasonable request to the corresponding author.

\begin{acknowledgements}
% NenuFAR
This paper is partially based on data obtained using the NenuFAR radio-telescope. The development of NenuFAR has been supported by personnel and funding from: Observatoire Radioastronomique de Nan\c{c}ay, CNRS-INSU, Observatoire de Paris-PSL, Universit\'{e} d'Orl\'{e}ans, Observatoire des Sciences de l’Univers en Région Centre, Région Centre-Val de Loire, DIM-ACAV and DIM-ACAV + of Région Ile-de-France, Agence Nationale de la Recherche.
% LOFAR
LOFAR, the Low Frequency Array designed and constructed by ASTRON, has observing, data processing, and data storage facilities in several countries, that are owned by various parties (each with their own funding sources), and that are collectively operated by the ILT foundation under a joint scientific policy. The ILT resources have benefitted from the following recent major funding sources: CNRS-INSU, Observatoire de Paris and Universit\'{e} d'Orl\'{e}ans, France; BMBF, MIWF-NRW, MPG, Germany; Science Foundation Ireland (SFI), Department of Business, Enterprise and Innovation (DBEI), Ireland; NWO, The Netherlands; The Science and Technology Facilities Council, UK.
% FR606
LOFAR station FR606 is hosted by the Nan\c{c}ay Radio Observatory and is operated by Paris Observatory, associated with the French Centre National de la Recherche Scientifique (CNRS) and Universit\'{e} d'Orl\'{e}ans.
% GMRT
We thank the GMRT staff for making these observations possible. The GMRT is run by the National Centre for Radio Astrophysics of the Tata Institute of Fundamental Research.
% ORN
The Nan\c{c}ay Radio Observatory (ORN) is operated by Paris Observatory, associated with the French Centre National de la Recherche Scientifique (CNRS) and Universit\'{e} d'Orl\'{e}ans. It is partially supported by the Region Centre Val de Loire in France.
% CDN
We acknowledge the use of the Nan\c{c}ay Data Centre computing facility (CDN -- Centre de Donn\'{e}es de Nan\c{c}ay). The CDN is hosted by the Observatoire Radioastronomique de Nan\c{c}ay in partnership with Observatoire de Paris, Universit\'{e} d'Orl\'{e}ans, OSUC and the CNRS. The CDN is supported by the R\'{e}gion Centre Val de Loire, d\'{e}partement du Cher.
% F.J., J.P.
This work has been supported by ANR-20-CE31-0010.
% T.R.
T.~R.\ acknowledges the support from the National Science Centre, Poland, grant No.\ 2023/49/B/ST9/01783.
% I.K.
I.~K.\ acknowledges the support of NAS of Ukraine by a Grant for Research of Young Scientists of the NAS of Ukraine (2025-2026, state registration number - 0125U002903, project code ''Dyspersiia'').
% I.K, V.Z., O.U.
I.~K., V.~Z., and O.~U.\ acknowledge the support of the Project: 101131928 – ACME – HORIZON-INFRA-2023-SERV-01 and the Ukrainian Program ''Scientific and scientific and technical (experimental) work in the priority area Radiophysical and optical systems for strengthening the defense capability of the state'' for 2025-2026 – ''Global monitoring of radio signals of natural and artificial origin of decametre-metre waves in the interests of cosmology and applied problems of defense capability'' (state registration number - 0125U000879).

\end{acknowledgements}


\bibliographystyle{aa} % style aa.bst
\bibliography{references} % your references Yourfile.bib

%%%%%%%%%%%%%%%%%%%%%%%%%%%%%%%%%%%%%%%%%%%%%%%%%%
%%%%%%%%%%%%%%%%%%%% APPENDICES %%%%%%%%%%%%%%%%%%
%%%%%%%%%%%%%%%%%%%%%%%%%%%%%%%%%%%%%%%%%%%%%%%%%%

\begin{appendix}

% switch to one-column layout
\clearpage
\onecolumn

\section{Observations}
\label{ap:observations}

\begin{table}[h]
\caption{Details of the uGMRT, NenuFAR, LOFAR FR606, and NRT data presented in this work.}
\label{tab:observations}
\begin{tabular}{lccccccccccc}
\hline\hline
PSR         & Date              & Start UT      & $t_\text{obs}$    & $\nu_\text{c}$    & $b$   & m &  $N_\text{chan}$  & $t_\text{samp}$   & $N_\text{ai}$ & $N_\text{ap}$ & $N_\text{p}$\\
            & (ymd)             & (hh:mm)       & (min)             & (MHz)             & (MHz) &       &   & ($\mu \text{s}$)   &   &   &\\
\hline
\multicolumn{12}{c}{uGMRT}\\
B0031$-$07  & 2023-12-05    & 13:19         & $4 \times 44$     & 650               & 200   & PA    & 2048  & 81.92 & 28 & 16    & 11\,198\\
B0329+54    & 2023-04-25        & 03:22         & 50 + 50           & 650               & 200   & PA    & 2048  & 81.92 & 27    & 15    & 8395\\
B0809+74    & 2023-04-25        & 10:09         & 50 + 50           & 650               & 200   & PA    & 2048  & 81.92 & 27,30 & 16,17 & 4643\\
B0823+26\tablefootmark{a}       & 2023-04-25    & 12:05             & 50 + 50           & 650   & 200   & PA    & 2048  & 81.92 & 30    & 17    & 11\,311\\
B0823+26    & 2023-12-05    & 17:16 & 50 + 50   & 650   & 200   & PA    & 2048  & 81.92 & 28    & 16    & 11\,303\\
B0834+06    & 2023-04-25        & 15:04         & 50 + 50           & 650               & 200   & PA    & 2048  & 81.92 & 29 & 17    & 4709\\
B0919+06    & 2024-05-05    & 12:14 & 55 + 55           & 650               & 200   & CD    & 512   & 40.96 & 29    & 17    & 15\,322\\
B0943+10    & 2023-12-05    & 19:22 & 51 + 51   & 650               & 200   & PA    & 2048  & 81.92 & 28    & 16    & 5576\\
B0950+08    & 2024-05-05        & 10:07         & 55 + 55           & 650               & 200   & CD    & 512   & 40.96 & 29    & 17    & 26\,026\\
B1112+50    & 2023-12-05        & 21:18         & 51 + 51           & 650               & 200   & PA    & 2048  & 81.92 & 28    & 16    & 3691\\
B1133+16    & 2023-04-25        & 16:06     & $2 \times 50$ + 15    & 650               & 200   & PA    & 2048  & 81.92 & 30 & 17    & 5809\\
B1237+25    & 2023-12-05        & 23:17         & 52 + 52           & 650               & 200   & PA    & 2048  & 81.92 & 28    & 16    & 4514\\
B1822$-$09  & 2023-04-24        & 22:22         & 56 + 56           & 650               & 200   & PA    & 2048  & 81.92 & 29    & 16    & 8743\\
\hline
\multicolumn{12}{c}{NenuFAR}\\
B0031$-$07  & 2023-12-04 & 18:32     & 82                    & 50            & 75     & --    & 384  & 655.36 & --    & 74   & 5190\\
B0329+54    & 2023-04-26        & 12:07     & 110                   & 54            & 75     & --    & 384  & 327.68 & --    & 78   & 9224\\
B0809+74    & 2023-04-25        & 10:02     & 112                   & 47            & 75     & --    & 384  & 327.68 & --    & 78   & 5219\\
B0823+26\tablefootmark{a,b} & 2023-04-25    & 12:26 & 92            & 55            & 75     & --    & 384  & 259.11 & --    & 78   & --\\
B0823+26    & 2023-12-05    & 03:02     & 80                    & 52            & 75     & --    & 384  & 655.36 & --    & 74   & 9085\\
B0834+06    & 2023-04-25        & 14:02     & 111                   & 50            & 75     & --    & 384  & 327.68 & --    & 78   & 5241\\
B0919+06    & 2024-04-23        & 18:02     & 110                   & 54            & 75     & --    & 384  & 327.68 & --    & 71   & 15\,285\\
B0943+10    & 2023-12-05        & 04:32     & 81                    & 51            & 75     & --    & 384  & 655.36 & --    & 74   & 4433\\
B0950+08    & 2024-05-05        & 17:02     & 113                   & 45            & 75     & --    & 384  & 327.68 & --    & 71   & 26\,828\\
B1112+50    & 2023-12-05        & 06:02     & 82                    & 49            & 75     & --    & 384  & 655.36 & --    & 74   & 2961\\
B1133+16    & 2023-04-25        & 16:02     & 113                   & 47            & 75     & --    & 384  & 327.68 & --    & 78   & 5694\\
B1237+25    & 2023-12-05        & 07:32     & 82                    & 49            & 75     & --    & 384  & 655.36 & --    & 74   & 3547\\
B1822$-$09  & 2023-04-27        & 03:02     & 110                   & 52            & 75     & --    & 384  & 327.68 & --    & 78   & 8603\\
\hline
\multicolumn{12}{c}{FR606 HBA}\\
B0031$-$07  & 2023-12-04       & 18:30     & 84                & 150            & 95     & --    & 488  & 327.68  & --    & 95    & 5360\\
B0329+54\tablefootmark{c}    & 2025-12-05    & 20:56 & 89 + 89   & 150   & 95    & --    & 488   & 327.68    & --    & 94    & 14\,974\\
B0809+74    & 2023-04-25        & 10:03     & 114              & 150            & 95     & --    & 488  & 655.36  & --    & 92    & 5312\\
B0823+26\tablefootmark{a}    & 2023-04-25    & 12:03 & 115     & 150            & 95     & --    & 488  & 327.68  & --    & 92    & 12\,989\\
B0823+26    & 2023-12-05        & 03:00     & 84               & 150            & 95     & --    & 488  & 327.68  & --    & 95    & 9518\\
B0834+06    & 2023-04-25        & 14:03     & 113              & 150            & 95     & --    & 488  & 655.36  & --    & 92    & 5328\\
B0919+06    & 2024-04-23        & 18:01     & 119              & 150            & 95     & --    & 488  & 655.36  & --    & 94    & 16\,612\\
B0943+10    & 2023-12-05        & 04:30     & 84               & 150            & 95     & --    & 488  & 327.68  & --    & 95    & 4604\\
B0950+08\tablefootmark{c}    & 2025-12-05   & 03:16 & 89 + 89  & 150            & 95     & --    & 488  & 327.68  & --    & 94    & 42\,344\\
B1112+50    & 2023-12-05        & 06:00     & 84               & 150            & 95     & --    & 488  & 327.68  & --    & 95    & 3050\\
B1133+16    & 2023-04-25        & 16:03     & 115              & 150            & 95     & --    & 488  & 655.36  & --    & 92    & 5806\\
B1237+25    & 2023-12-05        & 07:30     & 84               & 150            & 95     & --    & 488  & 327.68  & --    & 95    & 3655\\
B1822$-$09  & 2023-12-05        & 12:00     & 84               & 150            & 95     & --    & 488  & 327.68  & --    & 95    & 6569\\
\hline
\multicolumn{12}{c}{NRT L-band}\\
B0950+08    & 2026-03-12    & 21:51        & 63              & 1484            & 512    & --    & 128  & 64    & -- & -- & 14\,807\\
B1112+50    & 2026-04-05    & 21:39        & 59              & 1484            & 512    & --    & 128  & 64    & -- & -- & 2128\\
B1237+25    & 2026-02-27    & 01:29        & 66              & 1484            & 512    & --    & 128  & 124   & -- & -- & 2863\\
\hline
\end{tabular}
\tablefoot{
We list the observing time ($t_\text{obs}$), centre frequency ($\nu_\text{c}$), digitised bandwidth ($b$), uGMRT observing mode (m), which is either coherently dedispersed (CD) or phased-array (PA), the number of frequency channels ($N_\text{chan}$), sampling time ($t_\text{samp}$), number of antennas in the imaging data stream ($N_\text{ai}$), number of antennas (GMRT) or mini-arrays (NenuFAR \& FR606) in the phased array data stream ($N_\text{ap}$), and the number of single pulses recorded ($N_\text{p}$).
    \tablefoottext{a}{Pulsar was in its Q-mode for the entire duration.}
    \tablefoottext{b}{Only fold mode data were recorded.}
    \tablefoottext{c}{Replacement observation.}
}
\end{table}

\clearpage

\section{Integrated pulse profiles}
\label{ap:profiles}

\begin{figure}[!ht]
  \centering
  % B0031-07, B0329+54, B0809+74
  % nenufar
  \includegraphics[width=0.32\textwidth]{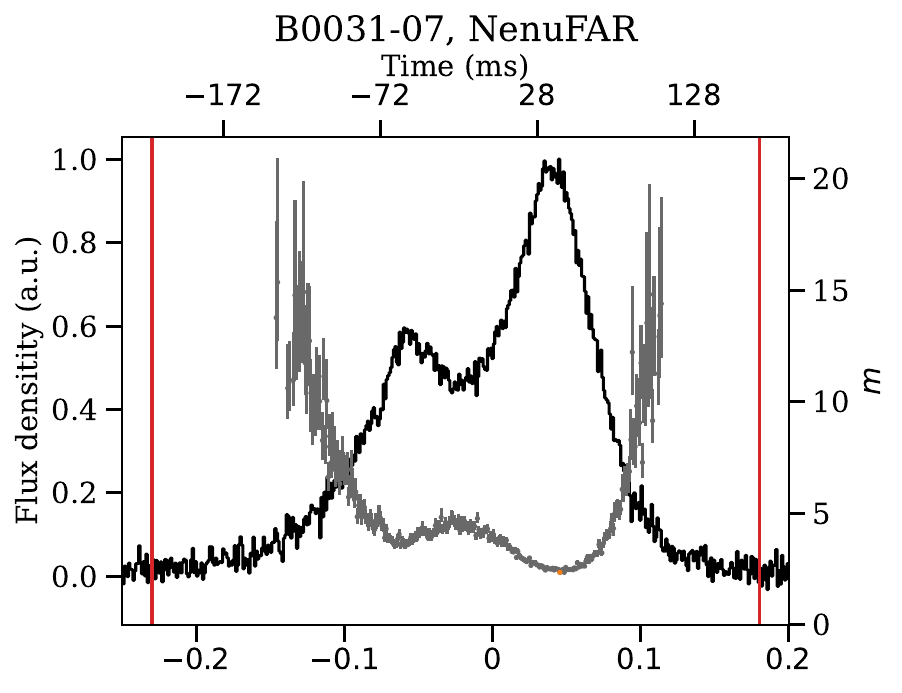}
  \includegraphics[width=0.32\textwidth]{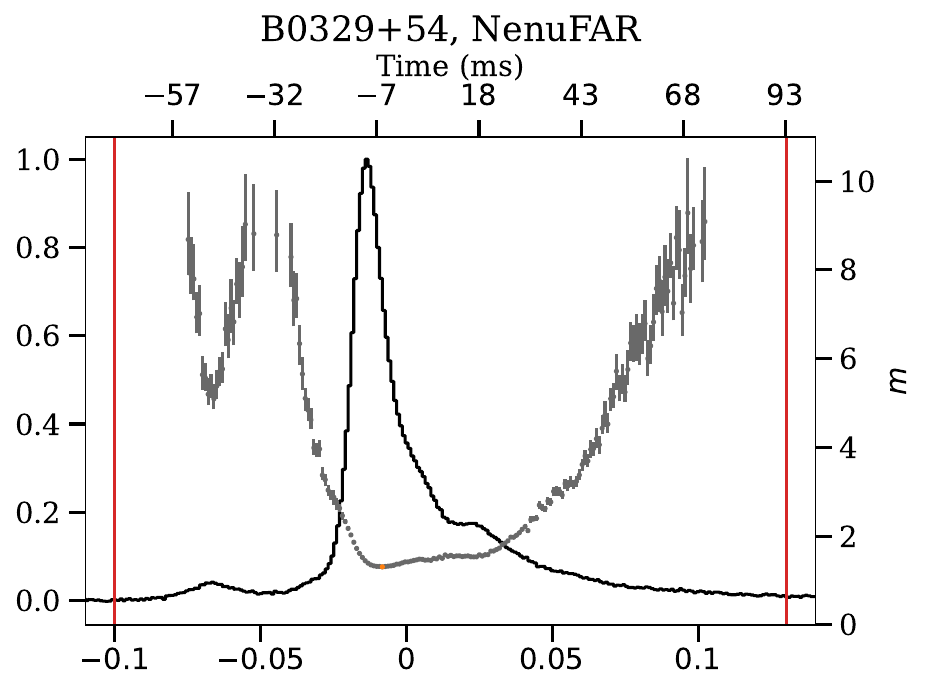}
  \includegraphics[width=0.32\textwidth]{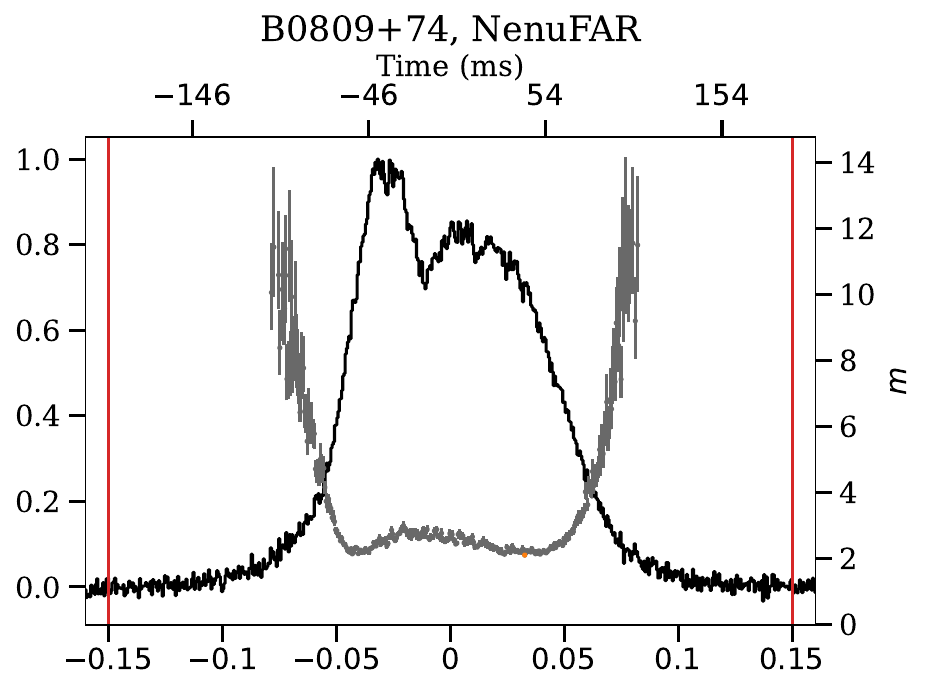}
  % fr606
  \includegraphics[width=0.32\textwidth]{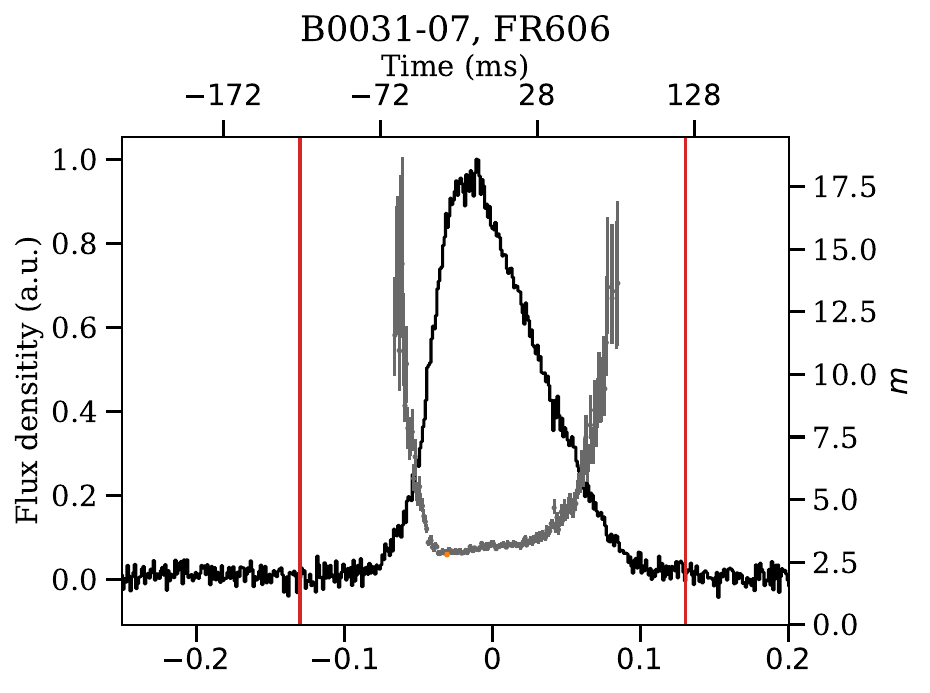}
  \includegraphics[width=0.32\textwidth]{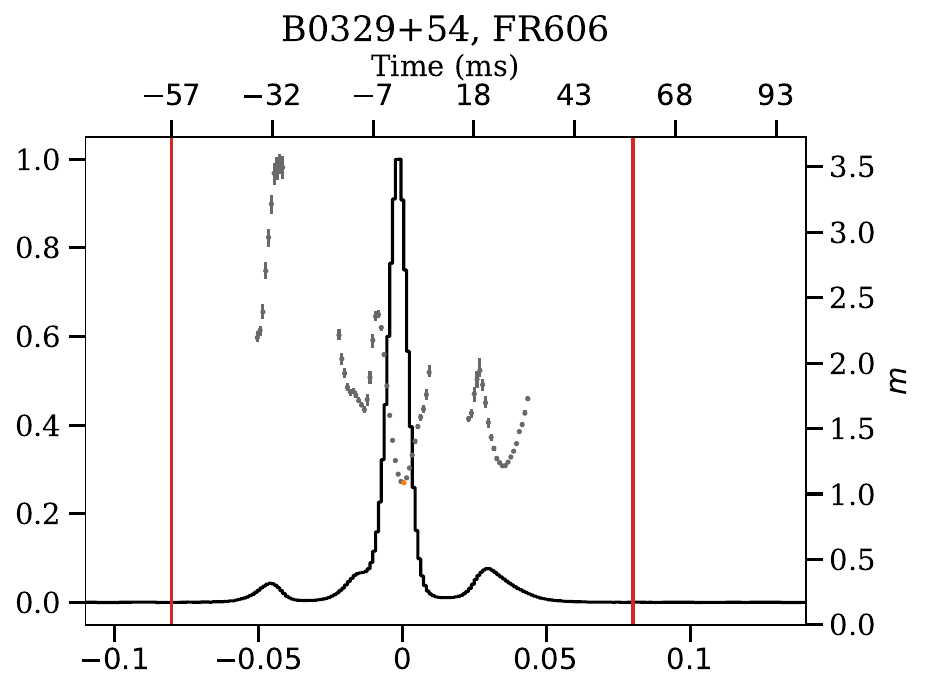}
  \includegraphics[width=0.32\textwidth]{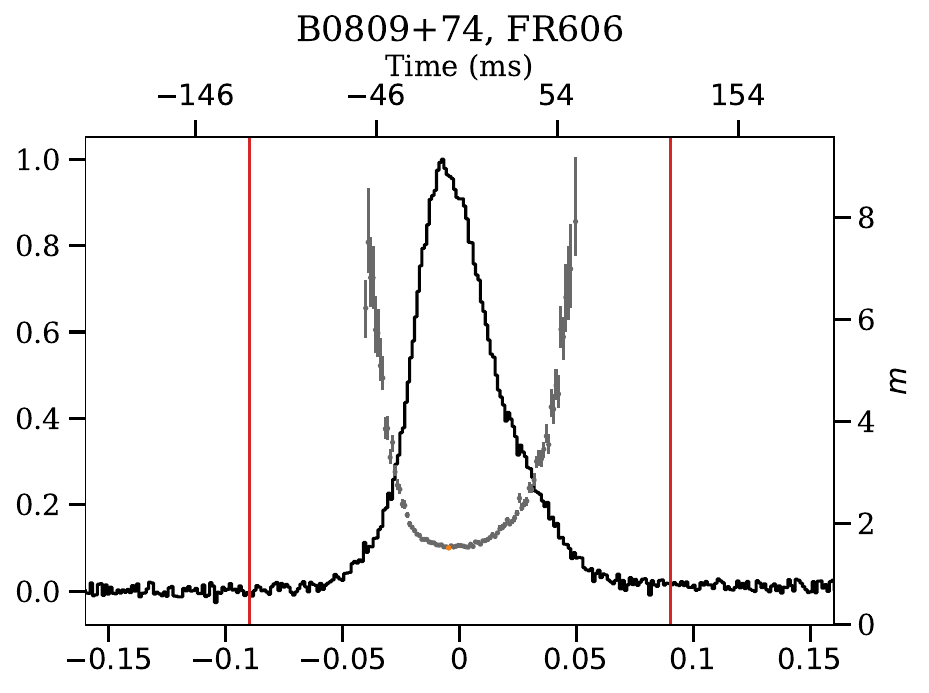}
  % gmrt
  \includegraphics[width=0.32\textwidth]{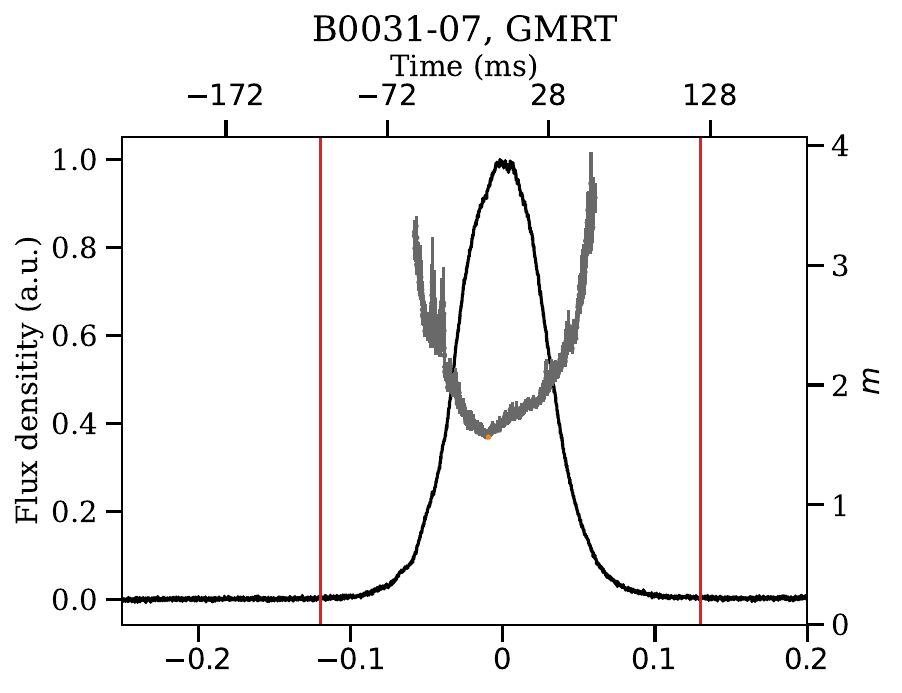}
  \includegraphics[width=0.32\textwidth]{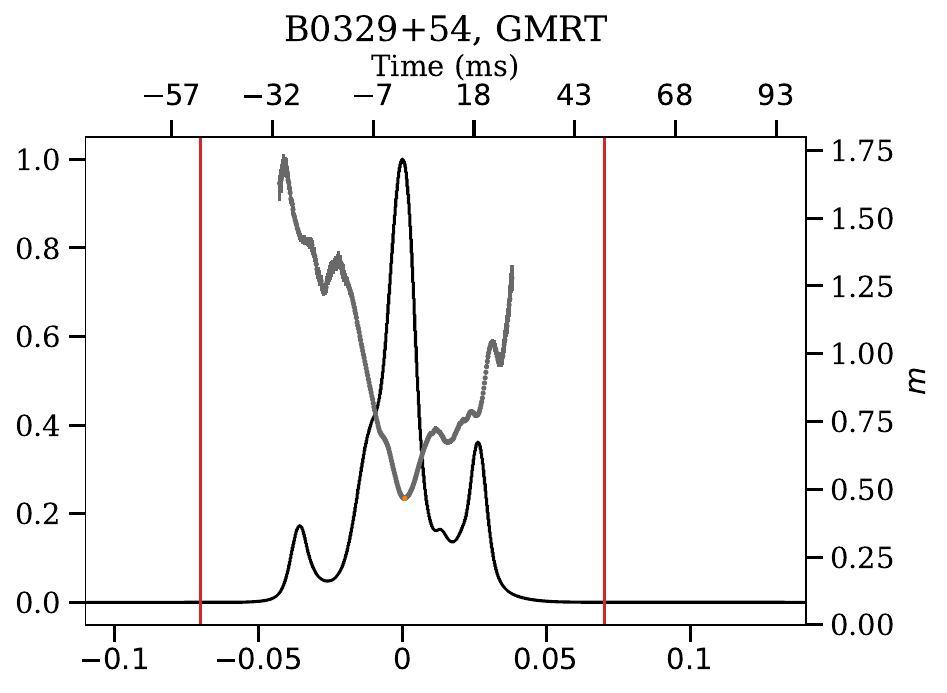}
  \includegraphics[width=0.32\textwidth]{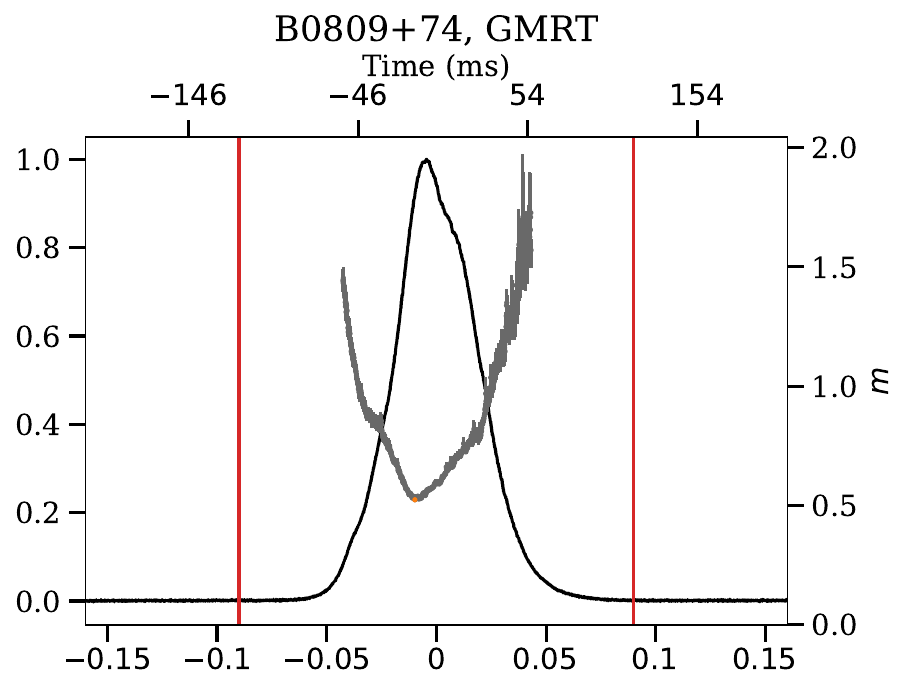}
  % nrt-l
  \includegraphics[width=0.32\textwidth]{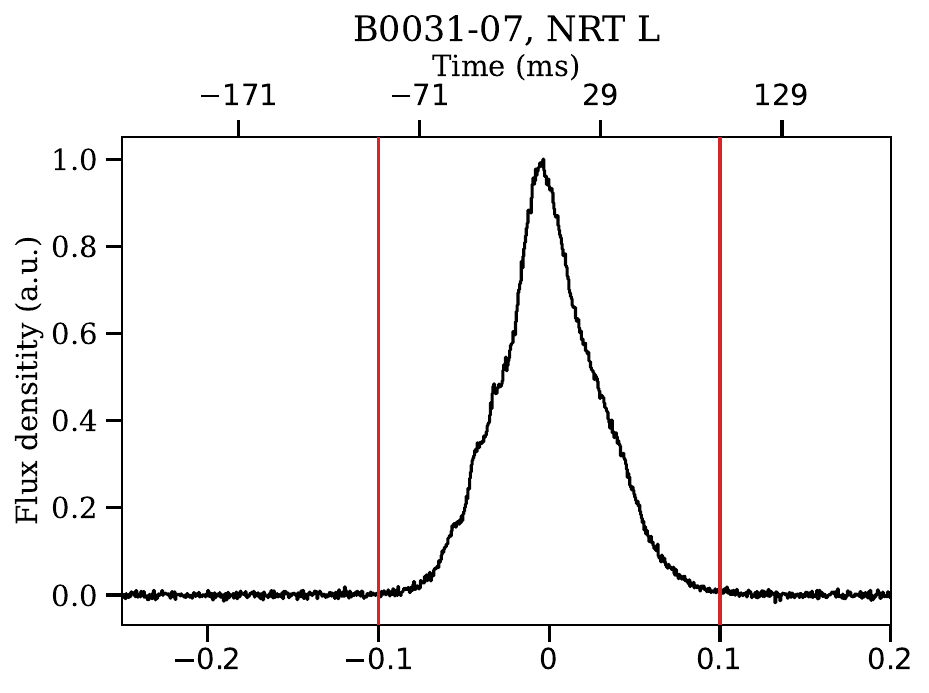}
  \includegraphics[width=0.32\textwidth]{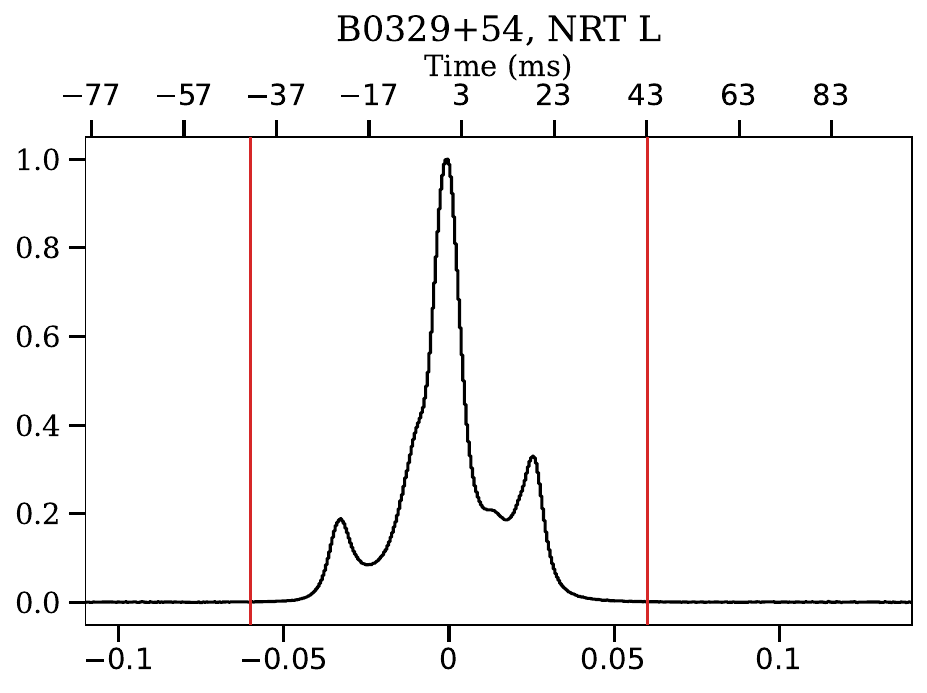}
  \includegraphics[width=0.32\textwidth]{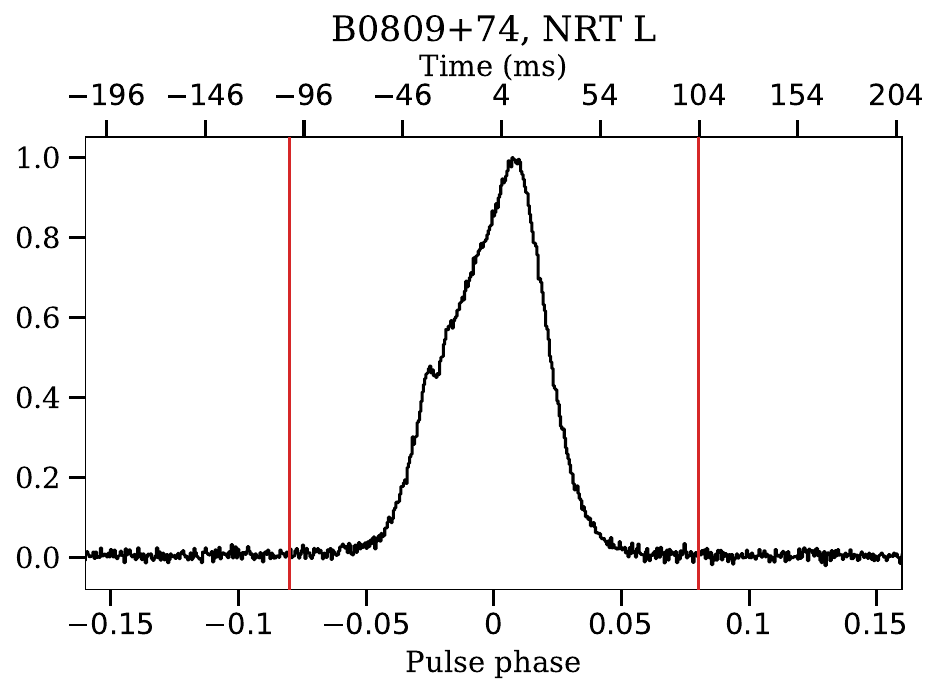}
  % nrt-s
  \includegraphics[width=0.32\textwidth]{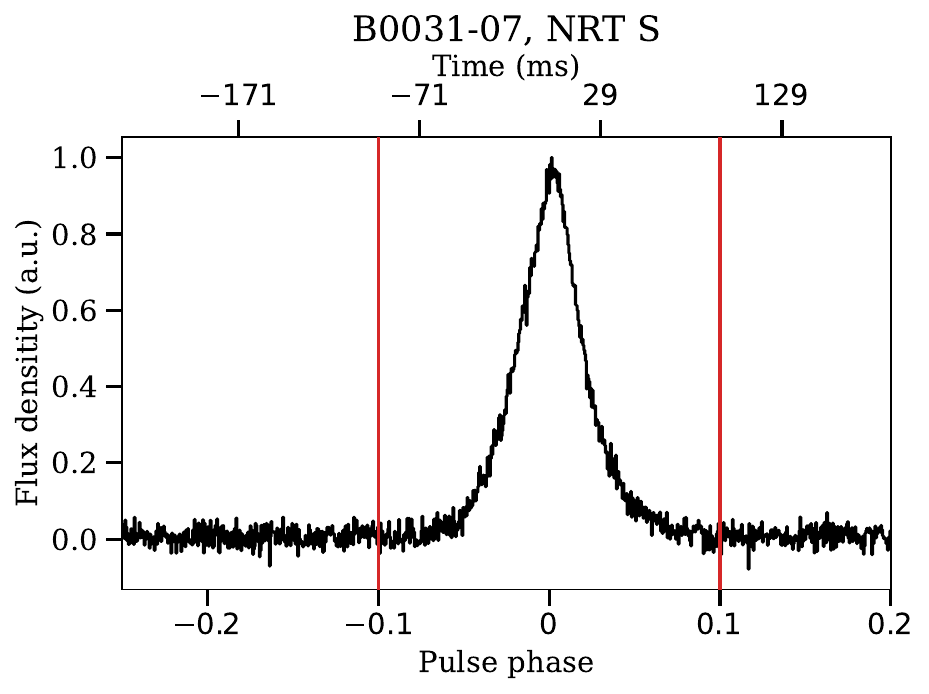}
  \includegraphics[width=0.32\textwidth]{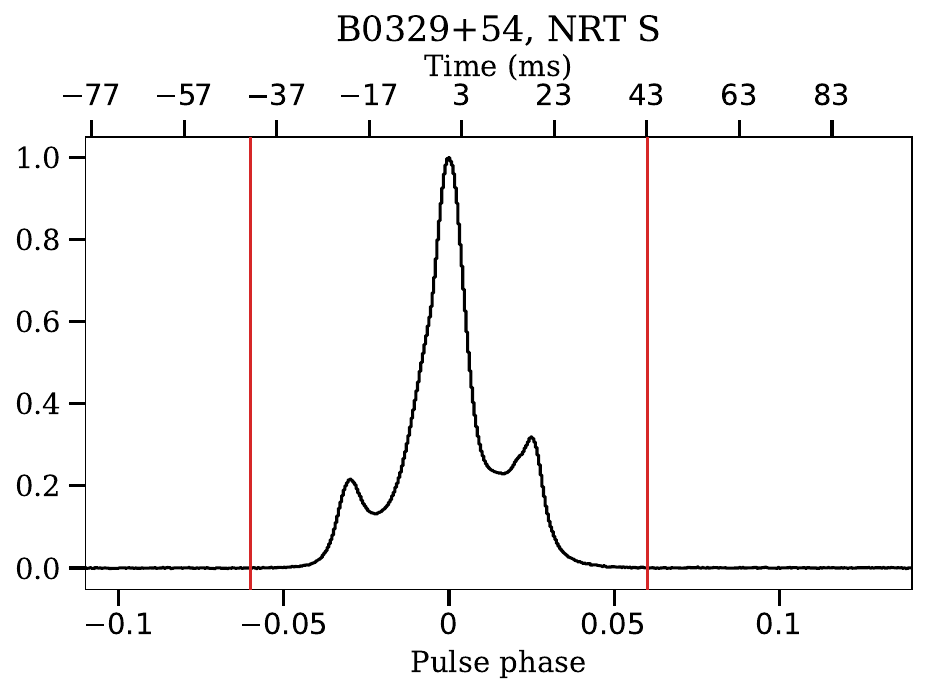}
  \makebox[0.32\textwidth]{\raisebox{0pt}[3cm][0pt]{}}
  \caption{Integrated pulse profiles and phase-resolved modulation indices of our pulsar sample at NenuFAR (50~MHz; first row), FR606 HBA (150~MHz; second row), uGMRT (650~MHz; third row), NRT L-band (1.5~GHz; forth row), and NRT S-band (2.5~GHz; bottom row).}
 \label{fig:profiles1}
\end{figure}

\clearpage

\begin{figure}[!ht]
  \centering
  % B0823+26 Q-mode
  % gmrt
  \includegraphics[width=0.32\textwidth]{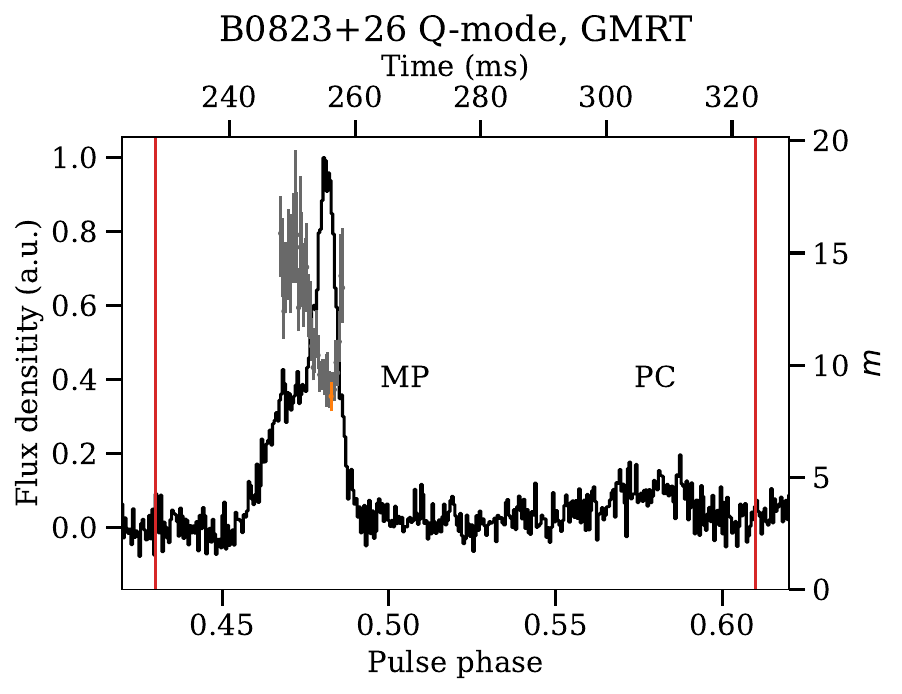}
  \caption{Continuation of Fig.~\ref{fig:profiles1}.}
 \label{fig:profiles2}
\end{figure}

\begin{figure}[!ht]
  \centering
  % B0823+26 B-mode mp, B0834+06, B0919+06
  % nenufar
  \includegraphics[width=0.32\textwidth]{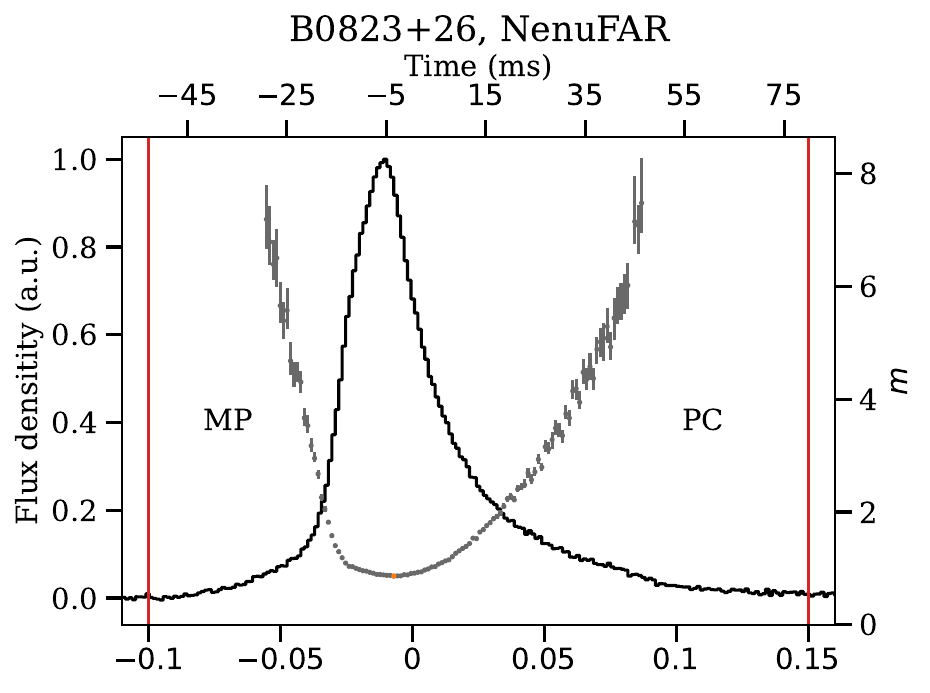}
  \includegraphics[width=0.32\textwidth]{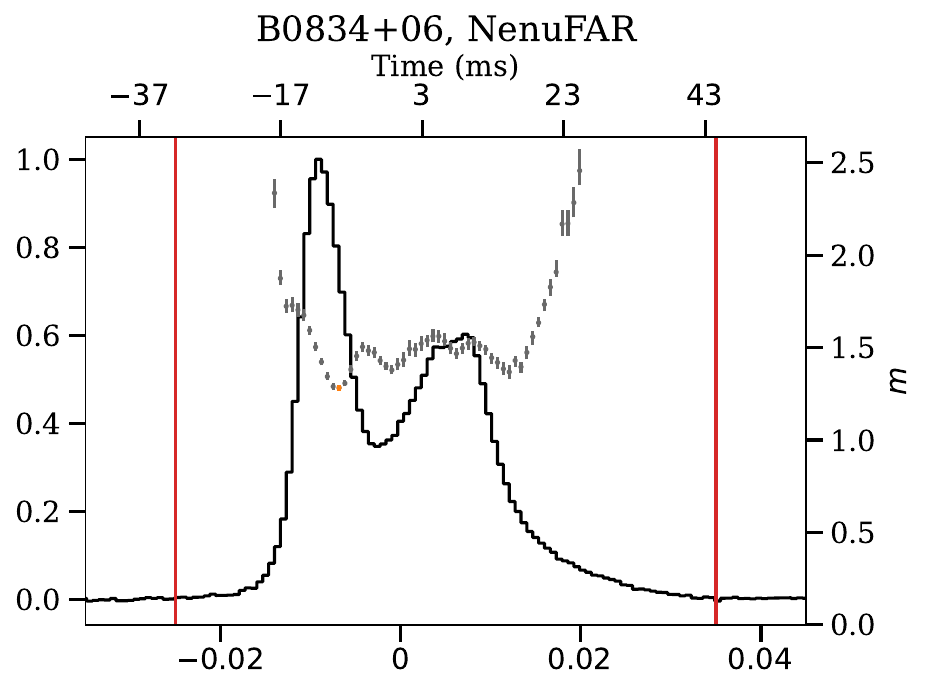}
  \includegraphics[width=0.32\textwidth]{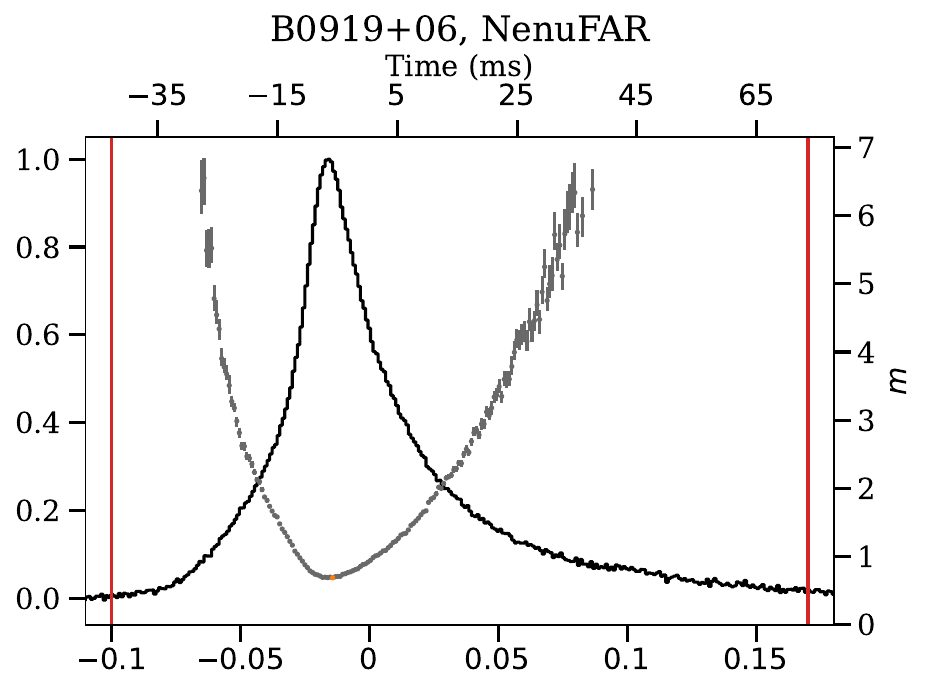}
  % fr606
  \includegraphics[width=0.32\textwidth]{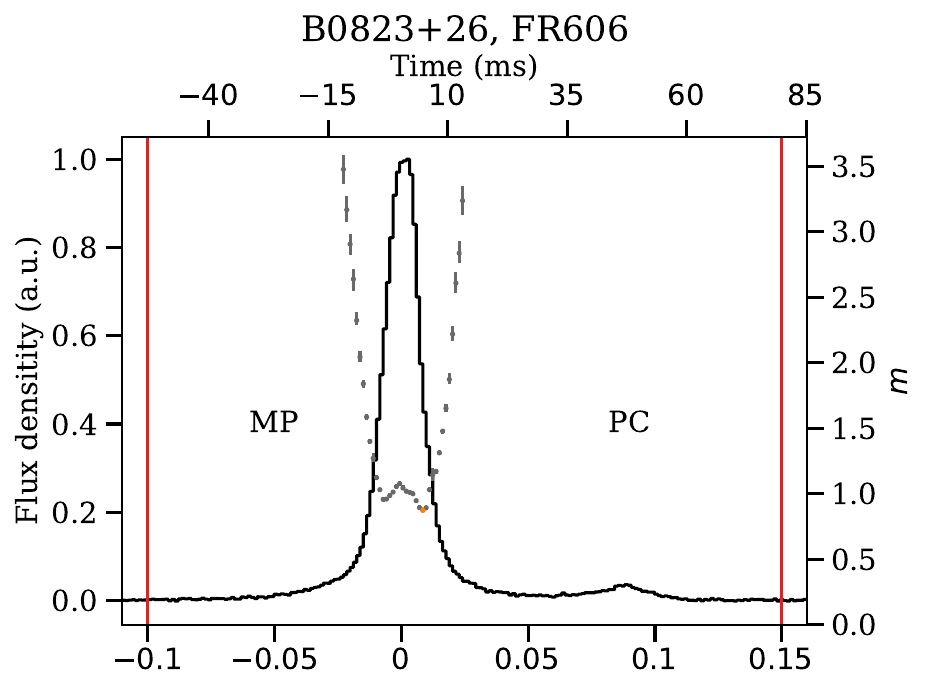}
  \includegraphics[width=0.32\textwidth]{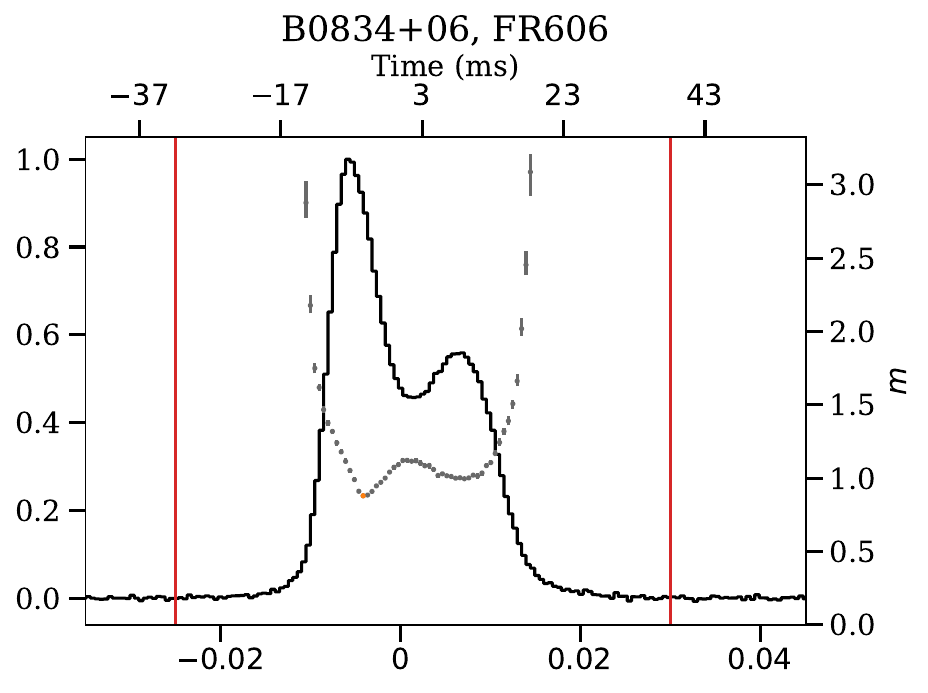}
  \includegraphics[width=0.32\textwidth]{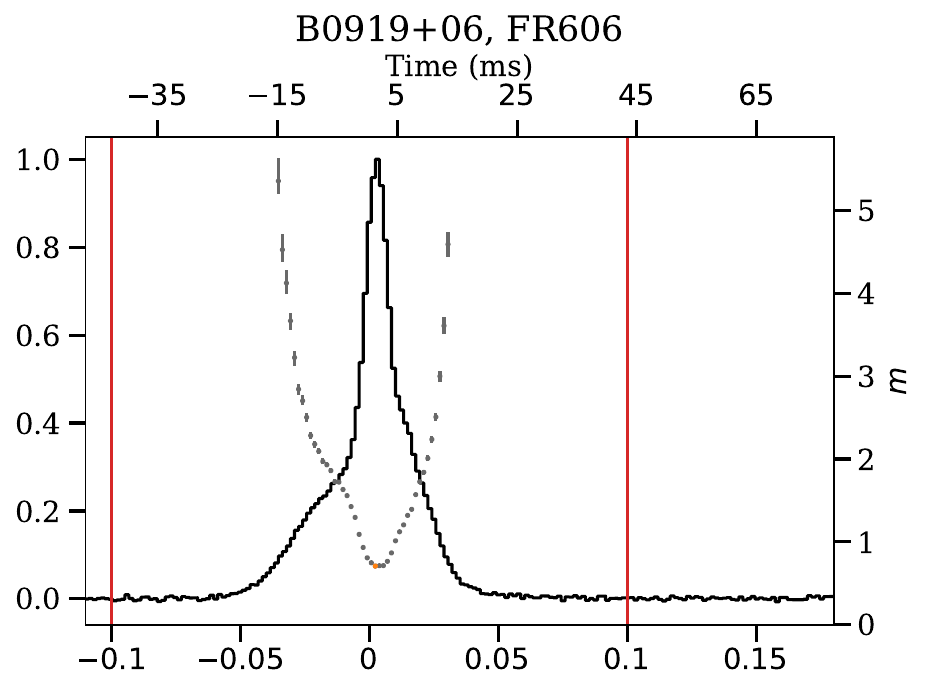}
  % gmrt
  \includegraphics[width=0.32\textwidth]{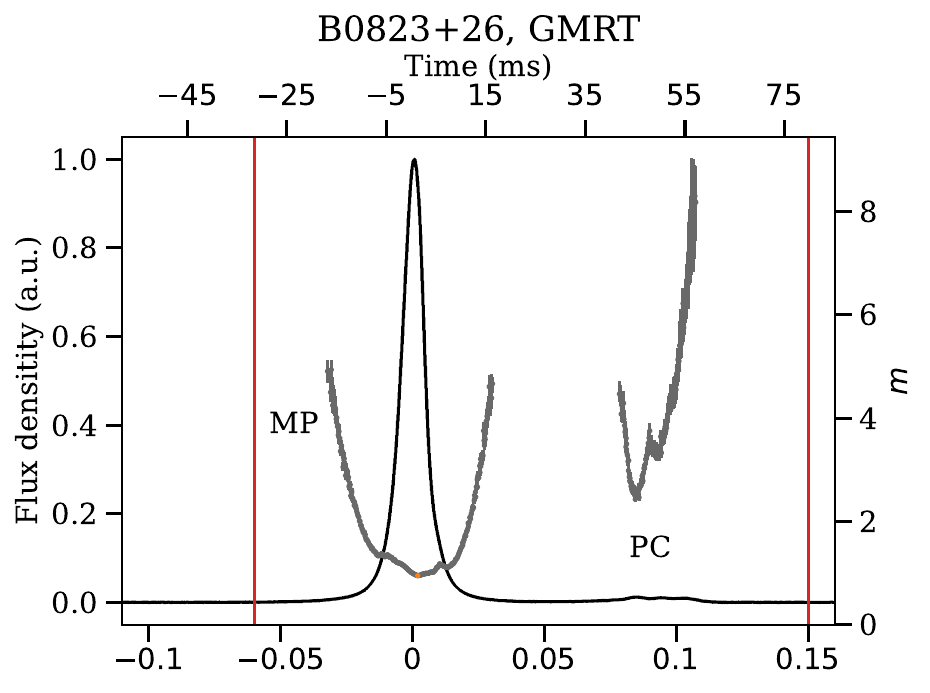}
  \includegraphics[width=0.32\textwidth]{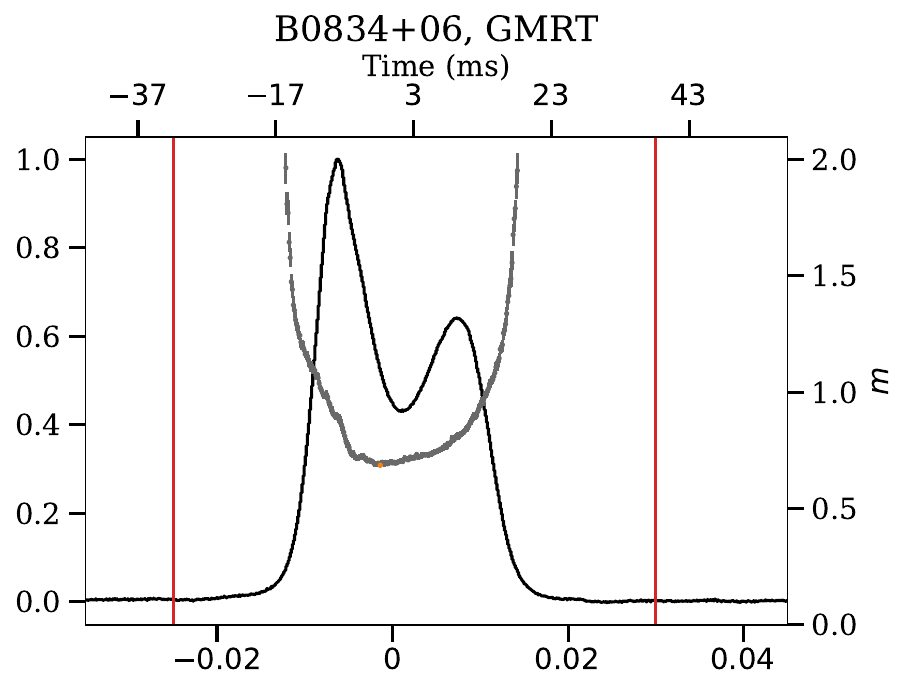}
  \includegraphics[width=0.32\textwidth]{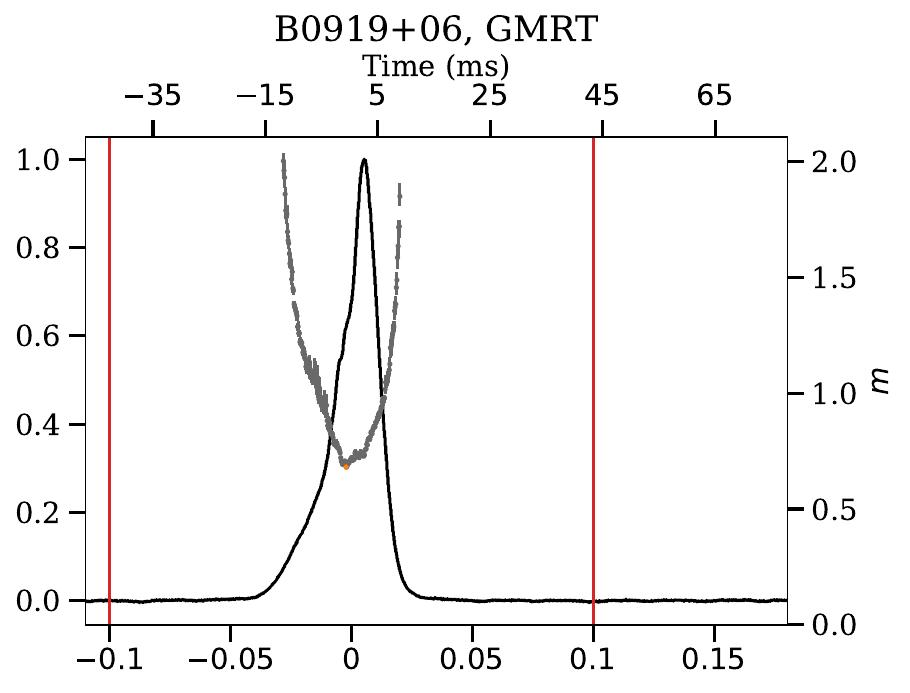}
  % nrt-l
  \includegraphics[width=0.32\textwidth]{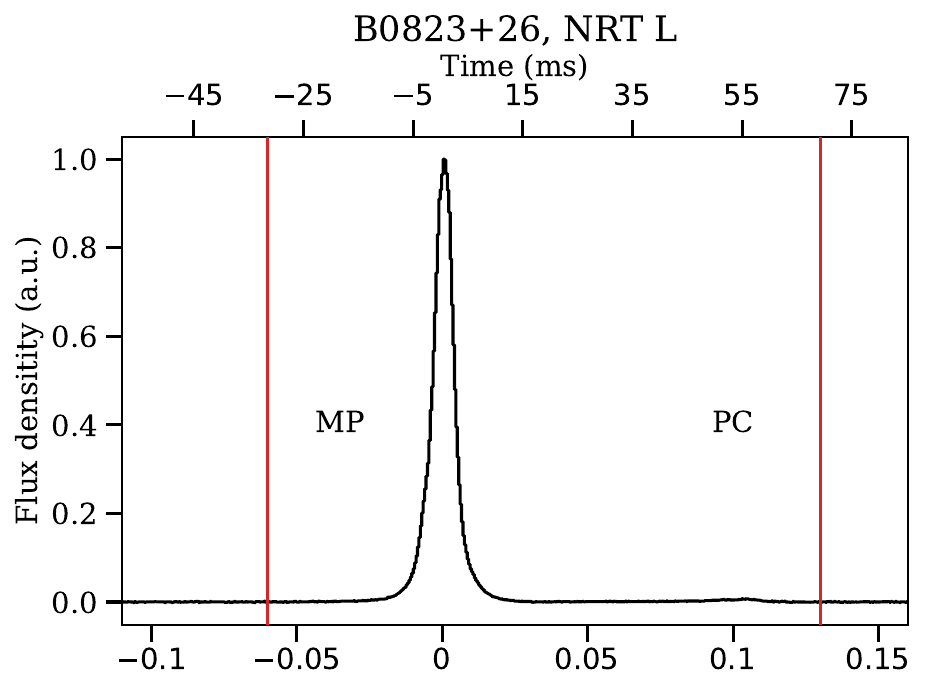}
  \includegraphics[width=0.32\textwidth]{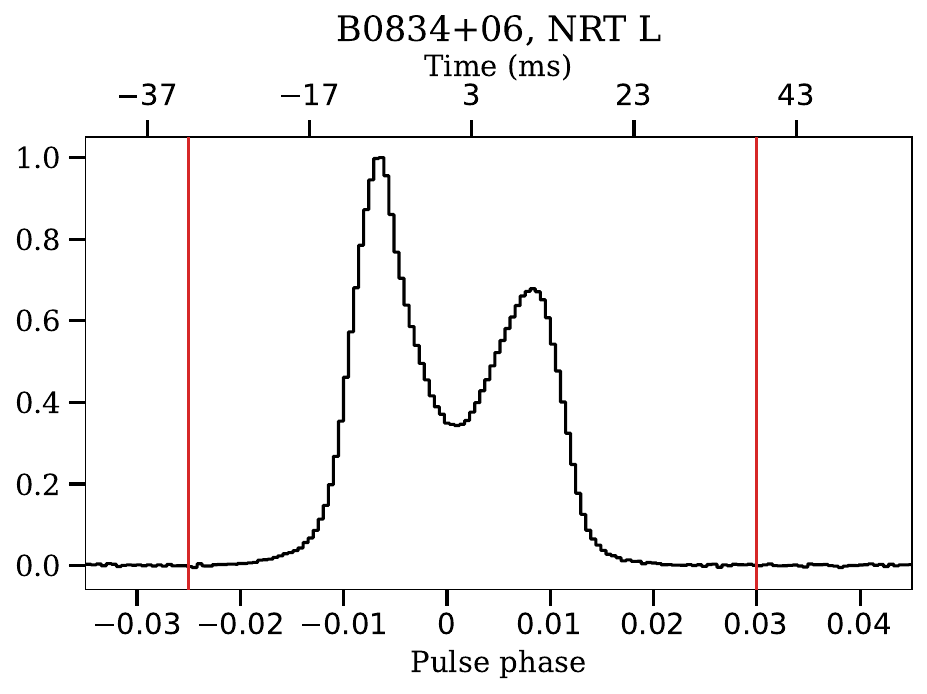}
  \includegraphics[width=0.32\textwidth]{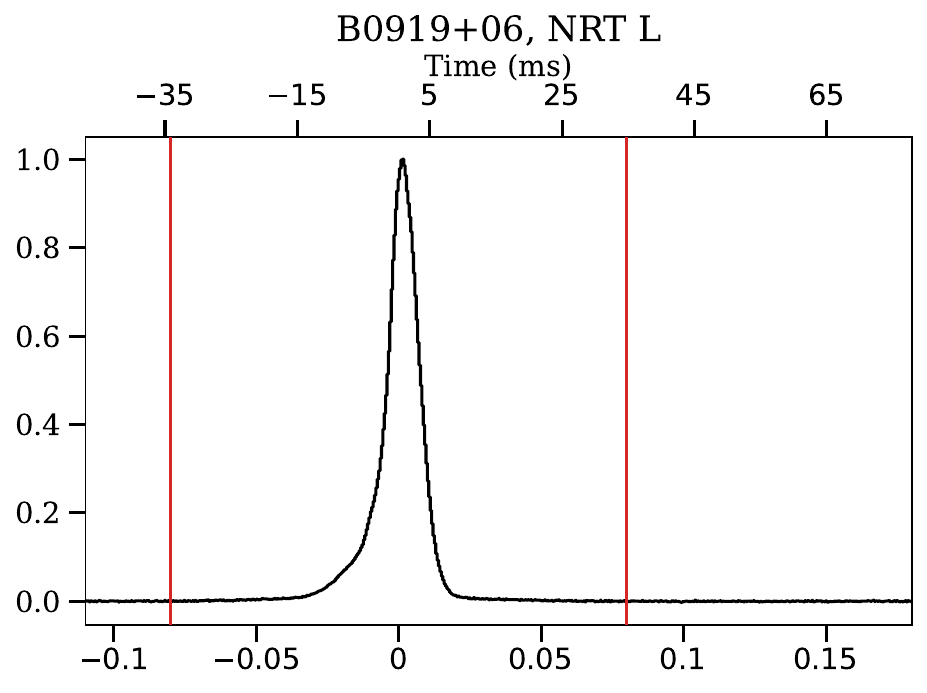}
  % nrt-s
  \includegraphics[width=0.32\textwidth]{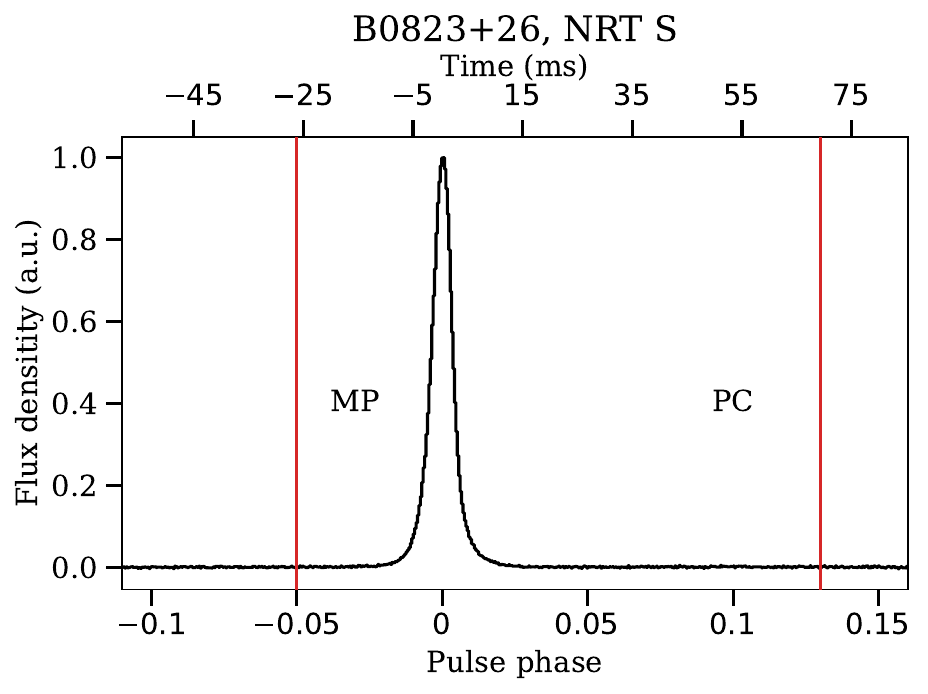}
  \makebox[0.32\textwidth]{\raisebox{0pt}[3cm][0pt]{}}
  \includegraphics[width=0.32\textwidth]{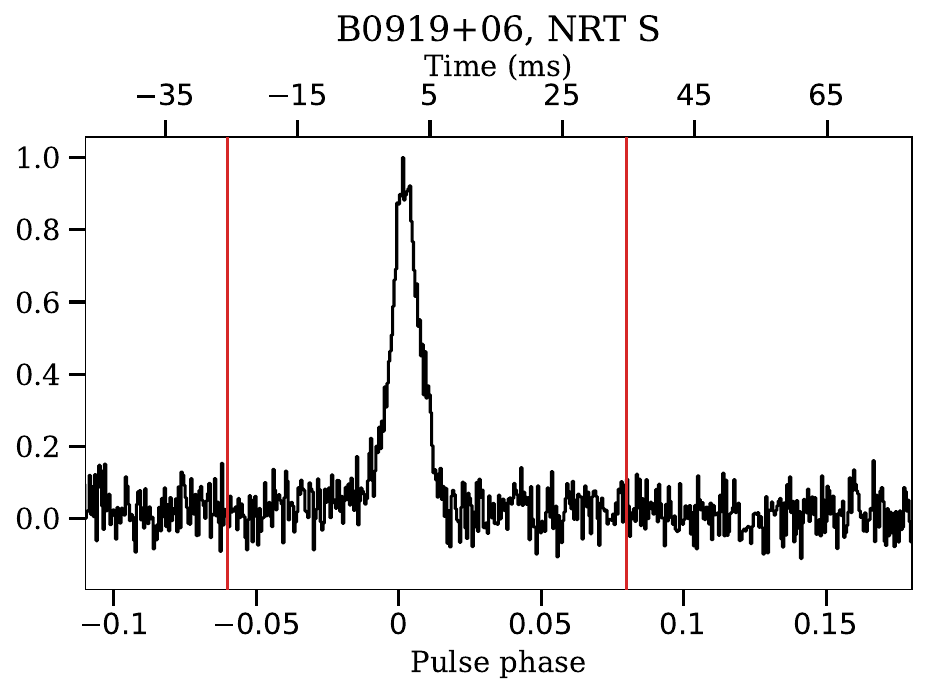}
  \caption{Continuation of Fig.~\ref{fig:profiles1}.}
 \label{fig:profiles3}
\end{figure}

\clearpage

\begin{figure}
  \centering
  % B0943+10, B0950+08 mp, B1112+50
  % nenufar
  \includegraphics[width=0.32\textwidth]{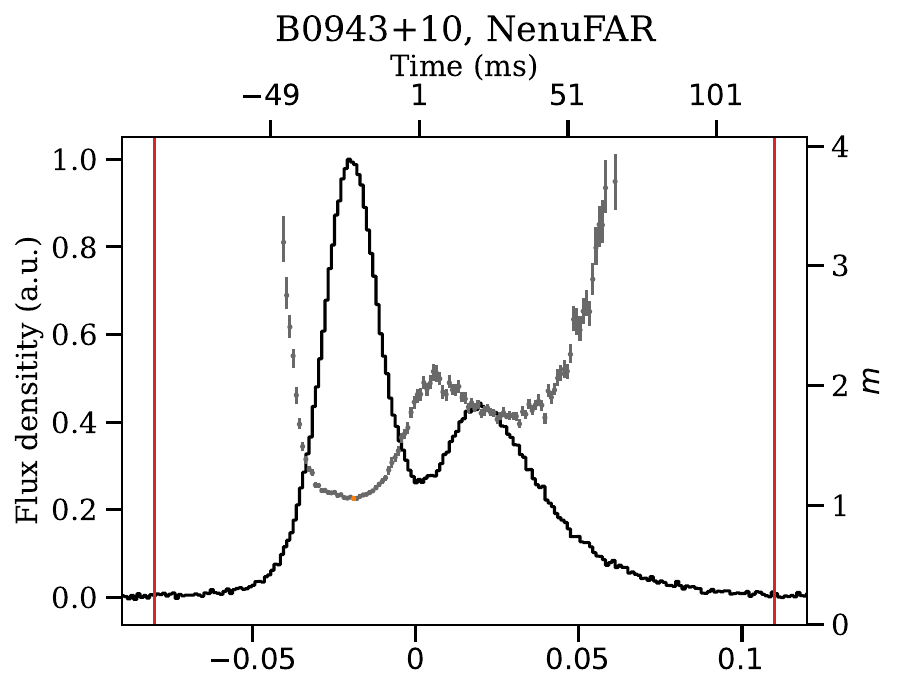}
  \includegraphics[width=0.32\textwidth]{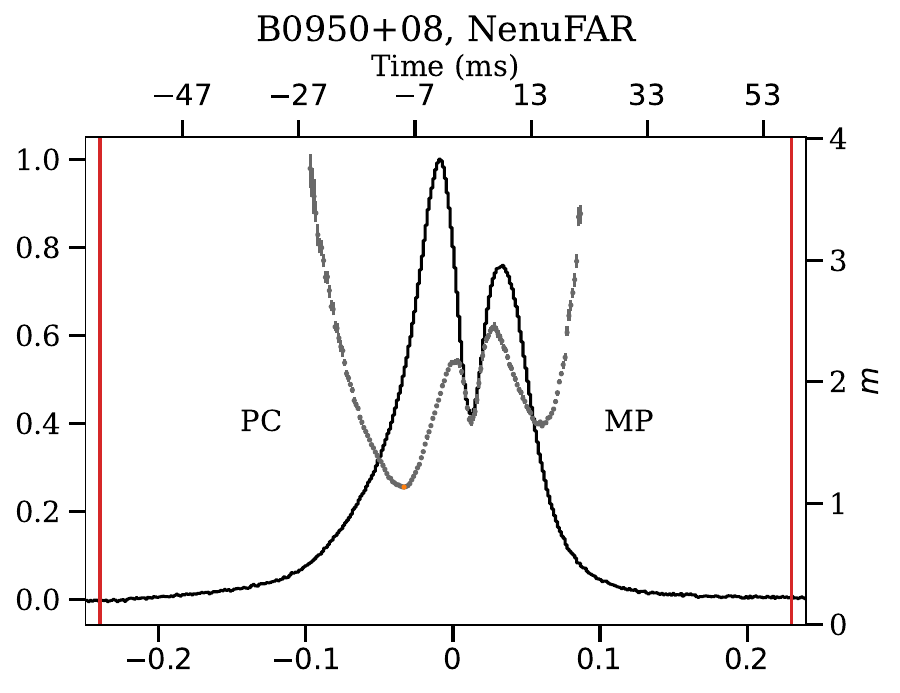}
  \includegraphics[width=0.32\textwidth]{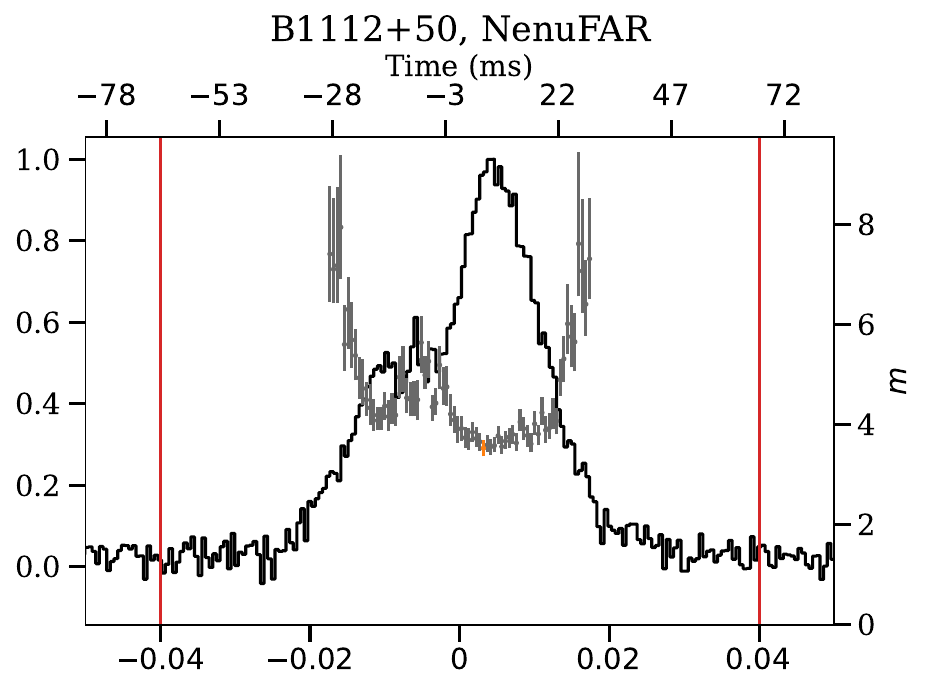}
  % fr606
  \includegraphics[width=0.32\textwidth]{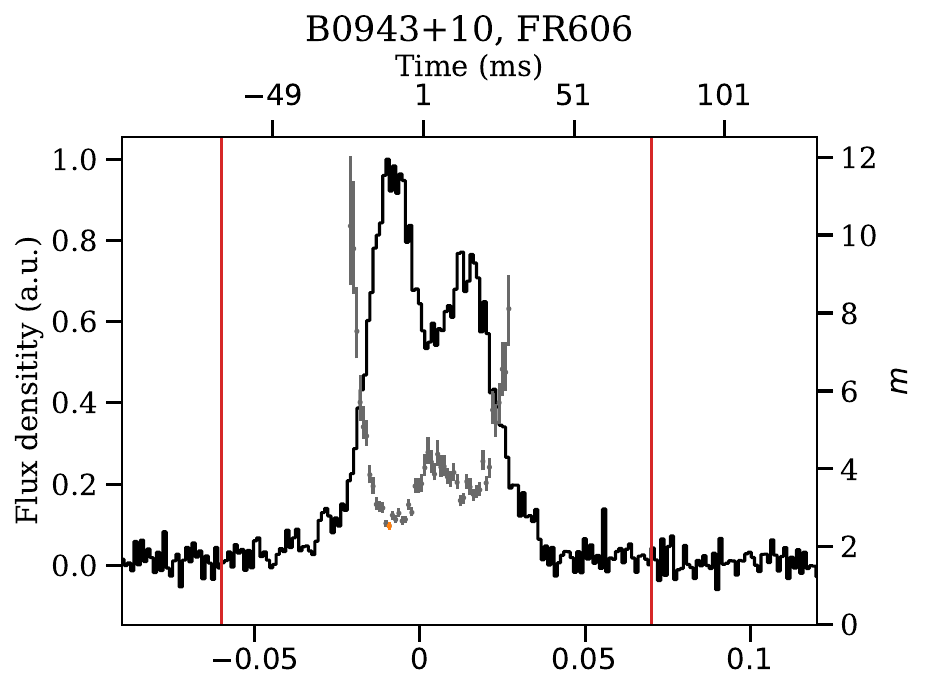}
  \includegraphics[width=0.32\textwidth]{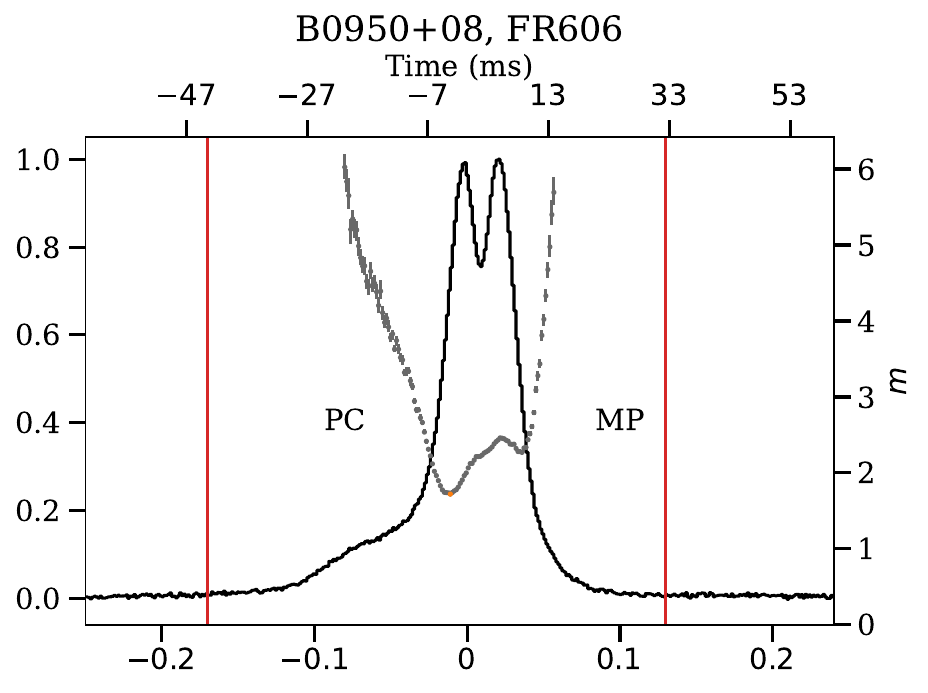}
  \includegraphics[width=0.32\textwidth]{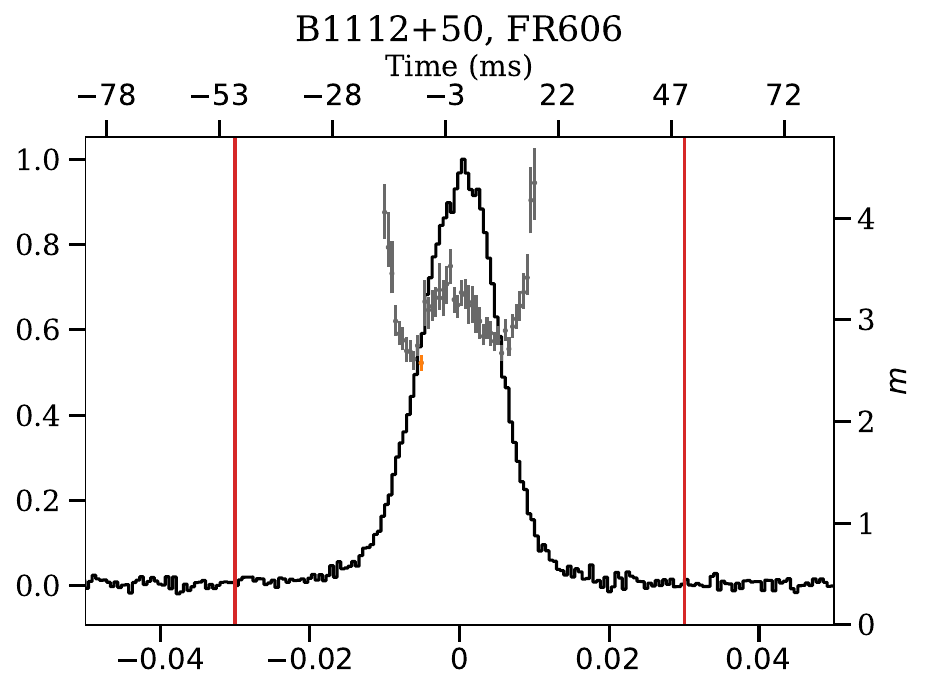}
  % gmrt
  \includegraphics[width=0.32\textwidth]{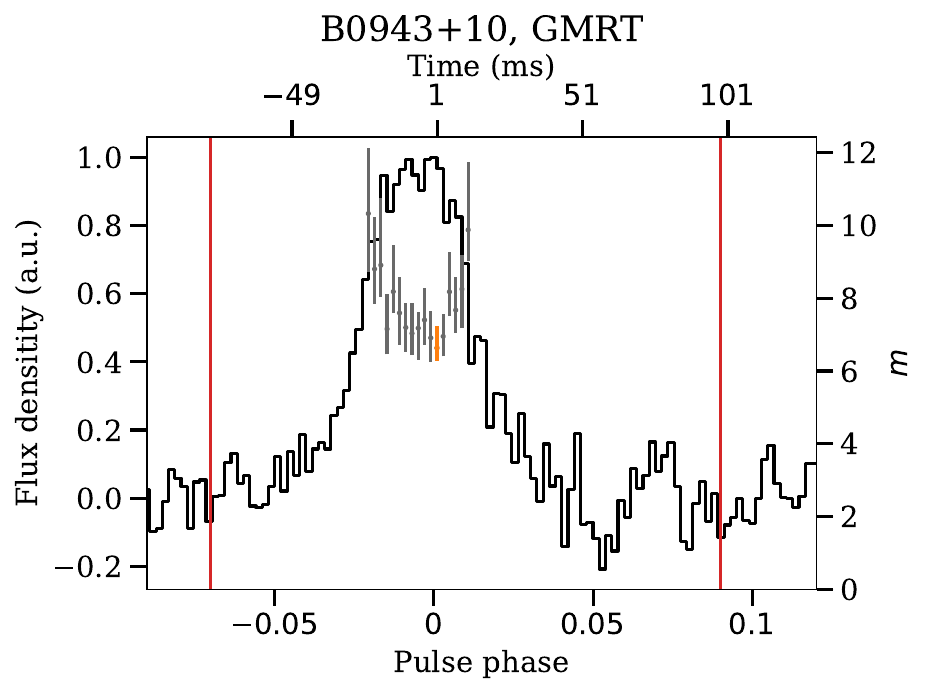}
  \includegraphics[width=0.32\textwidth]{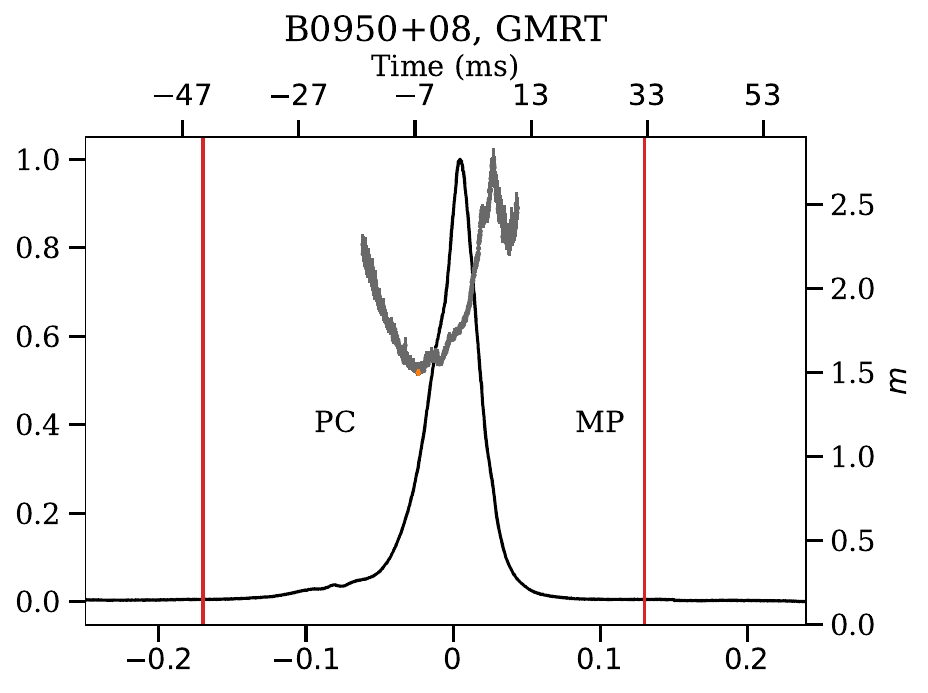}
  \includegraphics[width=0.32\textwidth]{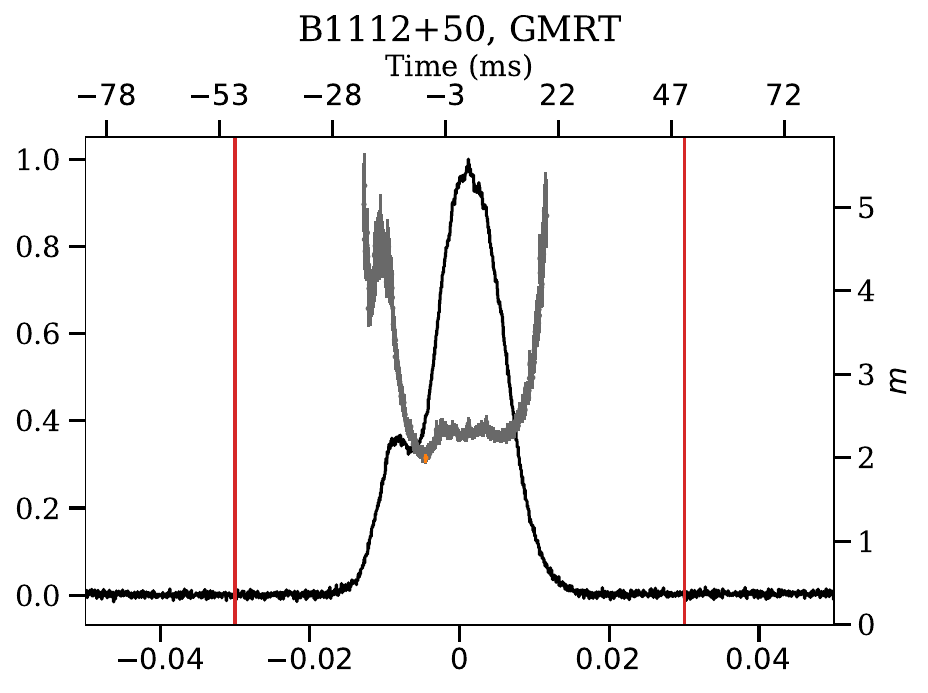}
  % nrt-l
  \makebox[0.32\textwidth]{\raisebox{0pt}[3cm][0pt]{}}
  \includegraphics[width=0.32\textwidth]{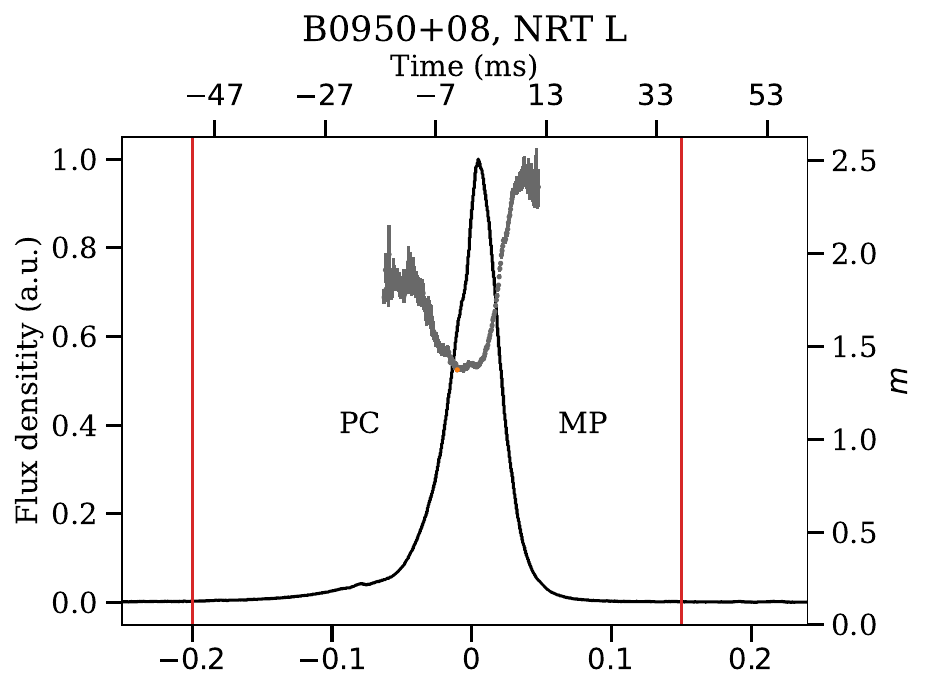}
  \includegraphics[width=0.32\textwidth]{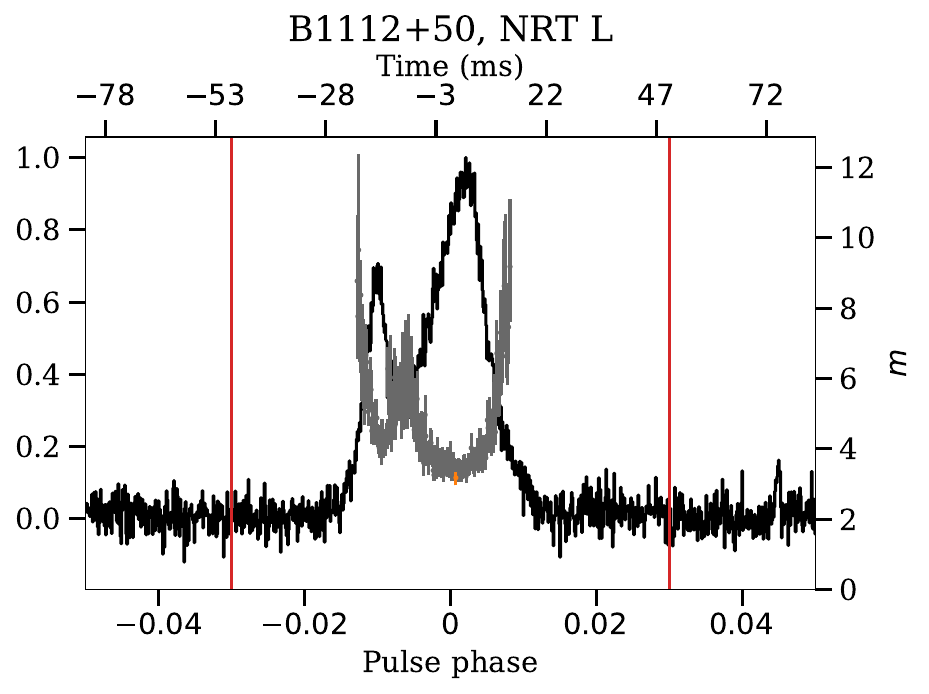}
  % nrt-s
  \makebox[0.32\textwidth]{\raisebox{0pt}[3cm][0pt]{}}
  \includegraphics[width=0.32\textwidth]{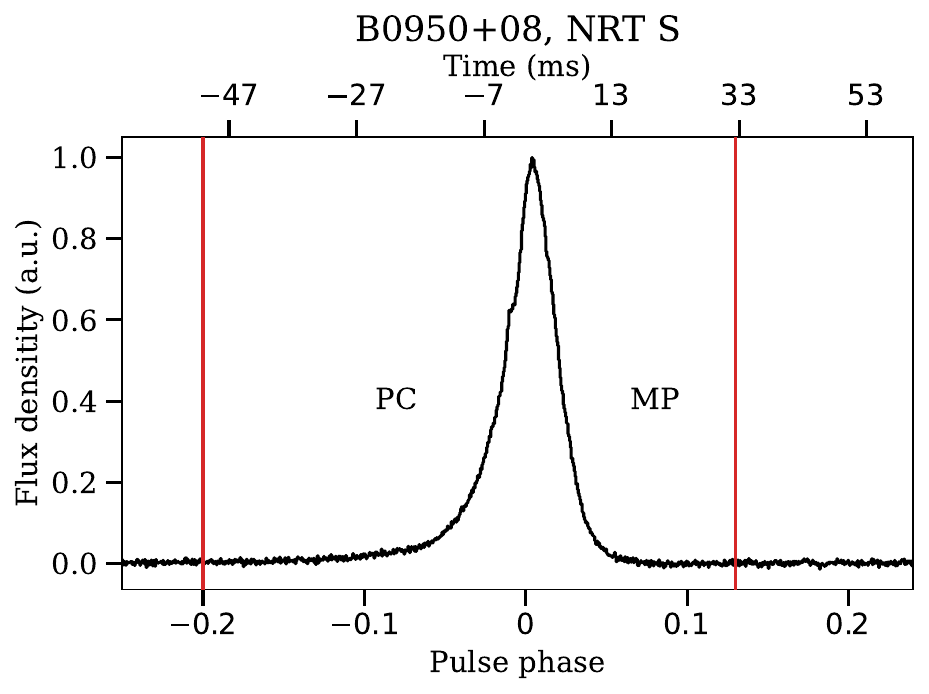}
  \makebox[0.32\textwidth]{\raisebox{0pt}[3cm][0pt]{}}
  \caption{Continuation of Fig.~\ref{fig:profiles1}.}
 \label{fig:profiles4}
\end{figure}

\clearpage

\begin{figure}
  \centering
  % B1133+16, B1237+25, B1822-09 MP
  % nenufar
  \includegraphics[width=0.32\textwidth]{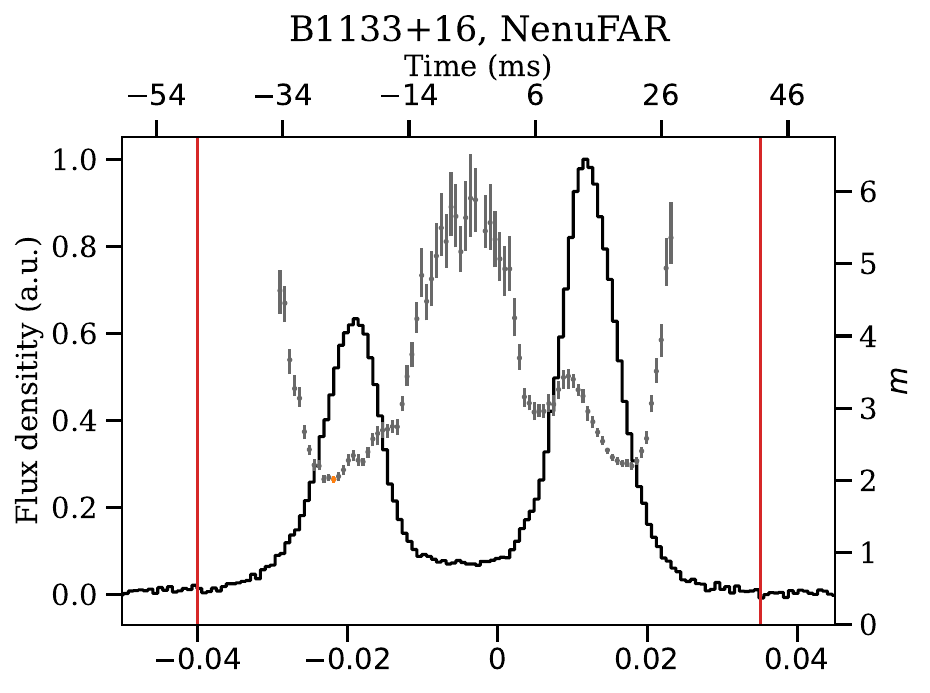}
  \includegraphics[width=0.32\textwidth]{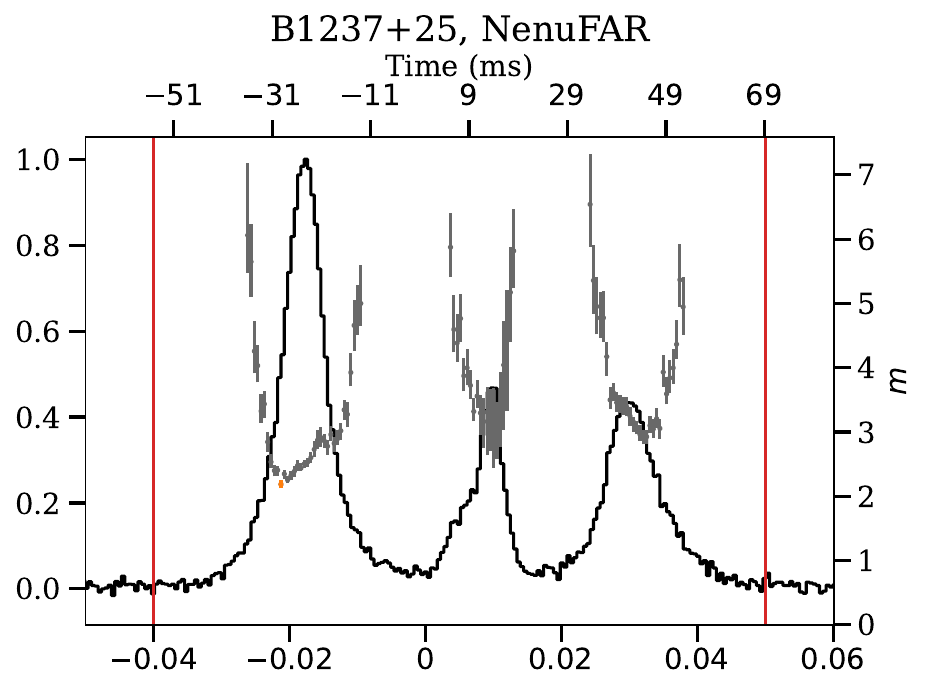}
  \includegraphics[width=0.32\textwidth]{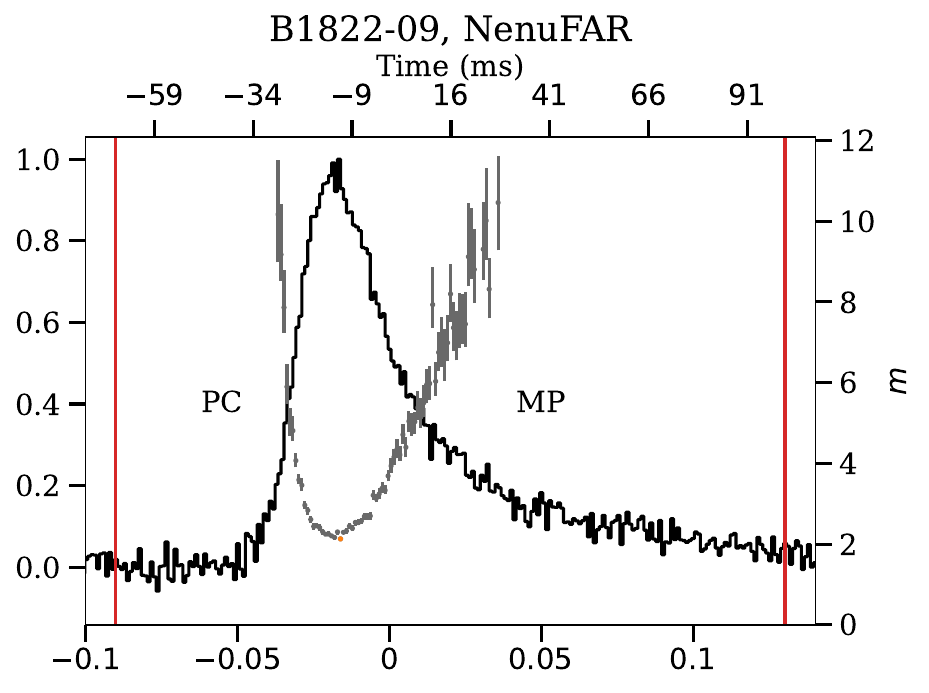}
  % fr606
  \includegraphics[width=0.32\textwidth]{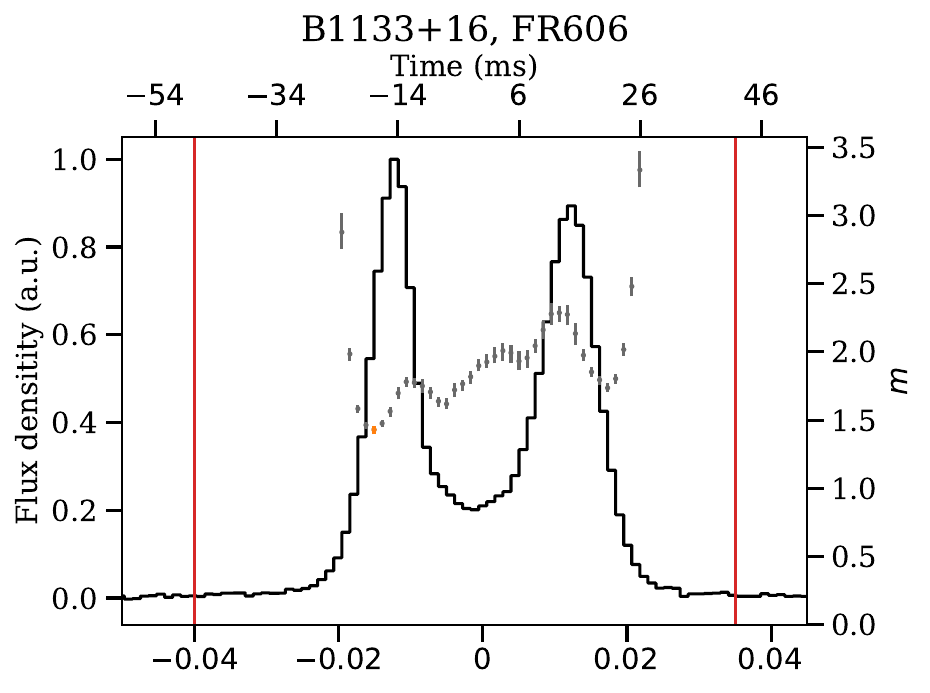}
  \includegraphics[width=0.32\textwidth]{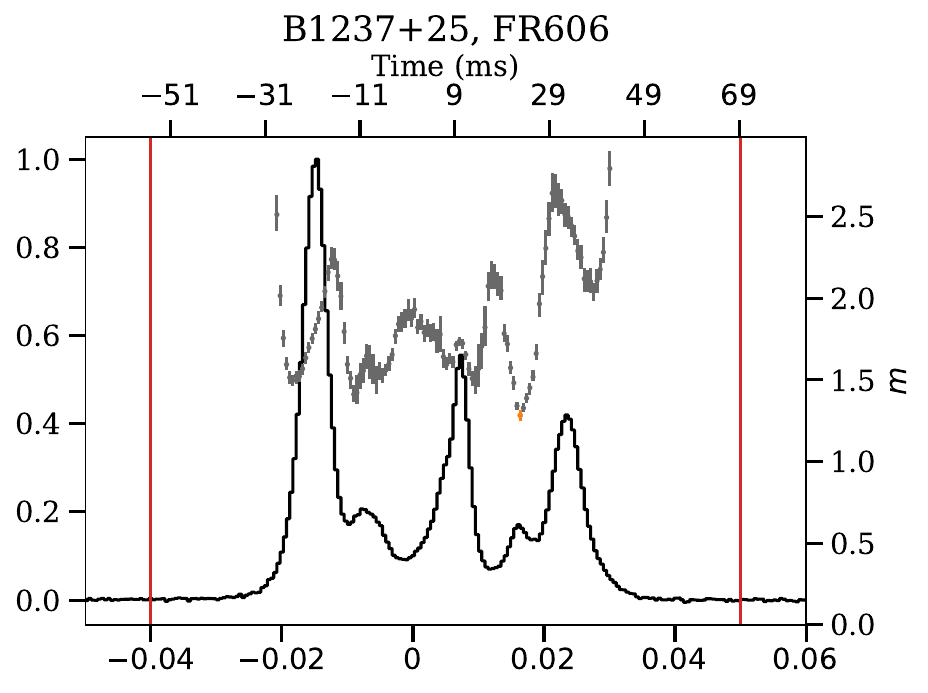}
  \includegraphics[width=0.32\textwidth]{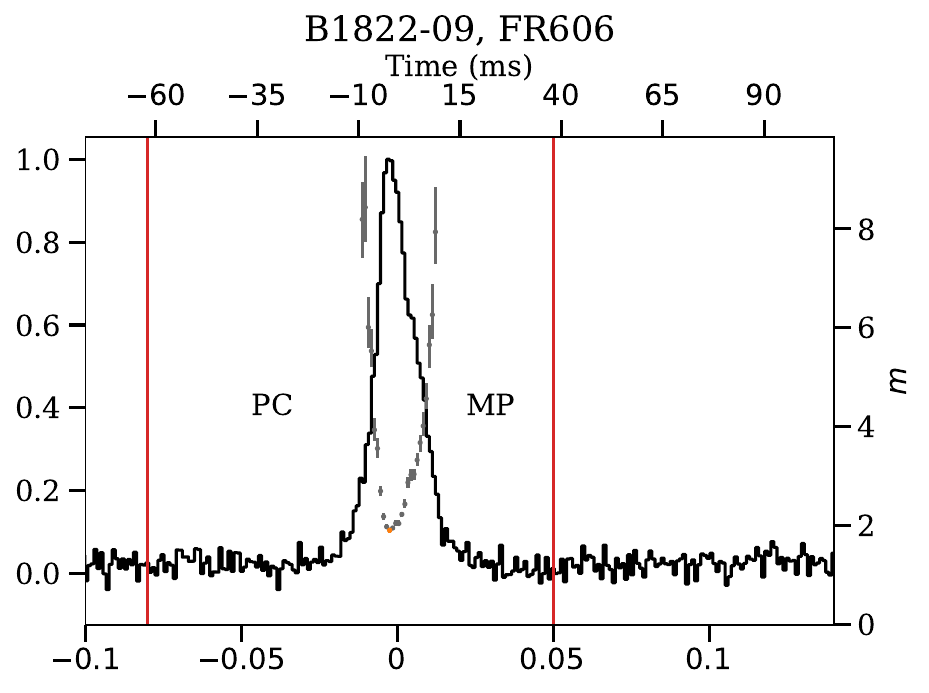}
  % gmrt
  \includegraphics[width=0.32\textwidth]{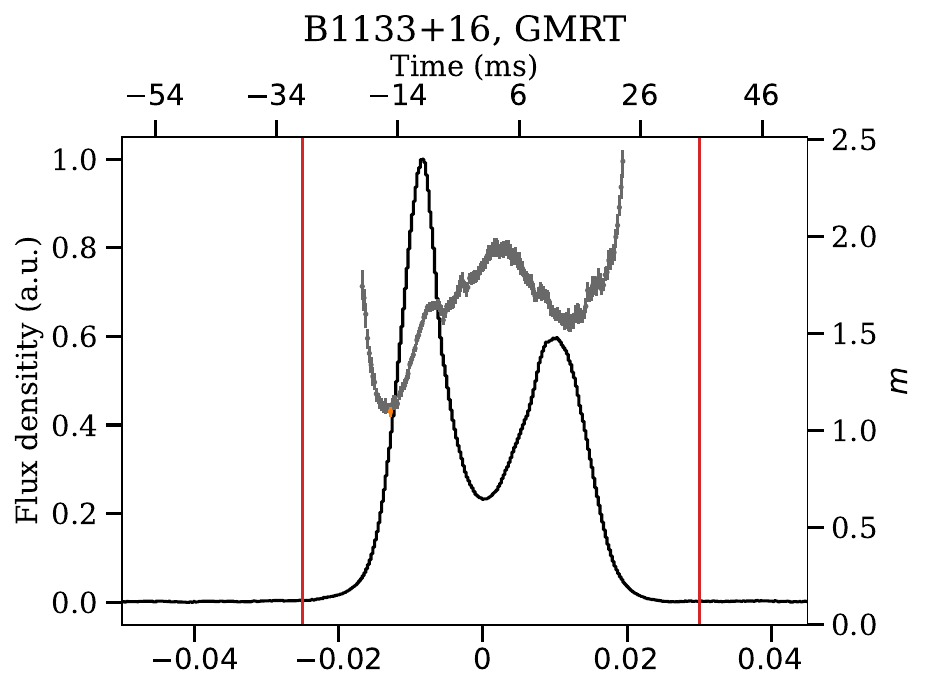}
  \includegraphics[width=0.32\textwidth]{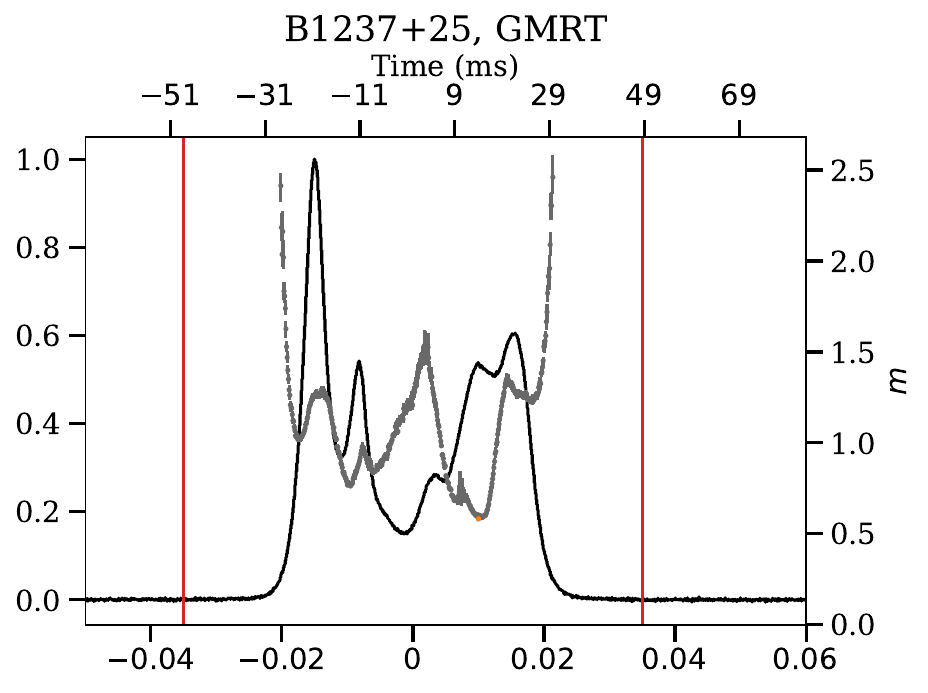}
  \includegraphics[width=0.32\textwidth]{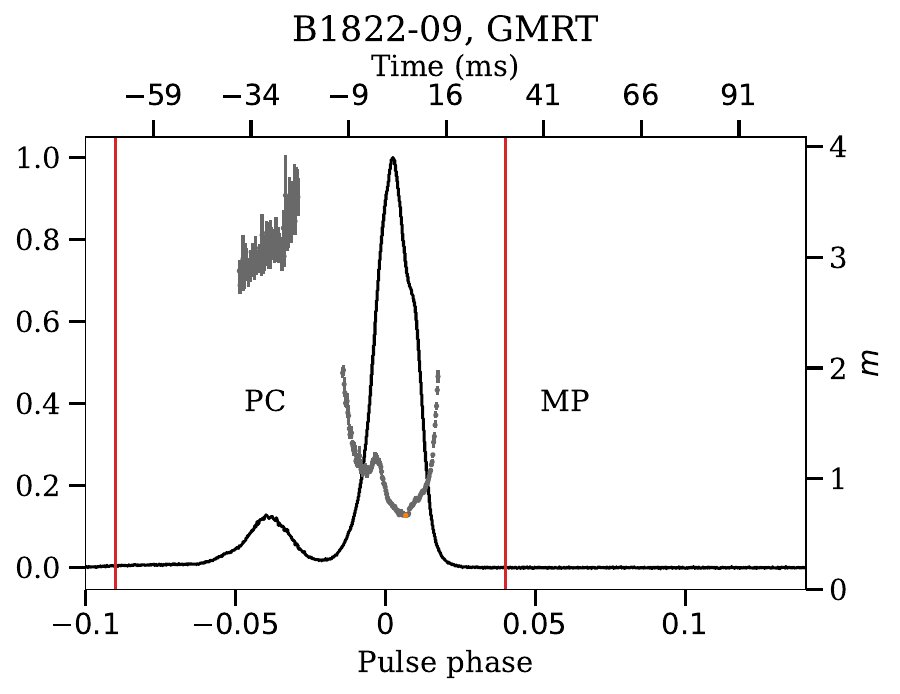}
  % nrt-l
  \includegraphics[width=0.32\textwidth]{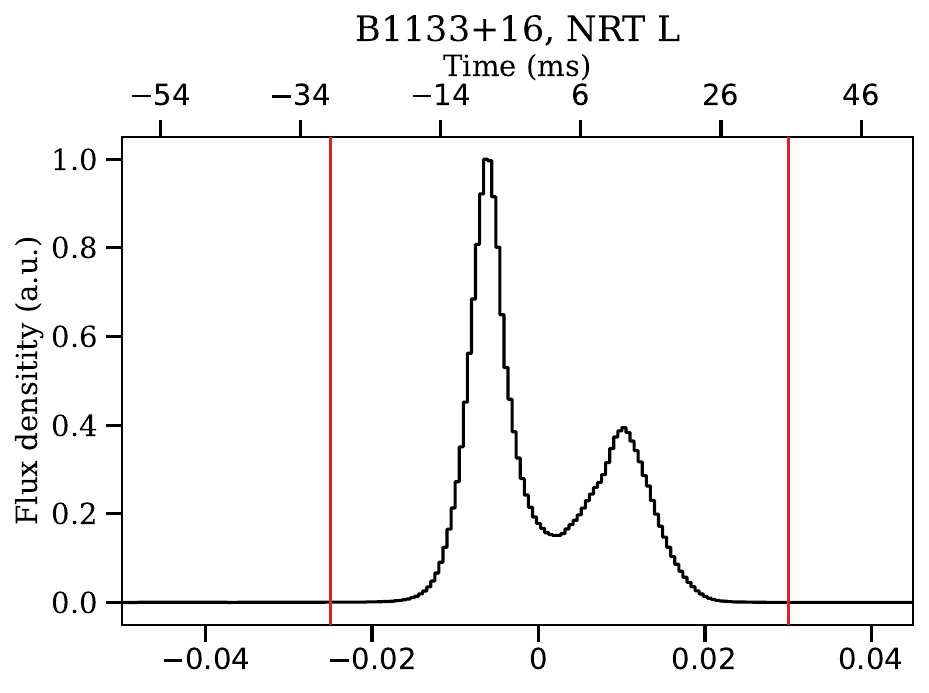}
  \includegraphics[width=0.32\textwidth]{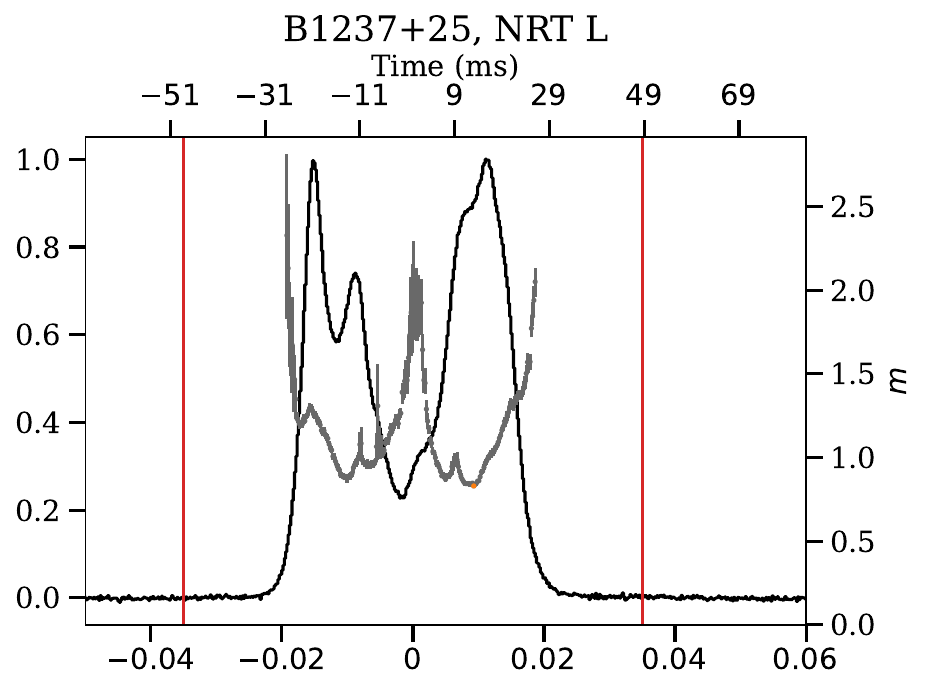}
  \makebox[0.32\textwidth]{\raisebox{0pt}[3cm][0pt]{}}
  % nrt-s
  \includegraphics[width=0.32\textwidth]{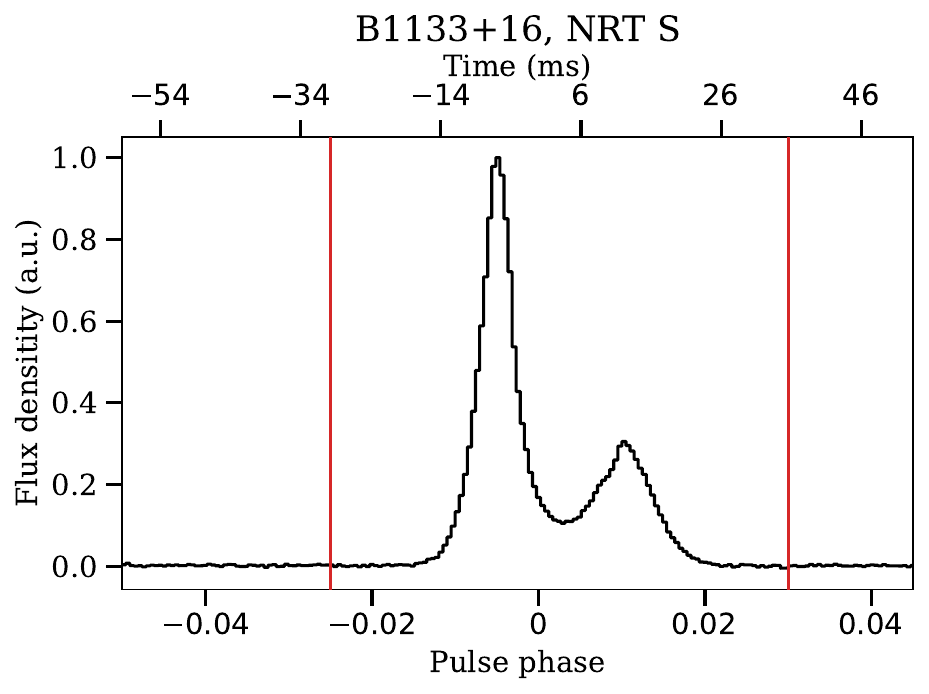}
  \includegraphics[width=0.32\textwidth]{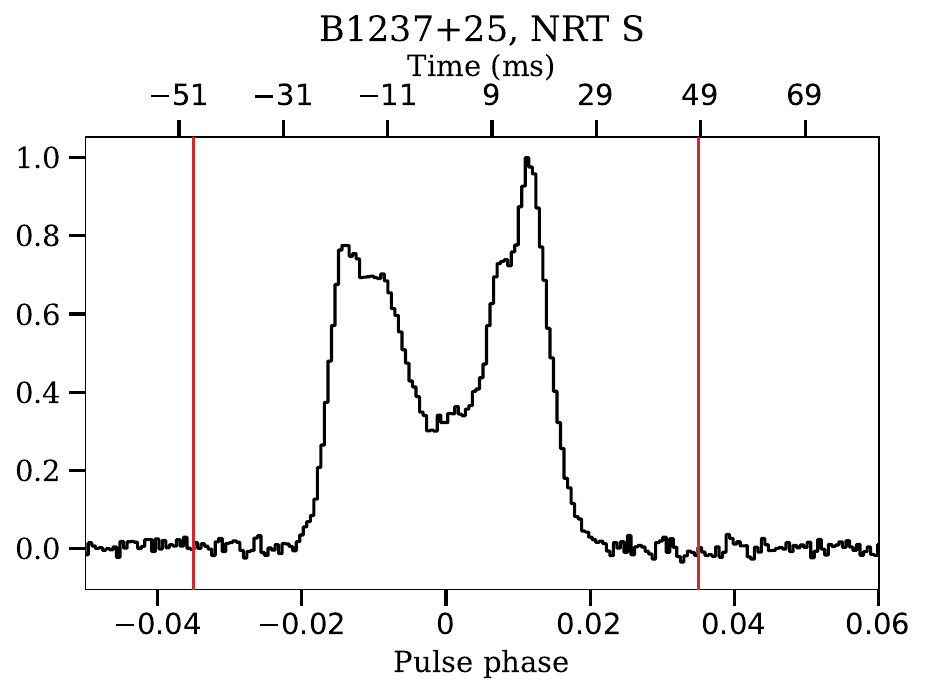}
  \makebox[0.32\textwidth]{\raisebox{0pt}[3cm][0pt]{}}
  \caption{Continuation of Fig.~\ref{fig:profiles1}.}
 \label{fig:profiles5}
\end{figure}

\clearpage

% revert to two-column layout
\twocolumn

\section{Pulse width estimators}
\label{ap:widthestimators}

\begin{figure}
  \centering
  \includegraphics[width=0.49\textwidth]{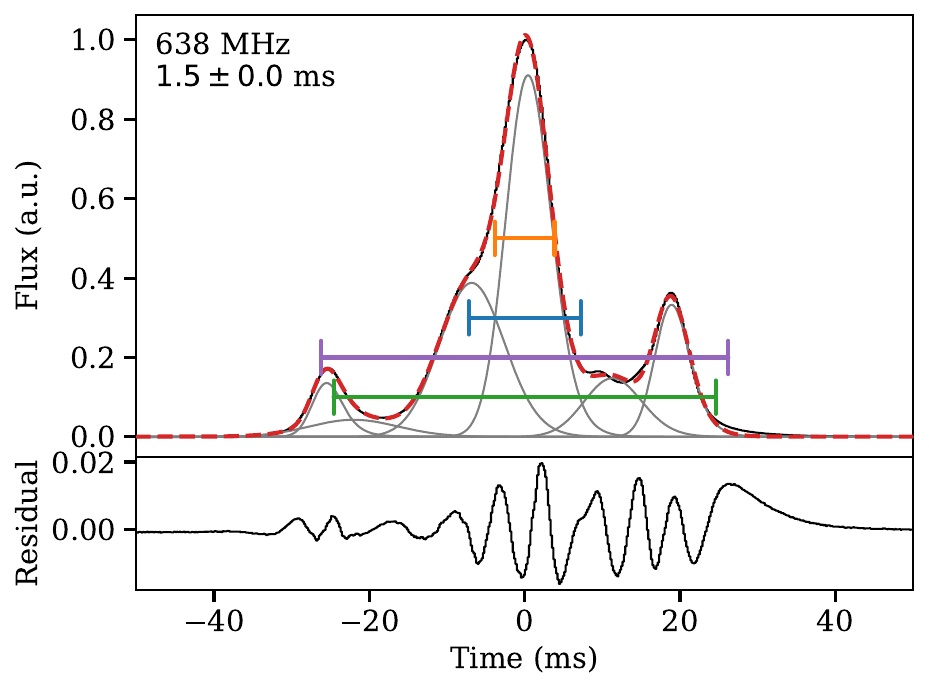}
  \caption{Example profile fit of PSR~B0329+54 at uGMRT frequencies with various pulse width estimators overlaid for illustration. The estimators are: $W_{50}$ (orange), $W_\text{eq}$ (blue), $W_\text{d4s}$ (purple), and $W_{10}$ (green). The width markers are centred at zero.}
 \label{fig:pulsewidthestimators}
\end{figure}

Estimating the width of a complex, multi-modal pulsar profile from real-world, noisy data is reasonably challenging. Our method included fitting a multi-component semi-analytical profile model consisting of the superposition of several exponentially-modified Gaussians to the data, where the exponential decay parameter represents the scatter broadening, and we simultaneously estimated the baseline noise. We then computed various pulse width estimators based on the best-fitting semi-analytical (noise-free) profile model.

Fig.~\ref{fig:pulsewidthestimators} shows an example profile fit with the pulse width estimators overlaid. The most obvious width estimators were the full width at half maximum (FWHM; $W_{50}$) and the full width at tenth maximum (FWTM; $W_{10}$). They are based on the profile's relative amplitude (amplitude-based) and disregard all profile components below a given level (50 or 10~\%). They are easy to compute and mildly sensitive to noise in the peak amplitude and near the threshold level. Low-amplitude profile components, such as conal `wings', are particularly problematic when they randomly exceed the threshold level due to noise and spectral evolution, imprinting significant jumps in pulse width. The boxcar equivalent width $W_\text{eq}$ is a pulse-energy or area-based estimator that incorporates all profile components, regardless of amplitude. The fact that it accounts for even the faintest profile wings makes it an ideal total pulse width estimator. On the downside, it requires careful baseline mean subtraction and is sensitive to noise at the peak and in the wings. Sloping or otherwise non-flat baselines are also problematic due to $W_\text{eq}$'s integrating nature. The second-moment or variance pulse width $W_\text{d4s}$ considers the pulse profile's entire tail or wing behaviour. Thus, it is an excellent estimator of the total pulse width, regardless of the amplitude threshold. It relies on accurate baseline mean subtraction and is mildly sensitive to baseline noise and changes in the mean centroid location due to its quadratic dependence on relative phase. Sloping baselines are problematic as well. Other estimators not considered here include percentage-of-power-based widths and those computed via autocorrelation or Fourier techniques.

Fig.~\ref{fig:pulsewidthestimators} illustrates that for a complex, multi-modal pulse profile with low-level wings $W_{50} < W_\text{eq} < W_{10} < W_\text{d4s}$. $W_{50}$ captures only the central core component, $W_{10}$ includes the leading and trailing (outer) conal components, and $W_\text{d4s}$ adds the profile tails below 10~\% amplitude for the fullest width estimation. As shown, the difference between $W_{10}$ and $W_\text{d4s}$ is small in practice, at least for non-degenerate profiles of slow pulsars.

Our profile modelling technique and pulse width estimators are implemented in the \texttt{scatfit} software as of version 0.5.2.

\clearpage

\end{appendix}

%%%%%%%%%%%%%%%%%%%%%%%%%%%%%%%%%%%%%%%%%%%%%%%%%%
%%%%%%%%%%%%%%% LIST OF OBJECTS %%%%%%%%%%%%%%%%%%
%%%%%%%%%%%%%%%%%%%%%%%%%%%%%%%%%%%%%%%%%%%%%%%%%%
\listofobjects

\end{document}